\documentclass[graybox]{svmult}

\usepackage{mathptmx}       % selects Times Roman as basic font
\usepackage{helvet}         % selects Helvetica as sans-serif font
\usepackage{courier}        % selects Courier as typewriter font
\usepackage{type1cm}        % activate if the above 3 fonts are
\usepackage{makeidx}         % allows index generation
\usepackage{graphicx}        % standard LaTeX graphics tool
\usepackage{multicol}        % used for the two-column index
\usepackage[bottom]{footmisc}% places footnotes at page bottom

\usepackage{sidecap}

\def\ltsima{$\; \buildrel < \over \sim \;$}
\def\lsim{\lower.5ex\hbox{\ltsima}}
\def\gtsima{$\; \buildrel > \over \sim \;$}
\def\gsim{\lower.5ex\hbox{\gtsima}}
\def\ga{\mathrel{\hbox{\rlap{\hbox{\lower4pt\hbox{$\sim$}}}\hbox{$>$}}}}
\def\la{\mathrel{\hbox{\rlap{\hbox{\lower4pt\hbox{$\sim$}}}\hbox{$<$}}}}

\newcommand{\HI}{\rm H\, {\scriptstyle I} }
\newcommand{\HII}{\rm H\, {\scriptstyle II} }

\newcommand{\HeIII}{\rm He\, {\scriptstyle III} }
\newcommand{\MHI}{\mathrm{H\,  I } }

\def\THI{\hbox{H~$\rm \scriptstyle I\ $}}
\def\THII{\hbox{H~$\rm \scriptstyle II\ $}} 
\def\THeI{\hbox{He~$\rm \scriptstyle I\ $}}
\def\THeII{\hbox{He~$\rm \scriptstyle II\ $}}
\def\THeIII{\hbox{He~$\rm \scriptstyle III\ $}} 

\newcommand{\apj}{ApJ}
\newcommand{\apjl}{ApJL}
\newcommand{\apjs}{ApJS}
\newcommand{\aj}{AJ}
\newcommand{\mnras}{MNRAS}
\newcommand{\nat}{Nature}
\newcommand{\pasj}{PASJ}
\newcommand{\araa}{ARA\&A}

\newcommand{\aap}{A\&A}

\newcommand{\prd}{PhRvD}
\newcommand{\physrep}{PhysRep}
\newcommand{\aaps}{A\&AS}

\newcommand{\nar}{New Astronomy Reviews}
\newcommand{\planss}{Planetary and Space Science}

\makeindex             % used for the subject index
\title*{The Epoch of Reionization}

\author{Saleem Zaroubi}
\institute{Saleem Zaroubi\\ Kapteyn Astronomical Institute, Landleven 12, 9747AD Groningen,
 The Netherlands\\
 and\\
  Physics Department,
 The Technion, 
 Haifa 32000, Israel \\
\email{saleem@astro.rug.nl}}

\begin{document}

\date{}

\maketitle

\abstract{
 The Universe's \emph{dark ages} end with the formation of the
first generation of galaxies. These objects start emitting ultraviolet
radiation that carves out ionized regions around them. After a sufficient
number of ionizing sources have formed, the ionized fraction of the gas
in the Universe rapidly increases until hydrogen becomes fully
ionized. This period, during which the cosmic gas went from neutral to
ionized, is known as the Universe's Epoch of Reionization . The Epoch of 
Reionization is related to many fundamental questions in cosmology, such as properties of
the first galaxies, physics of  (mini-)quasars, formation of very metal-poor stars and a slew of
other important research topics in astrophysics.  Hence uncovering it
will have far reaching implications on the study of structure
formation in the early Universe. This chapter reviews the current
observational evidence for the occurrence of this epoch, its key
theoretical aspects and main characteristics, and finally the various
observational probes that promise to uncover it. A special emphasis is
put on the redshifted 21~cm probe, the various experiments that are
currently being  either built or designed,  and what we can learn from them about the Epoch of Reionization.  
}

\section{Introduction}
\label{sec:intro}

The formation of the first galaxies marks a major transition in the evolution of structure
in the Universe.
These same galaxies with their zero metallicity Population III stars,  second generation Population II stars,
and  black hole driven sources (e.g., mini-quasars, x-ray binaries, etc.)
transformed the intergalactic medium from neutral to ionized.
This process, known as the Epoch of Reionization (EoR), is the central topic discussed
in this chapter.

As mentioned in  chapter \# 1 [by A. Loeb in this book], 
about 400,000 years after the Big Bang, the Universe's density
 decreased enough so that the temperature fell below 3000 K, allowing ions and electrons to
(re)combine into neutral hydrogen and helium -- the fraction of heavier elements wasf
negligible. Immediately afterwards, photons decoupled from baryons and
the Universe became transparent, leaving a relic signature known as
the cosmic microwave background (CMB) radiation. This event ushered
the Universe into a period of darkness, known as the \emph{dark
ages}. 

The \emph{dark ages} ended about 400 million years later, when the
first galaxies formed and start emitting ionizing radiation. 
Initially during the EoR, the intergalactic medium (IGM) is neutral except in regions
surrounding the first objects.  However, as this reionization progresses, an
evolving patchwork of neutral (\THI) and ionized hydrogen (\THII) regions
unfolds. After a sufficient number of UV-radiation emitting objects formed, 
the temperature and the ionized fraction of the gas in the Universe increase rapidly
 until eventually the ionized regions permeate to fill the whole Universe
\cite{barkana01, loeb01, bromm04, ciardi05, choudhury06,
furlanetto06a, morales10}.  

\begin{figure*}
\centering
\includegraphics[width=1.0\textwidth]{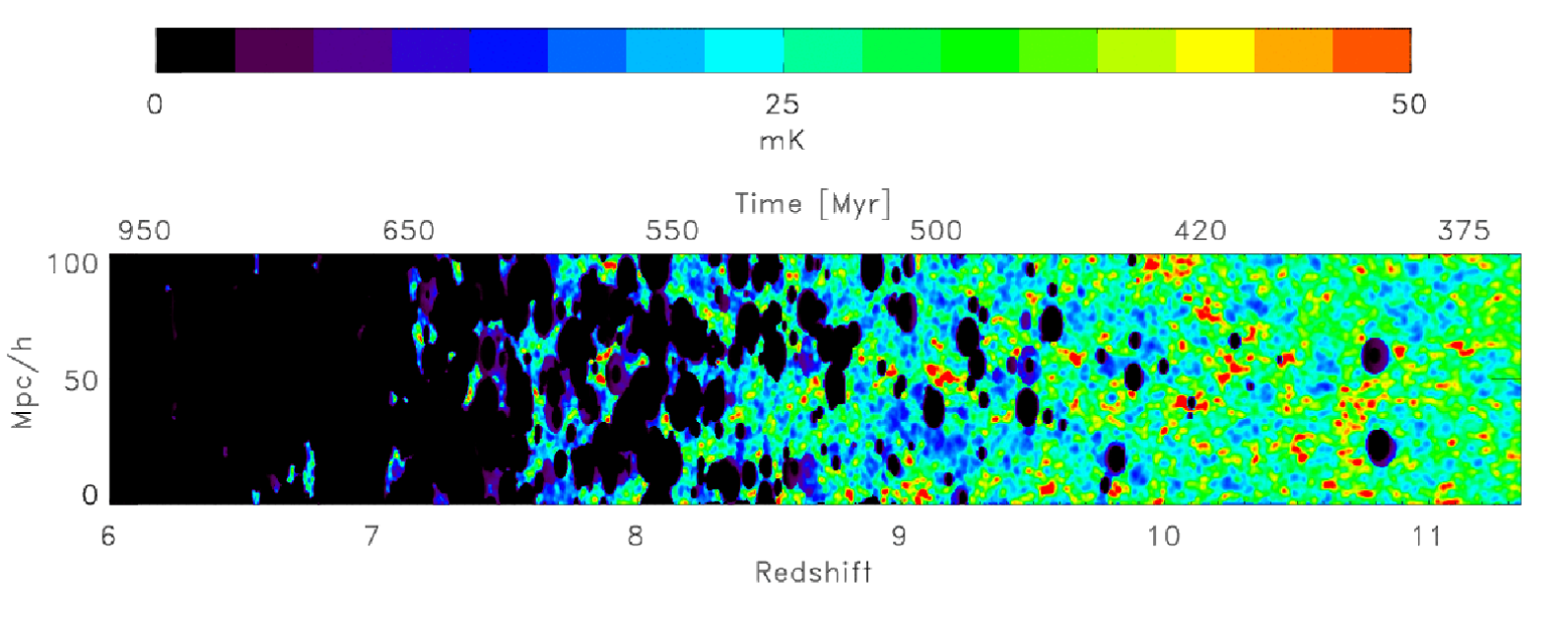} 
% \vspace{.2cm}
\caption{
This figure shows a slice through redshift of the 21 cm radiation in
which the reionization process progresses through the volume of a
cosmological simulation with radiative transfer \cite{thomas09}.
}
\label{fig:EoRslice}
\end{figure*}

The current constraints strongly suggest that the EoR
roughly occurs within the redshift range of $z \sim [6-15]$.
Figure~\ref{fig:EoRslice} shows a space-redshift slice of a
simulation of the progression of  reionization with time and how it
appears in 21~cm brightness temperature, which is proportional
to the density of neutral hydrogen (see section~\ref{sec:21cmProbe}).
 At high redshifts most of the gas is neutral,
hence, the signal is mostly sensitive to cosmological density
fluctuations, whereas at lower redshifts ionization bubbles start to
appear until they fill the whole Universe \cite{barkana01}.

The EoR is a watershed epoch in the history of the Universe. Prior to
it, the formation and evolution of structure was dominated by dark matter 
alone, while baryonic matter played a marginal role. The EoR marks the
transition to an era in which the role of cosmic gas in the formation
and evolution of structure became prominent and, on small scales, even
dominant.

The details of the reionization scenario I have laid out are yet to be
clarified.  For example, it is not known what controls the formation of the first
objects and how much ionizing radiation they produce, or how the
ionization bubbles expand into the intergalactic medium and what they ionize
first, high-density or low-density regions?. The answer to these
questions and many others that arise in the context of studying the
EoR needs knowledge of fundamental issues in cosmology, galaxy
formation, quasars and the physics of very metal poor stars; all including foremost
research in topics in modern astrophysics. Substantial theoretical
and observational efforts are currently dedicated to understanding the physical processes
that trigger this epoch and govern its evolution, and ramifications
on subsequent structure formation (c.f., \cite{barkana01,
bromm04, ciardi05, choudhury06, furlanetto06a}). However, despite
the pivotal role played by the EoR in cosmic history, observational
support for the proposed scenarios  is very scarce, and when available, is indirect and
model dependent.

\begin{figure}
\centering
\includegraphics[width=0.9\textwidth]{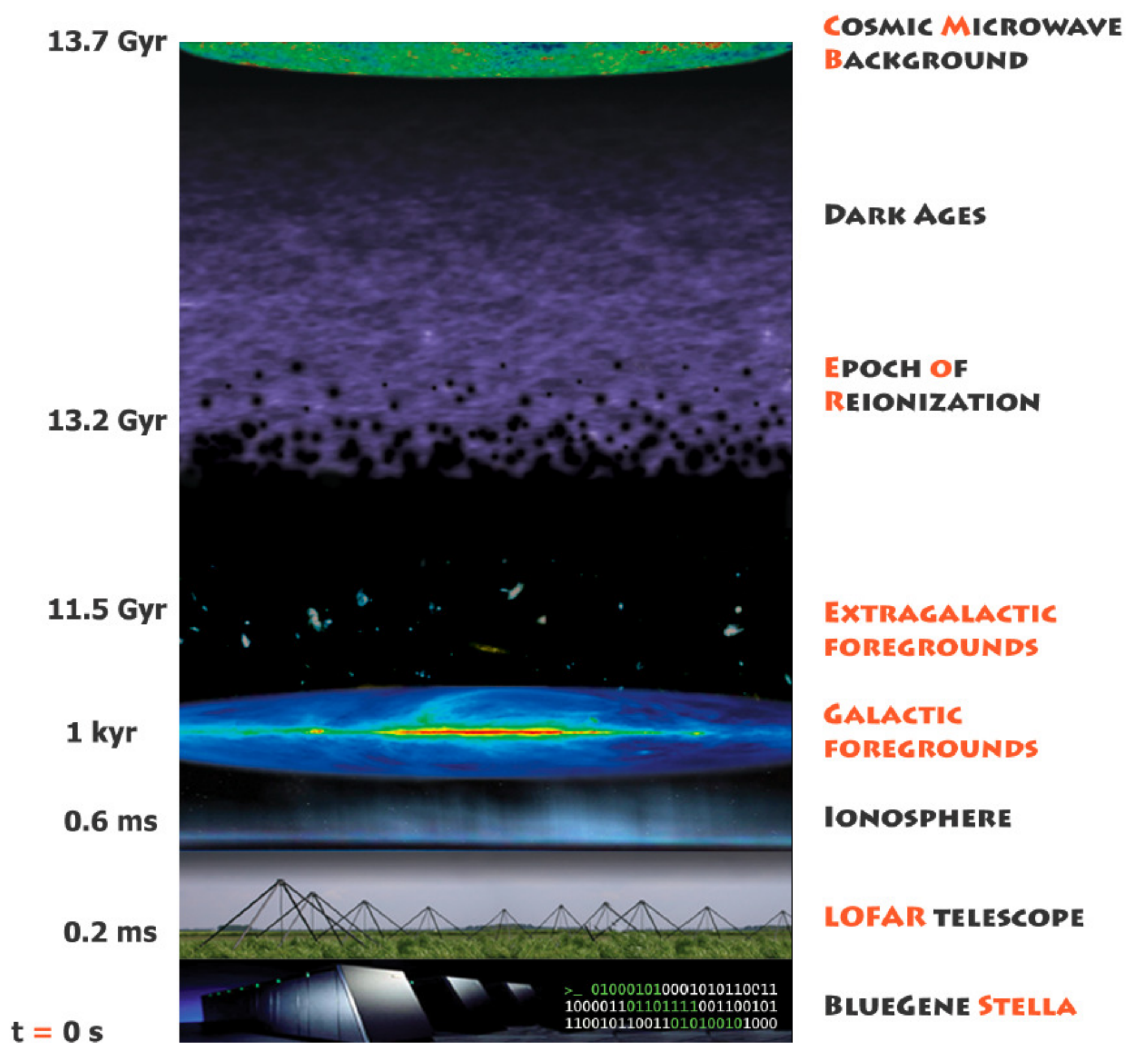}
\caption{{This figure shows a sketch of the likely development of the EoR.  About 500,000 years after the Big Bang  ($z \sim 1000$) hydrogen recombined and remained neutral for a few hundred million years during the \emph{dark ages}.  At a redshift, $z \sim 15$,
the first stars, galaxies and quasars began to form, heating and reionizing the hydrogen gas.  The neutral IGM can be observed with
LOFAR up to $z\approx 11.5$ through its redshifted 21cm spin-flip transition. However, many atmospheric, galactic and extra-galactic emission contaminate the 21~cm signal. \vspace{3 cm}}}
 \label{fig:EoRHist}
 \end{figure}
 
In principle, there are many different ways to observationally probe the EoR.
In this contribution, I mainly focus on the redshifted 21cm 
emission line from neutral hydrogen at high
redshifts. This is one of the most promising techniques for studying
the \emph{dark ages} and the EoR.
To date, there are a number of telescopes dedicated to measure this
faint radiation. In the short term, these consist of: The Low Frequency Array (LOFAR), 
the Murchison Widefield Array (MWA), 
Precision Array to Probe Epoch of Reionization (PAPER) and Giant Metrewave Radio Telescope (GMRT), 
while, on a somewhat longer time scales the Square
Kilometer Array (SKA).  One of the most challenging tasks in studying
the EoR is to extract and identify the cosmological signal from the data
 and interpret it correctly.  This is because the detectable signal in the
frequency range relevant to the EoR is composed of a number
of components -- the cosmological EoR signal, extragalactic and Galactic foreground, 
ionospheric distortions, instrumental response and noise --
each with its own physical origin and statistical
properties.

Figure~\ref{fig:EoRHist} shows a sketch of the likely evolution of reionization from the neutral hydrogen 
point of view. The figure emphasizes the other non-cosmological effects that are seen with the 21~cm experiments, e.g., foreground,
ionosphere and instrumental effects. The radio antennas seen at the bottom are LOFAR's Low Band Antennas.

 In this chapter I discuss various observational and theoretical aspects of the Epoch of Reionization. 
 In section~\ref{sec:ObsEvidence} the current observational scene is reviewed, specifically  focusing on
 the CMB data and the Lyman~$\alpha$ forest spectra. In sections~\ref{sec:EoRProcess} and \ref{sec:21cmProbe}, 
we discuss, respectively, the physics of the reionization process and the
 21~cm line transition and how it could be used to probe reionization. The redshifted 21~cm experiments their 
 potentials
  and the challenges  are discussed in section~\ref{sec:21cmObs}. Extraction and quantification of the information 
  stored in the redshifted 21~cm data using various statistics  is discussed in section~\ref{sec:statistics}. This chapter concludes with a brief summary
 (section~\ref{sec:summary}).

\section{Observational Evidence for Reionization}
\label{sec:ObsEvidence}

To date, the majority of observations related to the EoR provide
weak and model dependent constraints on reionization.
However, there are currently  a number of observations which could impose 
strong  constraints on reionization models, as discussed below.
It should be noted however that none of these observations constrains
the EoR evolution in detail.
 
 %%%%%%%%%%%%%% Lyman-alpha section %%%%%%%%%%%%%%%%%%%%

\subsection{The Lyman~$\alpha$ forest at $z\approx2.5-6.5$}

The state of the intergalactic medium (IGM) can be studied through the analysis of the Lyman-$\alpha$
forest. This is an absorption phenomenon seen  in the spectra
of background quasi-stellar objects (QSOs). The history of this field goes back
to 1965 when a number of authors \cite{gunn65, scheuer65}
predicted that an expanding Universe, homogeneously filled with gas, will
produce an absorption trough due to neutral hydrogen, known as the 
Gunn-Peterson trough, in the spectra of distant QSOs bluewards of the 
Lyman-$\alpha$ emission line of the quasar. That is,
 the quasar flux will be absorbed at the UV resonance line frequency of 1215.67~\AA\ . 
  Gunn \& Peterson  \cite{gunn65} found
such a spectral region of reduced flux, and used this measurement to
put upper limits on the amount of intergalactic neutral hydrogen. The
large cross-section for the Lyman~$\alpha$ absorption makes this
technique very powerful for studying gas in the intergalactic medium.

In the last 15 years two major advances occurred. The first was 
the development of high-resolution echelle
spectrographs on large telescopes (e.g., HIRES on the Keck and UVES on
the Very Large Telescope) that provided data of
unprecedented quality.  The second was the emergence of a theoretical 
paradigm within the context of cold dark matter (CDM) cosmology
that accounts for all the features seen in these systems
(e.g. \cite{bi92, cen94,  hernquist96,  machacek00, miralda-Escude96, theuns98,
 zhang95, zhang97}). According to this paradigm, the absorption is produced by
volume filling photoionized gas that contains most of the baryons
at redshifts at $z \sim 3-6$ and resides in mildly non-linear overdensities.

\begin{figure} \centering
\includegraphics[width=0.9\textwidth]{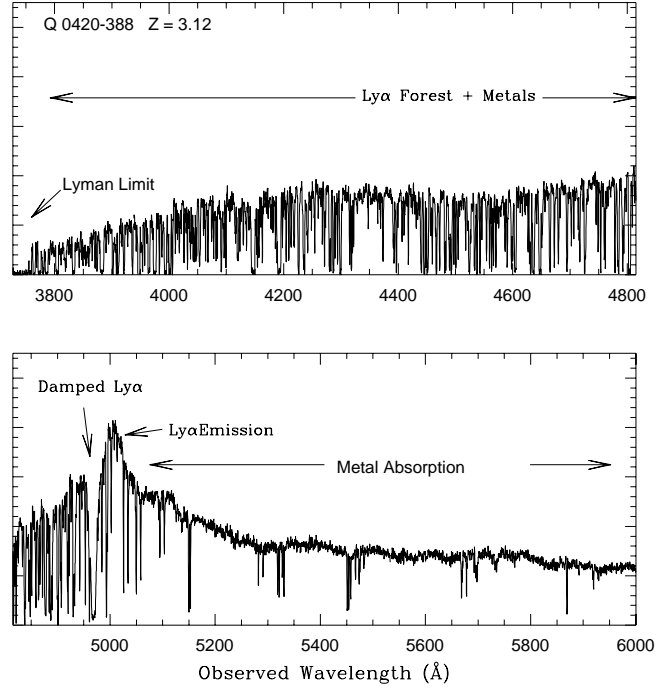}
\vspace{-2.5cm}
\caption{ High resolution spectrum of the $z=3.12$ quasar Q0420-388 obtained with the Las Campanas
echelle spectrograph by J. Bechtold and S. A. Shectman. The two panels cover the
whole wavelength range of the spectrum. The Lyman~$\alpha$  forest is clearly indicated 
in the upper panel of the figure, bluewards of the quasar rest frame Lyman~$\alpha$ emission feature. Remember, at the rest frame 
this feature should have a wavelength of 1215.67~\AA\ but since it is redshifted by a factor $1+z$ it appears at
a wavelength of about 5000~\AA. The Figure is courtesy of Jill Bechtold and appeared in~\cite{bechtold03}.
}
\label{fig:Lyaexample}
\end{figure}

 Figure~\ref{fig:Lyaexample} shows a typical example of the Lyman~$\alpha$ forest
seen in the spectrum  of the $z=3.12$ quasar Q0420-388. An interesting feature of
such spectra is the density of weak absorbing lines which increase with redshift due to
the expansion of the Universe. In fact, at redshifts above 4, the density of the absorption 
features become so high that it is hard to define them as separate absorption features.
Instead, one sees only the flux in between the absorption minima which appears as if they 
are emission rather than absorption lines.

The Lyman~$\alpha$ forest has turned out to be a treasure trove for studying the intergalactic medium and its properties
in both low and high density regions. In particular, it is very sensitive to the neutral hydrogen column density
and hence, to the neutral fraction as a function of redshift along the line of sight. In the following, we 
demonstrate how one could constrain the neutral fraction of hydrogen from the forest
and what the values obtained from the data are. For a review on the Lyman~$\alpha$ forest
the reader is referred to \cite{rauch98}.

We need to calculate the optical depth for absorption of Lyman~$\alpha$ photons.
A photon emitted by a
distant quasar with an energy higher than 10.196 eV  is continuously redshifted as it travels through the
intergalactic medium until it reaches the observer. At some
intermediate point the photon is redshifted to around 1216 \AA\ in
the rest-frame of the intervening medium, which may contain neutral
hydrogen. It can then excite the Lyman~$\alpha$ transition and be
absorbed. Let us consider a particular line of sight from the observer
to the quasar. The optical depth $\tau_\alpha$ of a photon is related
to the probability of the photon's 
transmission $e^{-\tau_\alpha}$. At a given
observed frequency, $\nu_0$, the Lyman~$\alpha$ optical depth is given by

\begin{equation} 
\tau_\alpha(\nu_0) = \int_{O}^{Q} n_{\HI} \sigma_\alpha dl/(1+z),
 \end{equation} 
 where $l$ is the comoving radial coordinate of some
intermediate point along the line of sight, $z$ is the redshift and
$n_{\HI}$ is the proper number density of neutral hydrogen at that
point. The limits of the integration, $O$ and $Q$, are the comoving
distance between the observer and the quasar, respectively.  The Lyman~$\alpha$
absorption cross section is denoted by $\sigma_\alpha$. It is a
function of the frequency of the photon, $\nu$, with respect to the rest-frame
of the intervening \THI at position $l$. The cross section is peaked when $\nu$ is equal to
the Lyman~$\alpha$ frequency $\nu_\alpha$. 
 The frequency $\nu$ is related to the observed frequency $\nu_0$ by
 $\nu=\nu_0(1+z)$, where  $1 + z$ is the redshift factor due
to the uniform Hubble expansion alone at the same position. Note that 
for the sake of simplicity here we ignore peculiar velocity effects. 

 Using $dl= c dt/a$, where $a$ is the Hubble scale factor and $t$ is
the proper time and Friedmann equation for a flat Universe with cosmological constant, we have,
 \begin{equation}
\tau_\alpha=\int \sigma_\alpha(\nu) n_{\HI} \frac{c
H_0^{-1} dz } {(1+z) \sqrt{\Omega_m(1+z)^3+\Omega_\Lambda}}.
 \end{equation}
 This optical depth should also
depend on the Lyman~$\alpha$ line profile function but here we assume that it is basically
a $\delta$-function centered at the frequency $\nu$. Considering $n_{\HI}= n_H x_{\HI}$, where $x_{\HI}$ is the neutral fraction of hydrogen, and integrating over this equation, one obtains
the following result:
 \begin{equation} 
 \frac{n_{\HI}}{n_H}= x_{\HI}\approx 10^{-4}\Omega_m^{1/2} h (1+z)^{\frac{3}{2}} \tau_\alpha.
 \end{equation} 
 Since the Lyman~$\alpha$ features mostly show mild
absorption probability ($\tau_\alpha \lsim 1$) this equation clearly
implies that at the mean density of the Universe at $\tau_\alpha$ of
about one the ionized fraction is on the order of $10^{-4}$.
Therefore, the fact that we observe the Lyman~$\alpha$ forest at all means
that the Universe is highly ionized at least until $z\approx 6$. This is the
most reliable and robust evidence that the Universe has in fact reionized.

Another important evidence relevant for reionization comes from high resolution spectroscopy 
of high redshift Sloan Digital
Sky Survey (SDSS) quasars \cite{fan03,fan06}. The SDSS has discovered about 19 QSOs
with redshifts around 6 that are powered by black holes with masses on the order of $10^9M_\odot$.
In a follow up observations with 10 meter class telescopes Fan et al. \cite{fan03,fan06} 
were able to obtain high resolution spectra of these objects.

\begin{figure}
\centering
\includegraphics[width=0.8\textwidth]{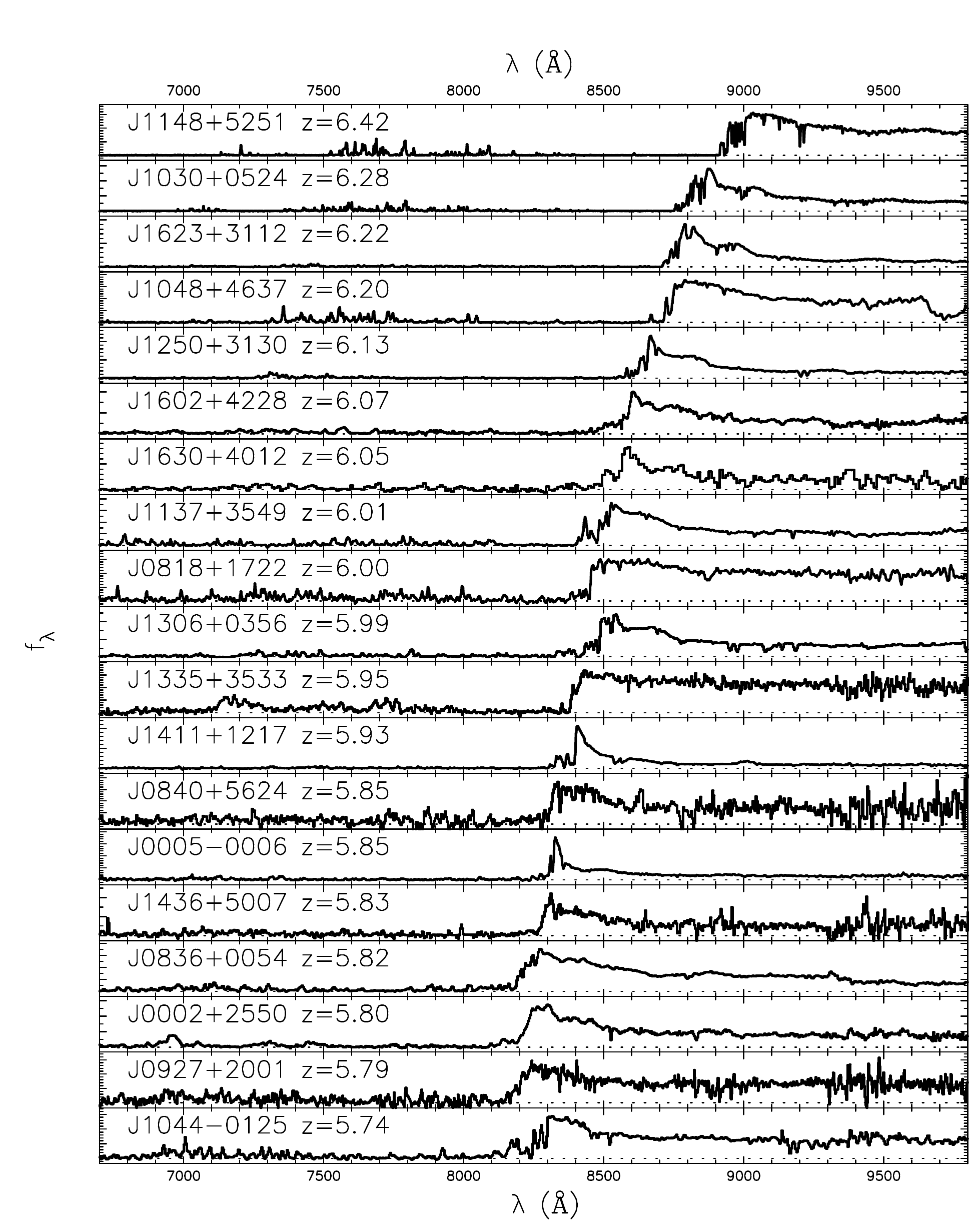}
\caption{ Spectra for high redshift SDSS quasars.
The Gunn-Peterson trough bluewards of the QSO Lyman~$\alpha$
emission that is clearly apparent in the highest redshift ones indicates
that the Universe has become somewhat more neutral at these redshifts. A similar behavior is also
seen bluewards of the QSO Lyman~$\beta$ region of the same spectra. The actual amount of increase in
neutral hydrogen implied by these spectra is not clear \cite{fan06}. \vspace{3 cm}}
\label{fig:Lya2}
\end{figure}

Fig.~\ref{fig:Lya2} shows the spectra of these high redshift quasars \cite{fan03, fan06}. 
Notice the complete absence of structure that some  of these spectra exhibit bluewards 
of the quasar Lyman~$\alpha$ restframe emission, especially those with redshift $z \gsim 6$. 
This is normally attributed to an increase in $\tau_\alpha$ as a
result of the decrease in the ionized fraction of the Universe. Notice also, that although
the trend with redshift is clear, it is by no means monotonic. For example, quasar J1411+3533 at $z=5.93$
shows an ``emptier" trough relative to quasars J0818+1722 at $z=6$. Such trend might be indicating
a more patchy ionization of the IGM at such redshifts.

\begin{figure} \centering
\includegraphics[width=0.8\textwidth]{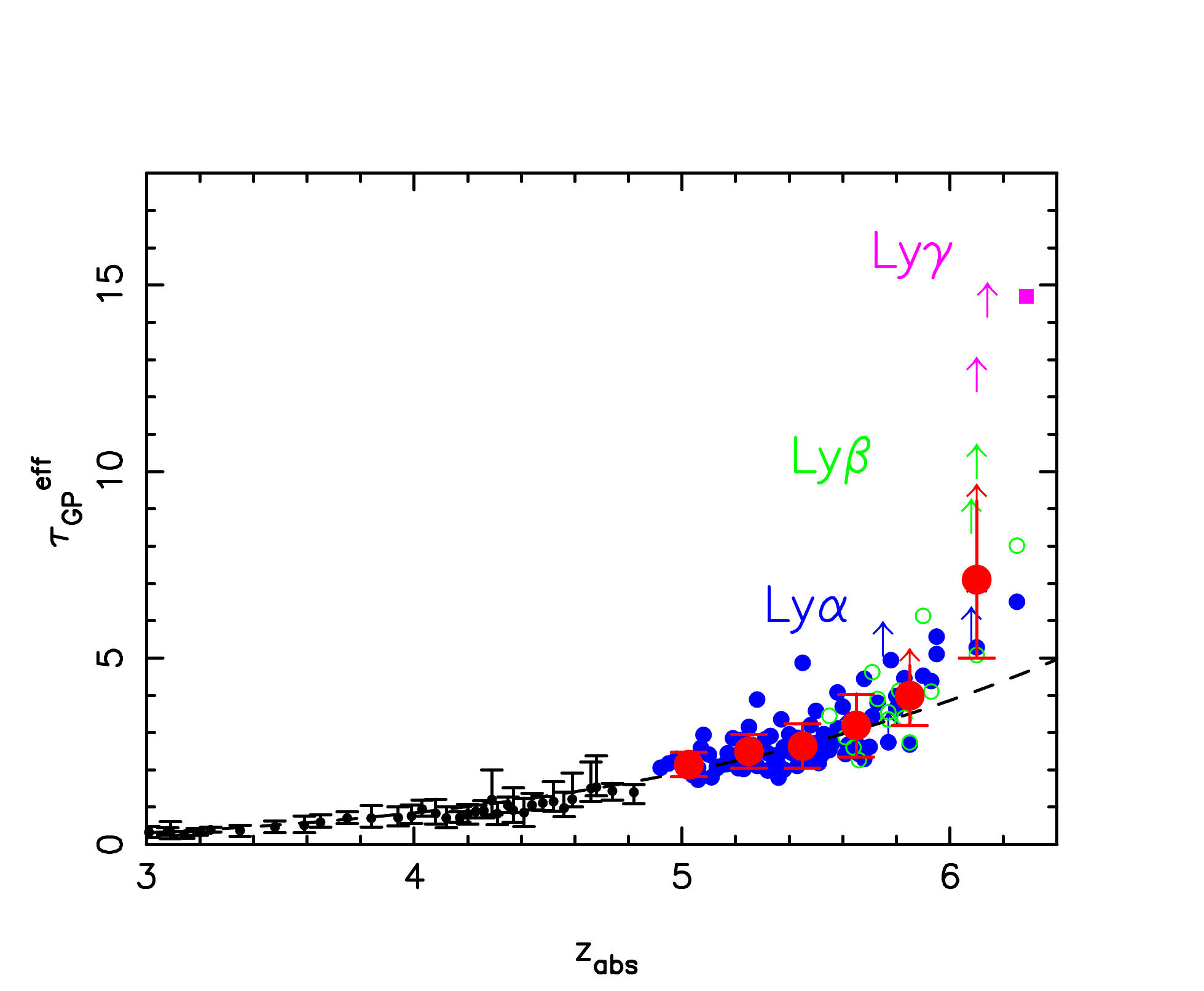}
\caption{  Evolution of the Lyman~$\alpha$, $\beta$ and $\gamma$ optical depth
from the high redshift Sloan quasars. The Lyman~$\beta$ and Lyman~$\gamma$ restframe wavelengths are 1026~\AA\ and 
972.5~\AA, respectively. The
Lyman~$\beta$ measurements are converted to Lyman~$\alpha$ Gunn-Peterson optical depth using
a conversion factor that reflects the difference in the cross section between the
two transitions, which is a factor of  5.27 lower in the case of  Lyman~$\beta$  (see~\cite{lidz02, cen02}) . 
The values in the two highest redshift
bins are lower limits, since they both contain complete Gunn-Peterson
troughs. The dashed line shows a redshift evolution of $\tau_\alpha
\approx (1+z)^{4.3}$. At $z>5.5$, the best-fit evolution has
$\tau_\alpha\approx (1 + z)^{>10.9}$, indicating an accelerated
evolution. The large filled symbols with error bars are the average
and standard deviation of the optical depth at each redshift. The sample
variance also increases rapidly with redshift. 
Figure taken from~\cite{fan06}. \vspace{0.5 cm} }
\label{fig:Lya3}
\end{figure}

Figure~\ref{fig:Lya3}
 shows the effective Lyman~$\alpha$ or Gunn-Peterson optical depth,
$\tau_{GP}^{eff}$, as a function of redshift as estimated from the
joint optical depths of Lyman~$\alpha$, $\beta$ and $\gamma$. From
this plot it is clear that the increase in the optical depth as a
function of redshift is much larger than expected (shown in the dashed line) from passive
redshift evolution of the density of the Universe.

The interpretation of the increase in the optical depth at $z\gsim
6.3$ has been the subject of some debate. All authors agree that this
is a sign of an increase in the Universe's neutral fraction at high
redshifts, marking the tail end of the reionization process. The
controversy is centered on the question of by how much the neutral
fraction increases. Some authors \cite{wyithe04a,wyithe04c, mesinger04} have
argued that the size of the so call Near Zone ionized by the quasar itself and set redwards of the 
Gunn-Peterson trough indicates that the
neutral fraction around the SDSS high redshift quasars is $\approx
10\%$. More recently is had been suggested that the
variations seen across various SDSS quasars indicate that the
ionization state of the IGM at these redshifts changes significantly
across different sightlines \cite{mesinger09}. However, given the intense radiation field around these quasars, it is not
possible to put general constraints on the neutral fraction of the IGM from quasars at redshift below 6.5 (see e.g., \cite{bolton07, wyithe08, maselli09, maselli07}).
Moreover, recently and with the discovery of the redshift
$z=7.1$  QSO ULAS J1120+0641~\cite{mortlock11} by the UKIDSS survey~\cite{lauwrence07} it has been 
argued that this quasar's Near Zone gives a clear evidence for an increase 
in the neutral fraction of hydrogen in the IGM at $z=7.1$~\cite{mortlock11, bolton11}. Note however that his conclusion relies on one quasar
and might change as more of such quasars at $z\gsim 7$ are discovered.

 There are more things that we can learn about reioinzation from the Lyman~$\alpha$
 forest that we will discuss later. But
 to summarize, the main conclusion from the Lyman~$\alpha$ optical 
 depth measurements is
that the Universe is highly ionized at redshifts below 6 (as seen in
Figure~\ref{fig:Lya2}), while at about z=6.3, the its neutral
fraction increases, forming the tail end of the
reionization process (see Figure~\ref{fig:Lya3}) . 
%%%%%%%%%%%%% CMB section %%%%%%%%%%%%%%

\subsection{The Thomson Scattering Optical depth for the Cosmic Microwave Background (CMB) Radiation}

This is a very evolved topic, discussed and reviewed by many
authors (e.g., \cite{peebles70, sunyaev72, bond84, ma95, hu97, aghanim08}). 
Here, I give a general review of the constraints provided by the CMB on
reionization. The CMB provides important information relevant
to the history of reionization. It is known that the Universe has indeed recombined and
became largely neutral at $z\approx 1100$. If recombination had been absent or
substantially incomplete, 
the resulting high density of free electrons would imply that photons could not escape
Thomson scattering until the density of the Universe dropped much further.
This scattering would inevitably destroy the
correlations at subhorizon angular scales seen in the CMB
data (see e.g.,~\cite{hu95, sugiyama95}). 

\begin{figure} \centering
\includegraphics[width=0.55\textwidth]{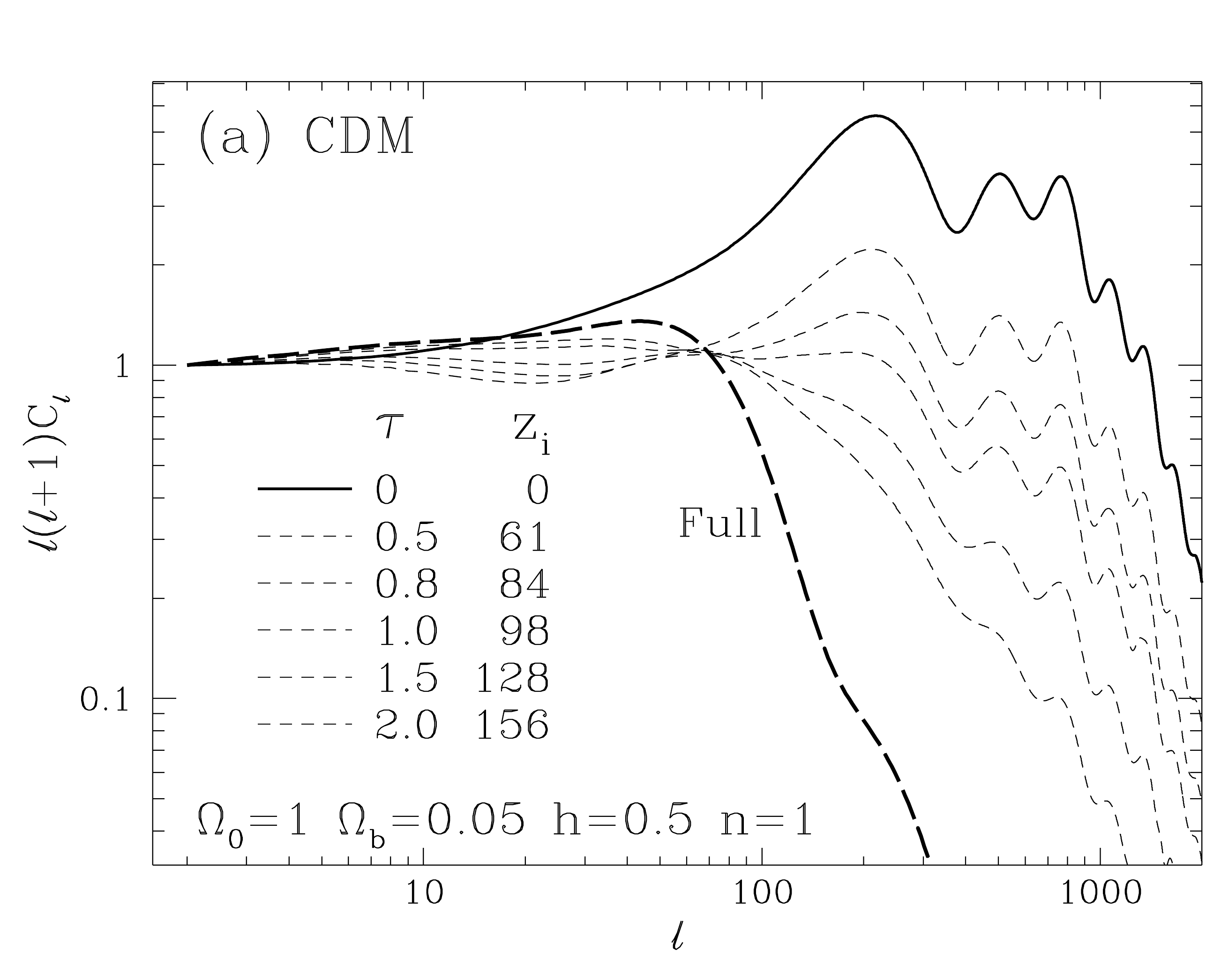}
\includegraphics[width=0.42\textwidth]{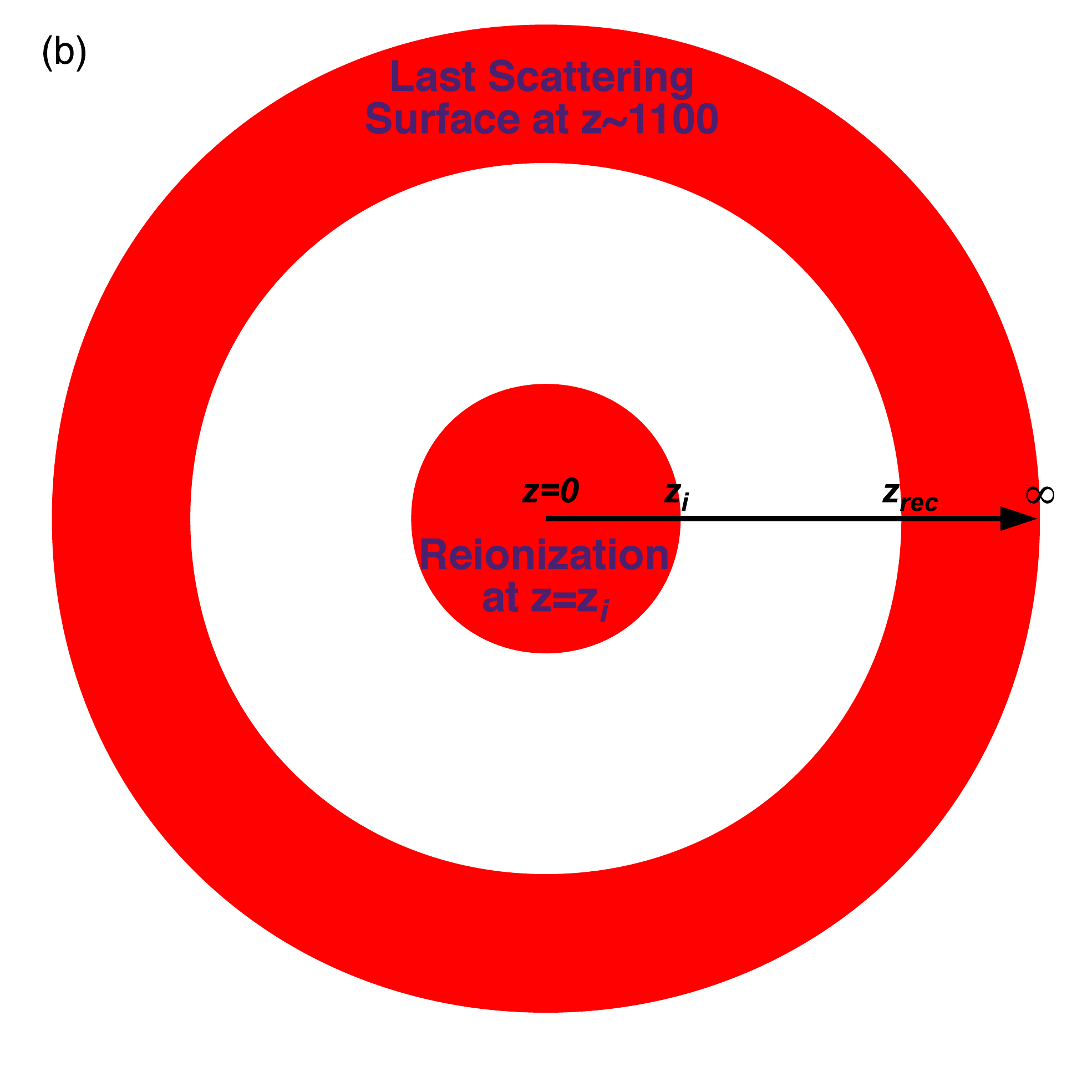}
\caption{ Left hand panel (a): The influence of reionization on the CMB
temperature angular power spectrum. Reionization damps anisotropy power 
as $e^{-2\tau}$ under the horizon (diffusion
length) at last scattering. The models here are fully ionized $x_e = 1.0$ out to a
reionization redshift $z_i$. Notice that with high optical depth, fluctuations at intermediate
scales are regenerated as the fully ionized (long-dashed) model shows. This figure is taken from Wayne Hu's PhD thesis \cite{hu95}. Right panel (b) shows the assumed reionization history used. It is obvious that since we are considering a uniform and  sudden
reionization model, a change in the reionization redshift, $z_i$, will translate uniquely to an optical depth for Thomson scattering.}
\label{fig:CMBpolar1}
\end{figure}

In order to calculate the effect of reionization on CMB photons, a function is often 
defined called the visibility function\footnote{Notice that this is a different ``visibility" than the one used 
in radio interferometry which we discuss in section~\ref{sec:21cmObs}.},
\begin{equation}
g(\eta) = -\dot{\tau} e^{-\tau(\eta)},
\label{eq:CMBvisibility}
\end{equation}
where $\eta \left(\equiv \int dt/a \right)$ is the conformal time, $a$ is the scale factor of the Universe and $\dot\tau$ is the derivative of the optical depth with respect to to $\eta$. The optical depth for Thomson scattering is 
given by  $\tau(\eta) = -\int_\eta^{\eta_0} d\eta \dot{\tau}=\int_\eta^{\eta_0} d\eta a(\eta) n_e \sigma_T$, where $\eta_0$ is the present time, $n_e$ is the electron density and $\sigma_T$ is the Thomson cross section. The visibility function gives the probability density that a photon 
had scattered out of the line of sight between $\eta$ and $\eta+\mathrm{d}\eta$. The influence of reionization on the 
CMB temperature fluctuations is obtained by integrating Equation~\ref{eq:CMBvisibility} along each sightline to estimate the  temperature fluctuation suppression due to the EoR. The suppression probability turns out to be
 roughly proportional to $1-e^{-\tau}$  \cite{zaldarriaga97}. Since 
the amount of suppression in the measured power spectrum is small, the optical depth for Thomson scattering must be small too \cite{page07}.
The left hand panel in Figure~\ref{fig:CMBpolar1} shows the influence of increasing the
value of $\tau$, the Thomson optical depth, on the CMB
temperature fluctuation power spectrum. The right hand panel shows the reionization history
of the Universe assumed  in the left panel. Since in this case a sudden global reionization is assumed,
there is one to one correspondence between the optical depth for Thomson scattering and 
the redshift of reionization. 

\begin{figure} \centering
\includegraphics[width=0.7 \textwidth]{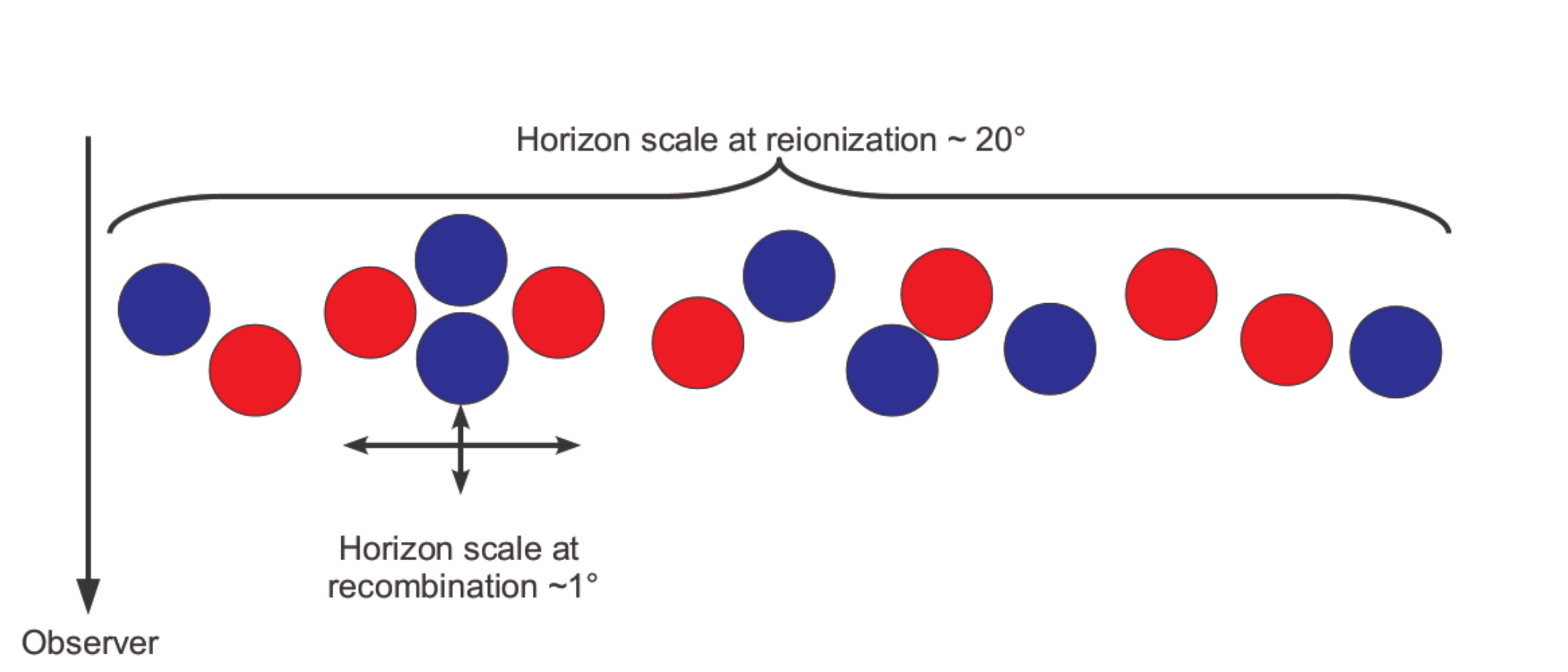}
\includegraphics[width=0.25 \textwidth]{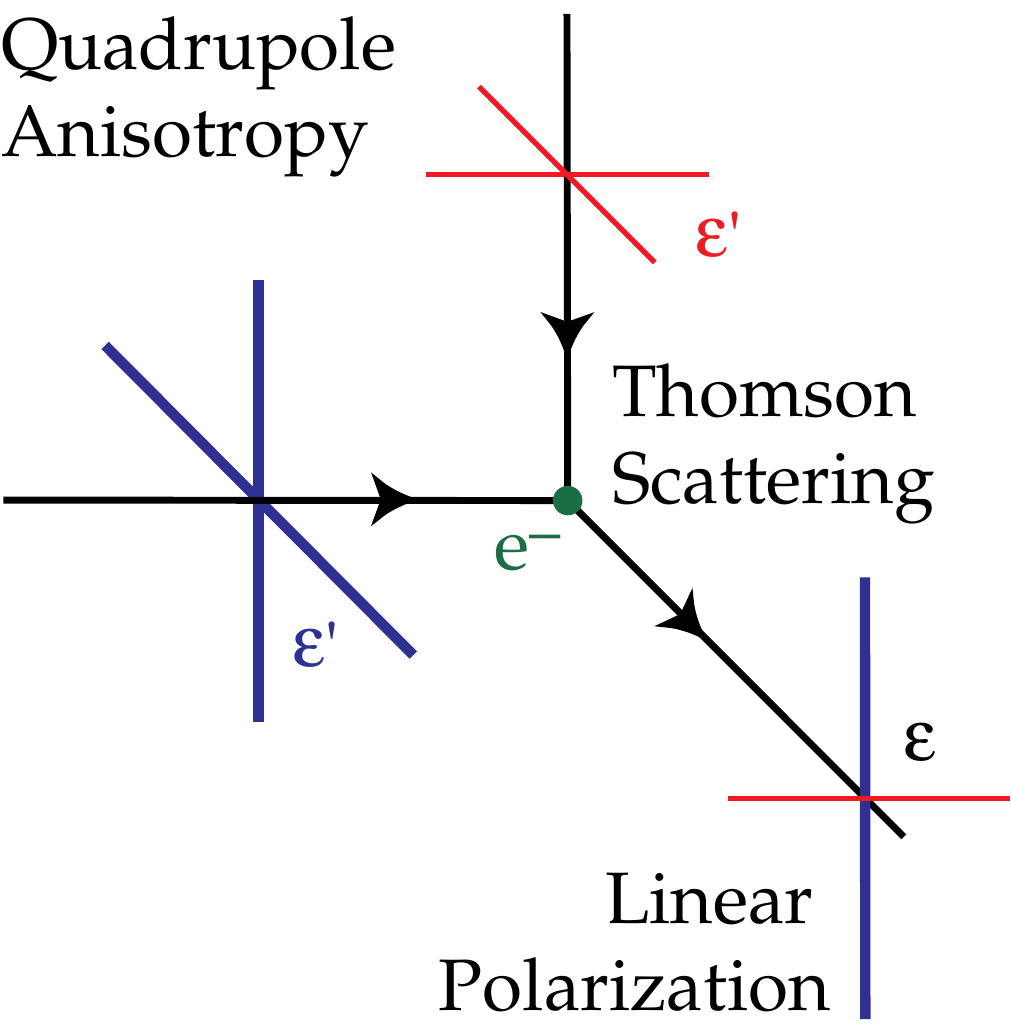}
\caption{Left hand panel: A sketch that shows why the CMB polarization is sensitive to the quadrupole momentum
of temperature fluctuations. Right hand panel: Thomson scattering of radiation with quadrupole
anisotropy generates linear polarization. The blue and red lines
represent cold and hot radiation.}
\label{fig:PolarSketch}
\end{figure}

TFruther information can be obtained from observations  of CMB via the polarization power
spectrum.  The polarization of the CMB emerges naturally from
 the Cold Dark Matter paradigm which stipulates  that small fluctuations 
 in the early universe grow, through gravitational instability,
into the large scale structure we see today (\cite{bond84, hu97, kamionkowski97, zaldarriaga97b}). Since, the temperature
anisotropies observed  in the CMB are the result of primordial
fluctuations, they would naturally polarize the CMB anisotropies. The degree of linear polarization of the
CMB photons at any scale reflects the quadrupole anisotropy in the
plasma when they last
scattered at that same scales. From this argument it is clear that the amount of polarization 
at scales larger than the horizon scale at the last scattering surface 
should fall down since there is no more coherent quadrupole contribution
due to the lack of causality. This is shown in the sketch
presented in the left hand panel in Figure~\ref{fig:PolarSketch}. The largest
scale at which a primordial quadrupole exists is the scale of the horizon at recombination,
which roughly corresponds to $1^\circ$. Therefore, any polarization
signature on scales larger than the horizon scale provides a
clear evidence for Thomson scattering at later stages where the
horizon scale is equivalent to the scale on which polarization has
been detected.

Furthermore, the polarized fraction of the temperature anisotropy must be  small,
normally one order of magnitude smaller than the anisotropy in the temperature.
This is simply because  these photons mush have passed through 
an optically thin plasma, otherwise they would not have reached us but they would have scattered 
and destroyed the sub-horizon (i.e., below $1^\circ$) correlation in the CMB, contrary to what we observe (see e.g., \cite{sugiyama95}).

The dependence Thomson scattering differential cross section
 on polarization is expressed as
\begin{equation} \frac{d \sigma_T}{d \Omega} =\frac{e^4}{m_e^2 c^4}
\vert \vec\epsilon\cdot\vec\epsilon'\vert^2
\end{equation} where $e$ and $m_e$ are the electron charge and mass
and $\vec\epsilon\cdot\vec\epsilon'$ is the angle between the incident
and scattered photons. The right hand panel of Figure~\ref{fig:PolarSketch}
shows how the Thomson scattering produces polarization of the CMB photons.
If the CMB photons scatter later due to reionization and the incident radiation has a
quadrupole moment, then it will be scattered in a polarized manner on the scale 
roughly equivalent to the horizon scale at the redshift of scattering. That is why the
scale at which the large scale polarization is detected gives information
about the reionization redshift.

\begin{figure} \centering
 \includegraphics[width=10cm]{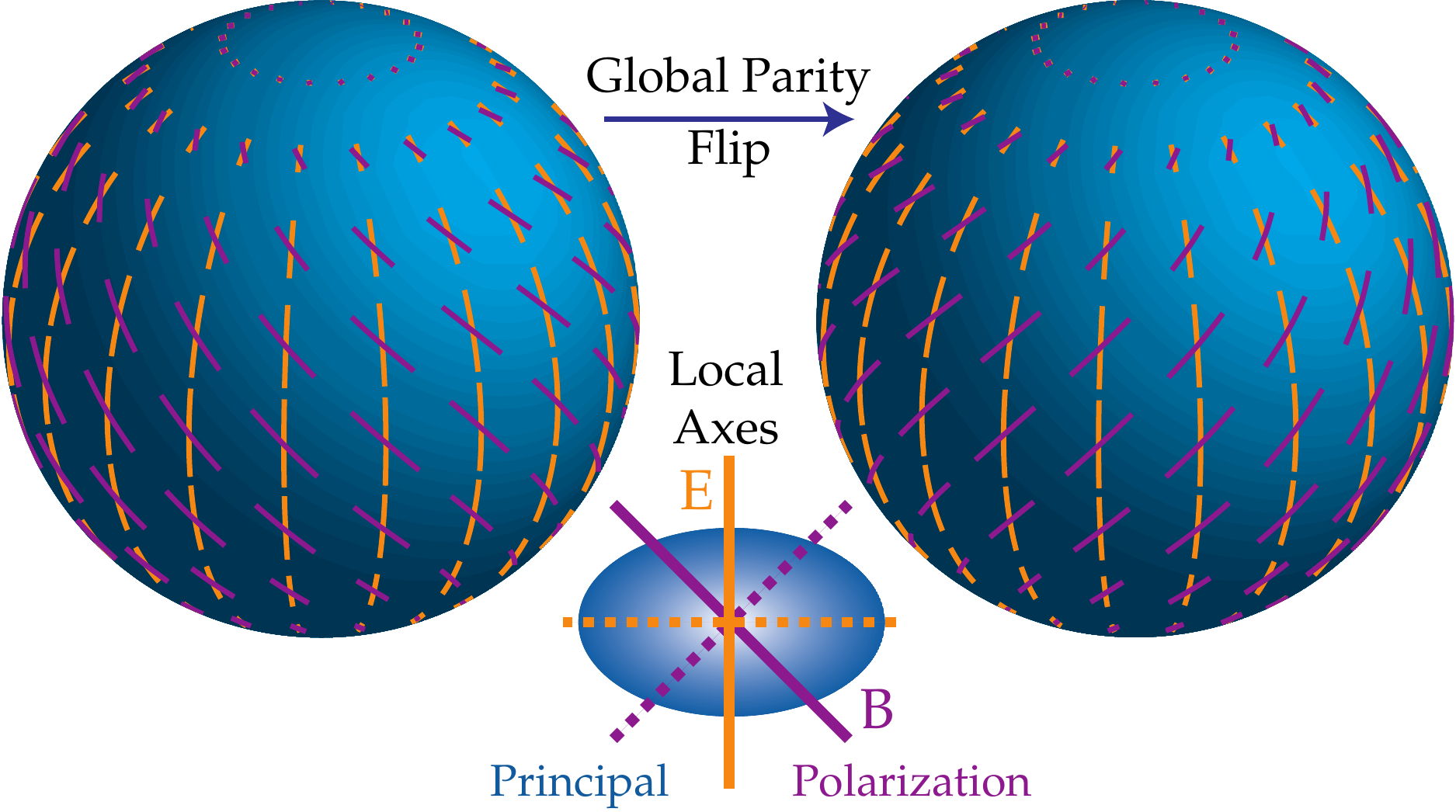}
 \caption{ The E and B polarization modes are
distinguished by their behavior under parity transformation.  The
local distinction is that the E mode is aligned with the principle
axis of polarization whereas the B mode is 45$^\circ$ crossed with it
(this figure is taken from Hu and White~\cite{hu97}). \vspace{0.5 cm}}
 \label{fig:polarparity}
 \end{figure}

The polarization field of the CMB photons is usually described in
terms of the so called ``electric'' (E) and ``magnetic'' (B) components
which can be derived from a scalar or vector field. The harmonics of
an E-mode have $(-1)^\ell$ parity on the sphere, whereas those of 
B-mode have $(-1)^{\ell+1}$ parity. Under parity transformation, i.e.,
$\hat n\rightarrow -\hat n$, the E-mode thus remains unchanged for
even $\ell$ , whereas the B-mode changes sign and vice versa.  
Fig.~\ref{fig:polarparity} illustrates such (a)symmetry under parity 
transformation for the
simple case of $\ell=2,\, m=0$~\cite{hu97}.

\begin{figure} \centering
 \includegraphics[width=0.8 \textwidth]{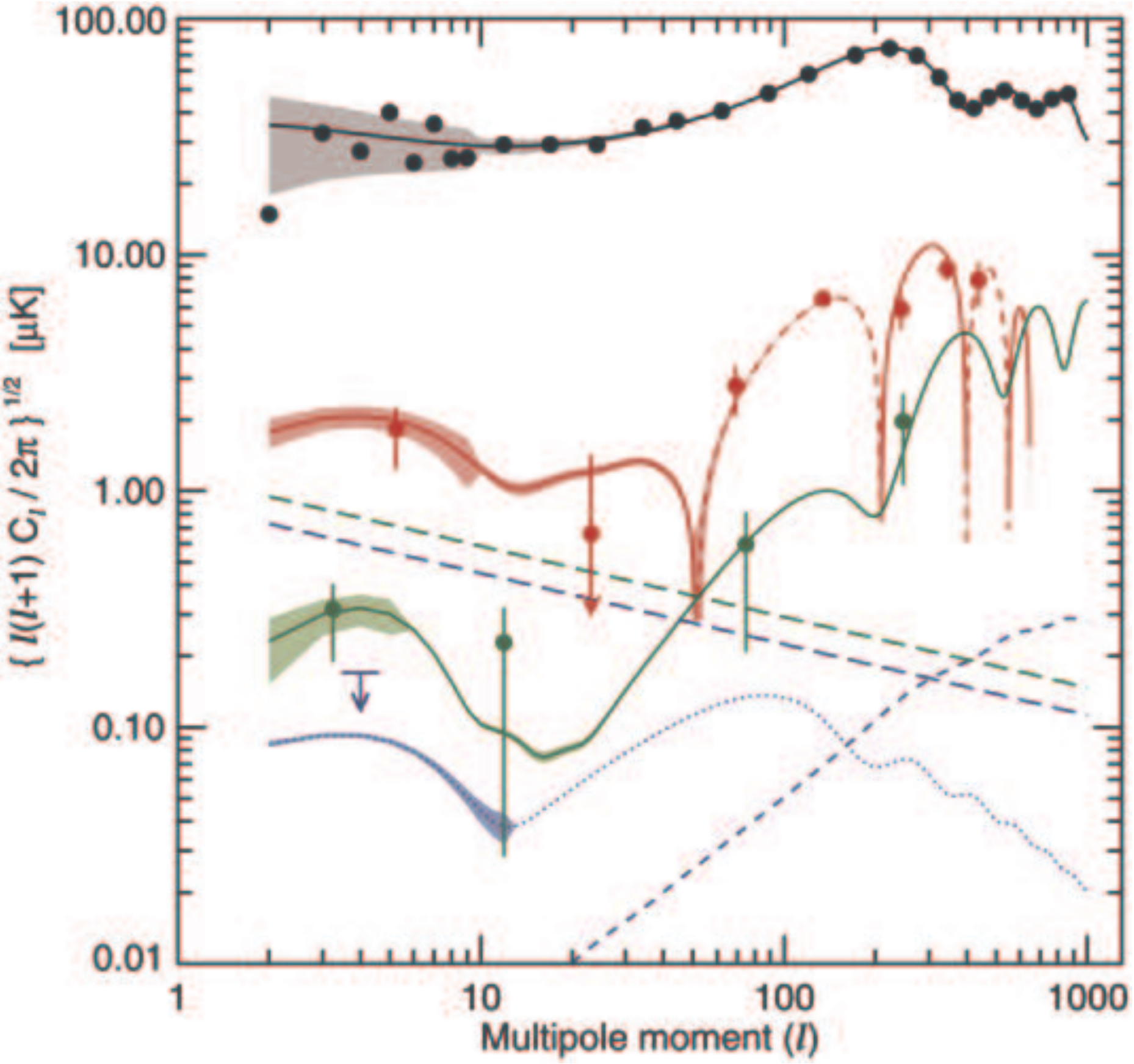}
 \caption{ The temperature and E-mode polarization power
and cross-power spectra as measure by the WMAP satellite \cite{page07}.  
Plots of signal for TT (black), TE (red ), and EE ( green) for the
best-fit model. The dashed line for TE indicates areas of anticorrelation. For more details
about  this Figure we refer the reader to the Page et al.~\cite{page07}. Notice the excess
power on large scales caused by reionization seen in the TE and EE power spectra.
\vspace{1 cm} }
 \label{fig:CMBpolar2}
 \end{figure}

 Various physical processes lead to different effects on the CMB
polarization. Most of these effects are expected to produce E mode
polarization patterns on the CMB. However, gravitational waves in the
primordial signal and gravitational lensing of the CMB on its way to
us produce a B mode polarization patterns. A large scale E mode polarization signal could only be caused by the
process of reionization. The main reason for this is that large scale
polarization could not be caused by causal effects on the last
scattering surface which has a 1$^\circ$ scale whereas reionization,
which occurs much later, has no such restriction.
Figure~\ref{fig:CMBpolar2} shows the measured and predicted CMB
angular power and cross-power spectra from the WMAP 3$^{rd}$ year
data. The existence of large scale correlation in the E-mode is a
strong indication that the Universe became ionized around redshift $z\approx 10$. The
argument in essence is mostly geometric, namely it has to do with the
scale of the E-mode power spectrum as well as the line of sight
distance to the onset of the reionization front along a given
direction. Some authors have also argued that one can have somewhat more detailed 
constraints on reionization from the exact shape of the CMB E-mode polarization large scale
 bump \cite{holder03, lewis06, mortonson08}. Unfortunately however, the large cosmic variance at 
 large scales limits the amount of possible information one can extract. Still, the Planck surveyor 
 is expected to be able to retrieve some of the large scale bump shape.
 
From Figure~\ref{fig:CMBpolar2} one can also deduce the optical depth for
Thomson scattering, $\tau$, caused by the scattering of the
CMB photons off free electrons released by reionization  to be  $0.087\pm 0.017$ \cite{dunkley09}. 
This  could be turned
into a constraint on the global reionization history through the
integral,
 \begin{equation} 
 \tau=\int_0^{z_{dec}} \sigma_T n_{e}
\frac{c H_0^{-1} dz } {(1+z) \sqrt{\Omega_m(1+z)^3+\Omega_\Lambda}}.
 \end{equation} 
 Here $z_{dec}$ is the decoupling redshift,
$\sigma_T$ is the Thomson cross section, $\mu$ is the mean molecular
weight and $n_e$ is the electron density.  This formula works for the optical depth along each sight
line but also for the mean electron density, i.e., mean reionization history, of the Universe.

An important point to notice here is that, in order to turn $\tau$ into a measurement of the reionization 
redshift, one needs a model for $n_e$ as a function of redshift. Hence, one has to be careful when
using the reionization redshift given by CMB papers as in most cases a gradual reionization is assumed.
Sudden reionization gives a one to one correspondence between the measured optical depth and the
reionization redshift, e.g., the WMAP measurement optical depth implied $z_i=11.0 \pm 1.4$ . 

However, sudden reionization is very unlikely and most models predict a more gradual
evolution of the electron density as a function of redshift. Furthermore, in such scenarios the redshift of reionization
is not clearly defined, therefore authors refer instead to the redshift at which half of the IGM volume is ionized, $z_{x_{HI}=0.5}$.
Obviously, in the case of sudden reionization the two redshifts coincide, $z_i=z_{x_{HI}=0.5}$.
It is also important to notice that in the case of sudden reionization the WMAP measured Thomson optical depth
does not imply that the redshift at which half the IGM is ionized is the same as $z_i$ and in most cases one 
obtains $z_{x_{HI}=0.5}< z_i$ \cite{thomas09}.
 
 The patchy nature of the reionization process will also leave an imprint at arcminute scales on the CMB
 sky. Such an imprint will be mostly caused by the reionization bubbles that form during the EoR. However,
 the strength of the reionization signal at small scales is found to be smaller than that caused by gravitation lensing
 and is very hard to extract unless the experiment has a very high signal-to-noise at such small scales \cite{dore07}.
 
 \subsection{The Intergalactic Medium at $z \lsim 6$} 
 
 There are a number
of other observations that put somewhat less certain constraints on
reionization. Those constraints come mostly from detailed analysis of
high resolution Lyman~$\alpha$ forest data and from the observation
of high redshift Lyman break galaxies. Here we present the two
``strongest" of those constraints.  

\subsubsection{IGM Temperature Evolution}

\begin{figure} \centering
\includegraphics[width=0.8\textwidth]{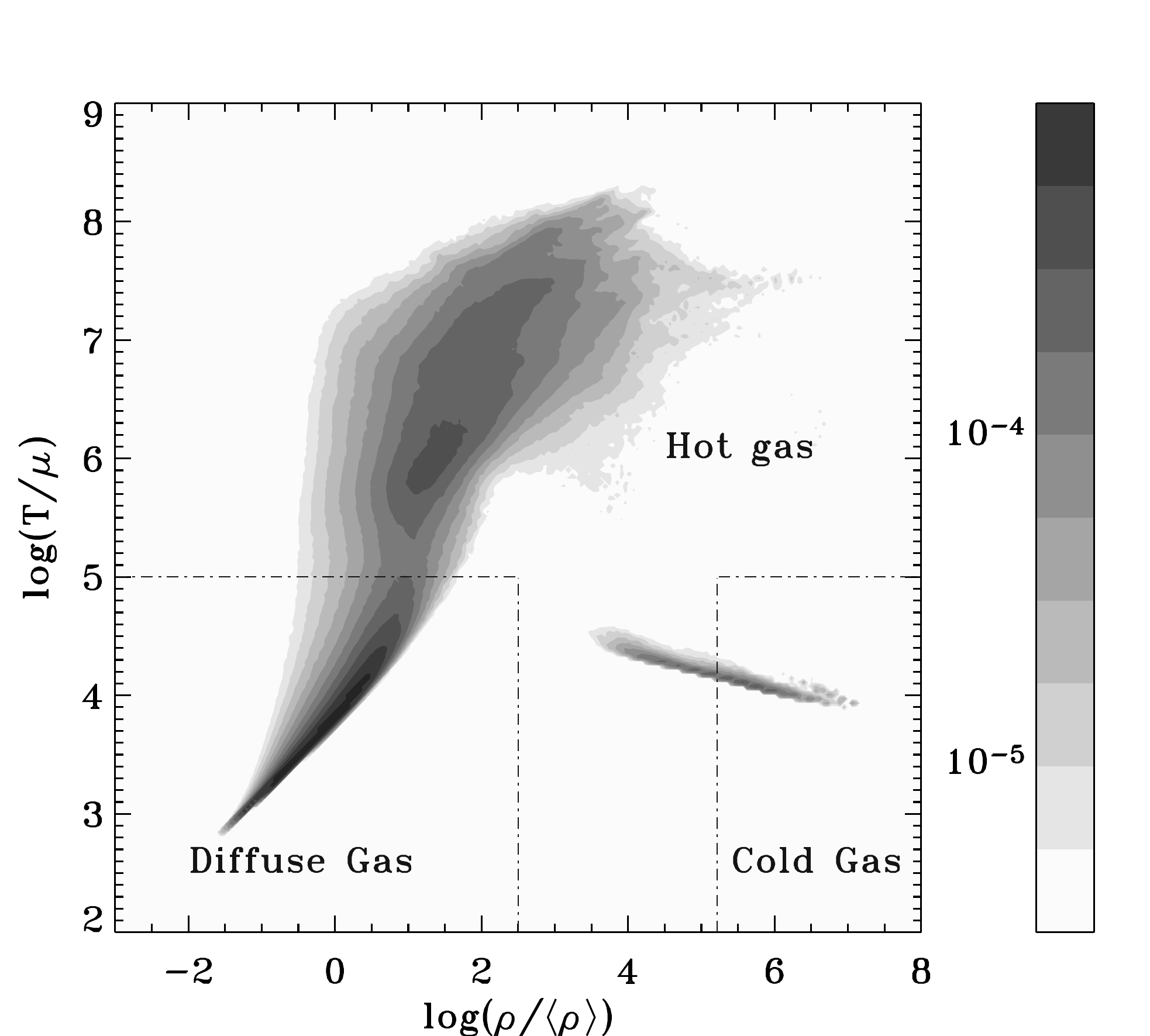}
\caption{ The different baryon phases in the $\rho -T$
diagram. Gray contours show a mass-weighted histogram: the baryon mass
fraction at a given density and temperature. Each region corresponds
to a given phase (diffuse background, hot, or cold gas)   \cite{rasera06}.   
\vspace{0.2 cm}}
\label{fig:IGMphase}
\end{figure}

 Another constraint on the reionization history comes from studying the
 thermal history of the IGM. Due to its low density, the intergalactic medium cooling time is long
and retains some memory of when and how it was last heated, namely, reionized. Hence, measuring
the IGM temperature at a certain redshift ($\gsim 3.5$) allows us to reconstruct,
under certain assumptions, its thermal history up to the reionization phase where the IGM has been substantially heated. Such a measurement has been carried out by a number of authors using
high resolution Lyman~$\alpha$ forest data, especially using the very low column 
density absorption lines.  The width of these
absorption features carries information about the temperature of the
underlying IGM. This temperature obviously varies with density and
with other parameters like the background UV flux. Based on both theoretical
arguments \cite{hui97} and on numerical simulations \cite{theuns98} in the
linear and quasilinear regime, the temperature-density relation follows the simple power law,
\begin{equation} T=\bar T \left(\frac{\rho}{\bar\rho}\right)^{\gamma-1},
\label{eq:IGMtemp}
\end{equation} 
where $\bar T$ is the temperature of the IGM at the mean
density of the Universe and $\gamma$ is the adiabatic power law index.
Figure~\ref{fig:IGMphase} shows the so called phase diagram, i.e., the relation between the temperature and density, 
obtained from a cosmological hydrodynamical simulation \cite{rasera06}.  The relation between the density and temperature at 
the low density end of the diagram, marked as diffuse background, follows a power law.
The hot phase at intermediate densities where cooling is not efficient, is driven by shock heating.
At high densities, cooling becomes very efficient and drives the gas temperature. At high redshifts
more than 90\% of the gas is in the diffuse phase.

Given the validity of equation~\ref{eq:IGMtemp} at low densities, it is meaningful 
to define an IGM temperature as the gas temperature at the mean density, $\bar T$.
Such a measurement has been performed by a number of authors at $z\approx 3-4$ 
\cite{lidz10, schaye00, theuns02b, zaldarriaga02} and recently at $z\approx 6$ by \cite{bolton10}.  

The usefulness of this temperature to
constrain the reionization history was first realized by \cite{theuns02, hui03} who
 used the measured temperature around redshift $3$ to set
$z\approx 9$ as an upper limit for the reionization process. Bolton
et al. (\cite{bolton10}) have recently confirmed these findings with higher
redshift quasars. 
That is, the measured temperatures of the IGM at redshift $z\approx 3$ and $z\approx 6$ 
are too high for the bulk of reionization to have occurred at redshift $\gsim 10$. 

\begin{figure} \centering
\includegraphics[width=0.9 \textwidth]{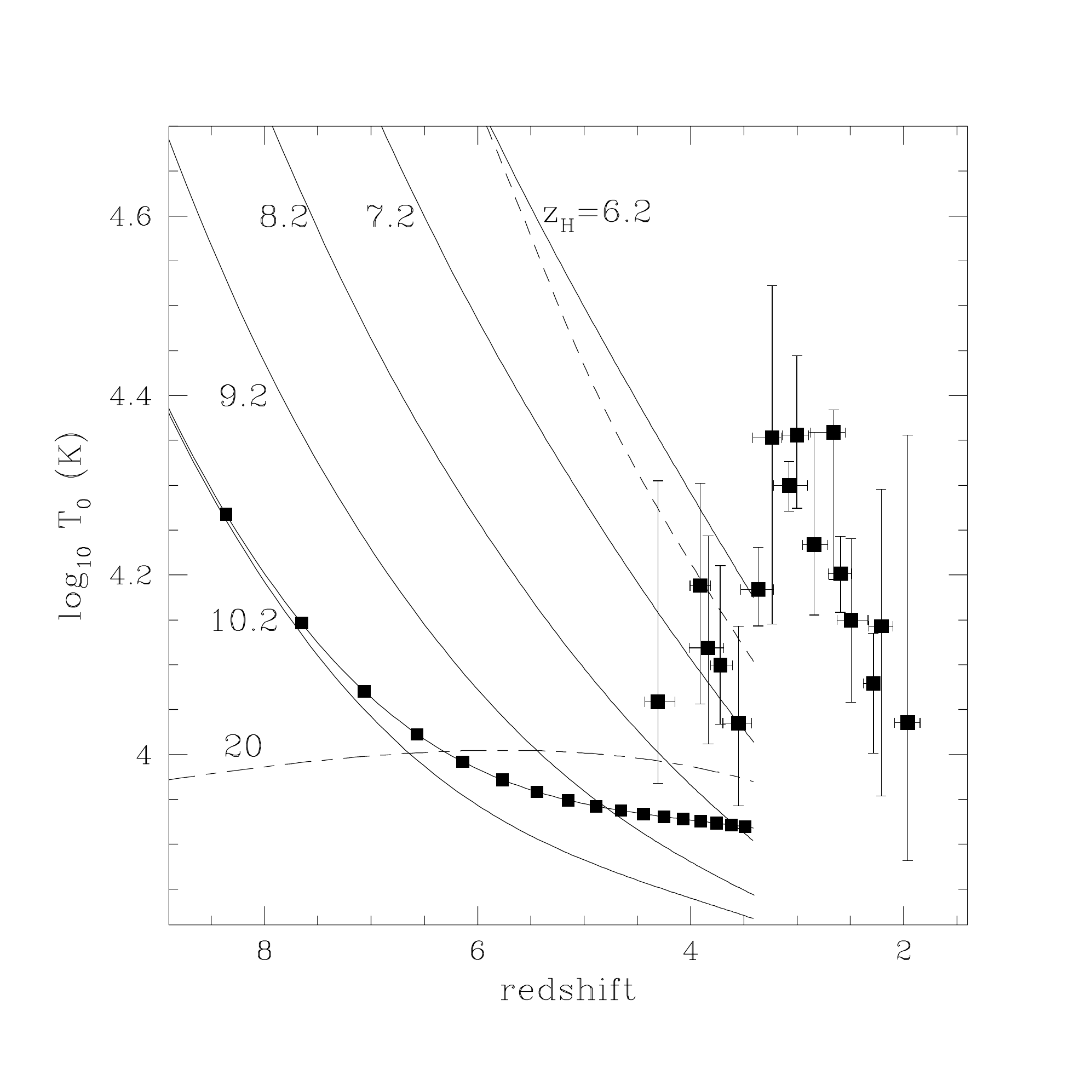}
\caption{ Temperature evolution of the IGM above redshift
3.4. The solid curves indicate the evolution of the temperature at the
mean density for various \THI reionization redshifts $z_H$, as
indicated. The temperature after hydrogen reionization is assumed to
be $T_0 = 6 \times 10^4$ K, and the hydrogen photoionization rate is
$\Gamma = 10^{-13} s^{-1}$ ($\Gamma= 10^{-14} s^{-1}$, short-dashed
curve). The \THeII photoionization rate is adjusted so that the
\THeIII abundance is $x_{\HeIII}\approx 0.1$ at $z=3.5$. The solid
curve connecting the filled squares indicates $z_H = 10.2$ and a
higher \THeII photoionization rate, $x_{\HeIII}(z = 3.5) =
0.6$. Finally, the long-dashed curves has $z_H = 20$ but a still
higher \THeII photoionization rate, $x_{\HeIII}(z = 3.5) = 0.95$. If
He is mostly singly ionized at $z \approx 3.5$, then the rapid
decrease in $T_0$ after reionization places an upper limit of $z_H <
9$ on the redshift of hydrogen reionization. The filled squares with error bars 
show the measured IGM temperature as as function of redshift.  
This figure taken from \cite{theuns02}.
\vspace{0.5 cm}  }
\label{fig:IGMT1}
\end{figure}

After reionization, the evolution of the IGM mean temperature $\bar T$ is given by
\begin{equation}
 \frac{1}{\bar T} \frac{d\bar T}{dt} - \frac{1}{\mu}
\frac{d\mu}{dt} = -2 H+\frac{\mu \Delta_\epsilon}{\frac{3}{2} k_B \bar T},
\end{equation} 
where $H$ is the Hubble parameter, $k_B$ is the Boltzmann
constant, $\mu$ is the mean molecular weight, and $\Delta_\epsilon$ is
the effective radiative cooling rate (in units of ergs g$^{-1}$
s$^{-1}$). $\Delta_\epsilon$ is negative (positive) for net
cooling (heating) and includes photoelectric heating and cooling via
recombination, excitation, inverse Compton scattering, collisional
ionization, and bremsstrahlung. Without cooling/heating processes the cooling
rate is set by adiabatic cooling, namely, Hubble expansion.  This equation enables us to calculate
the temperature evolution as a function of redshift. Measuring the IGM
temperature at a given redshift will allow us to extrapolate back in
time until we reach a temperature of $6 \times 10^4$K which is the
temperature at which hydrogen ionizes. Figure~\ref{fig:IGMT1} 
 demonstrates this procedure~\cite{theuns02}.

Obviously, the weak point of this argument is the assumption that one knows the cooling/heating function
of the IGM at every redshift up to the time of reionization. Still, this is a useful argument and certainly
any model for the reionization history would have to explain the temperature we measure at lower redshifts.

\subsubsection{Number of Ionizing Photons per Baryon}

Another constraint that comes mostly from the Lyman~$\alpha$ forest but
also from the recently discovered galaxies at $z \gsim 7$ is the
number of ionizing photons per baryon.  Using physically motivated
assumptions for the mean free path of ionizing photons, Bolton and
Haehnelt (\cite{bolton07}) turned the measurement of the photoionization rate into an
estimate of the ionizing emissivity.  They showed that the inferred
ionizing emissivity in comoving units, is nearly constant over the
redshift range $2-6$ and corresponds to $1.5-3$ photons emitted per
hydrogen atom over a time interval corresponding to the age of the
Universe at $z=6$.  Completion of reionization at or before $z=6$ requires
therefore, either an emissivity which rises towards higher redshifts or one which
remains constant but is dominated by sources with a rather hard
spectral index, e.g., mini-quasars.  

With the installation of the WFC3 camera aboard the Hubble Space Telescope,  searches for high redshift
galaxies at $z=6-10$ have improved dramatically. In particular, a number of authors \cite{oesch10, bouwens10, bunker10, mclure10}
 have reported detection of very high redshifts galaxies using
the Lyman-break drop-out technique. The most striking result of these studies is 
the low number of galaxies found beyond redshift $\approx 6$, making it very hard for these 
galaxies to ionize the Universe. This conclusion depends however on assuming 
a luminosity function for galaxies at these redshifts, a function that is very poorly known.
More surprising is the very steep drop in the number of 
galaxies at redshift $z\approx 9$ \cite{bouwens11} which makes it even harder to 
explain reionization with such galaxies. 

\begin{figure} \centering
\includegraphics[width=0.9 \textwidth]{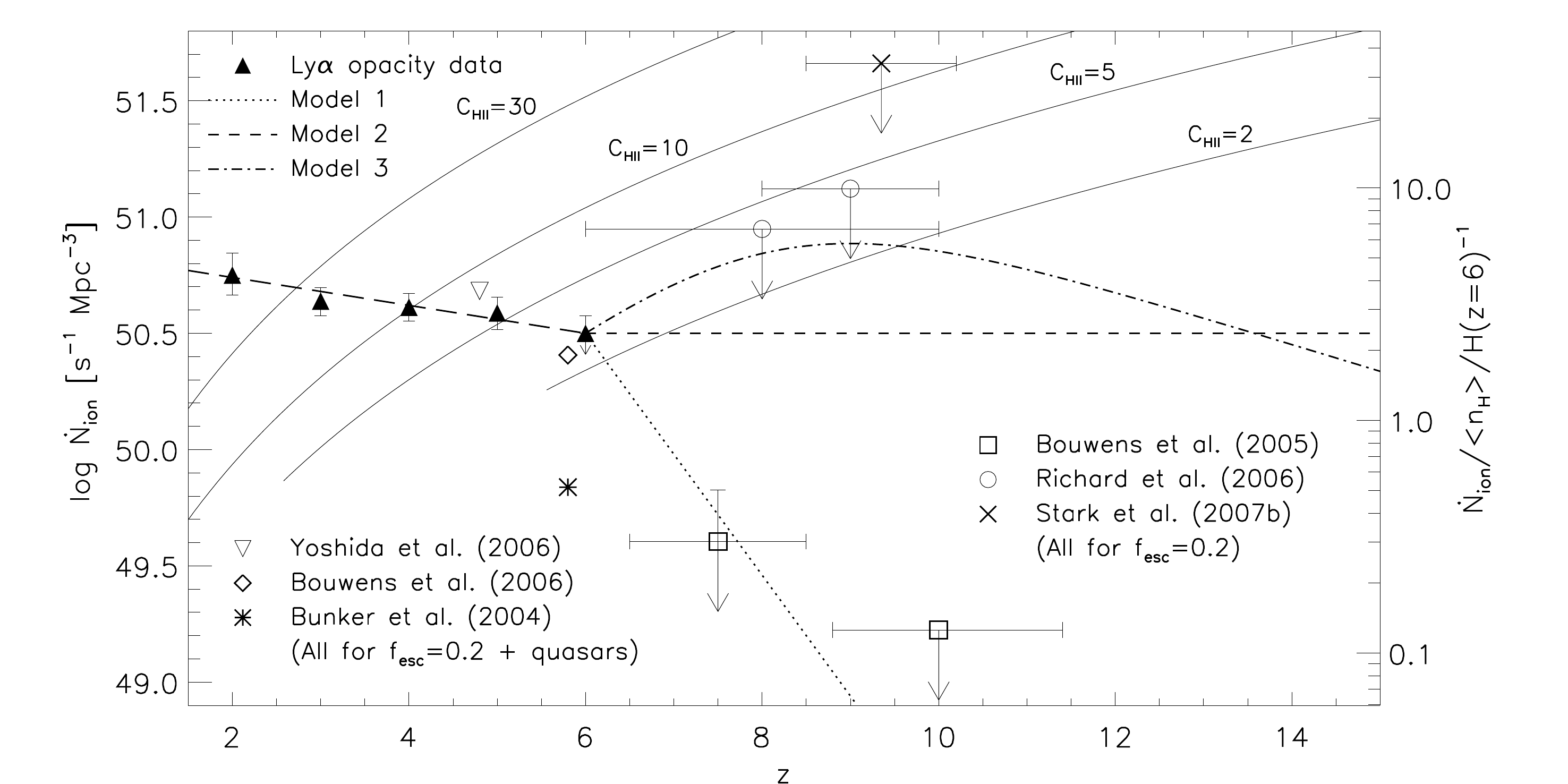}
\caption{ Observational constraints on the emission rate
of ionizing photons per comoving Mpc, $\dot N_{\rm ion}$, as a
function of redshift. The scale on the
right-hand vertical axis corresponds to the number of ionizing photons
emitted per hydrogen atom over the Hubble time at $z=6$. The filled
triangles give an estimate of $\dot N_{\rm ion}$ based on the
constraints obtained from
the Lyman~$\alpha$ effective optical depth from \cite{bolton05} .
The inverted triangle at $z=5$ and the diamond and star at $z=6$
correspond to estimates of $\dot N_{\rm ion}$ based on the Lyman limit
emissivities of LBGs and quasars.  The data have
been slightly offset from their actual redshifts for clarity.  An
escape fraction of $f_{\rm esc}=0.2$ has been assumed in this
instance.  At $z>6$, the open squares and circles are derived from the
upper limits on the comoving star formation rate per unit volume
inferred by \cite{bouwens05, richard06},
respectively.  The cross is derived from the number density of
Lyman~$\alpha$ emitters estimated by \cite{stark07}.  Three
simple models for the evolution of $\dot N_{\rm ion}$ are also shown
as the dotted, short dashed and dot-dashed lines.  The solid lines
correspond to the emission rate of ionizing photons per unit comoving
volume, $\dot N_{\rm rec}$, needed to keep the IGM ionized for various
\THII clumping factors. This figure is taken from \cite{bolton07}, see also~\cite{calverley11}.  }
\label{fig:ionphoton}
\end{figure}

 The last two observational findings have led some authors to claim that the reionization
is photon starved, i.e., has a low number of ionizing photons per baryon, which results
in a very slow and extended reionization process \cite{bolton07, calverley11}.
Figure~\ref{fig:ionphoton} shows the number density of ionizing photons (left-hand vertical axis)
and number of ionizing photons per baryon (right-hand vertical axis) as a function 
of redshift. The number of ionizing photons per baryon at redshift 6 is of the order of 2.
More recent results deduced from Lyman-break galaxies are consistent with this figure
and show an even lower ratio of ionizing photons per baryon at higher redshifts.

\subsection{Other Observational Probes}

In addition to the probes that we discussed so far, there are a
large number of other observational probes that could potentially add
valuable input to the reionization models.  Examples of such 
probes are cosmic infrared and soft x-ray backgrounds \cite{dijkstra04},
Lyman~$\alpha$ emitters~\cite{ouchi09}, high redshift QSOs~\cite{mortlock11} and GRBs\cite{bromm06}, metal abundance at high redshift~\cite{rudie12}, etc.
However, such probes currently provide very limited constraints
on the EoR.

In the coming chapters we will focus on the very large effort currently
made to measure the diffuse neutral hydrogen in the IGM as a function
of redshift up to $z \gsim 11$ using the redshifted 21~cm emission line. This probe will give the most direct and detailed 
evidence on the reionization process.

\section{The Reionization process}
\label{sec:EoRProcess}

The inflationary process that occurred very early in the Universe has created the initial tiny fluctuations
in matter density field. The high density peaks in these fluctuations field are the seeds around which  galaxies 
form. The formation process is initially driven by gravitational instability alone but later gas physics, cooling, heating, radiation
processes and feedback effects play an important role as well \cite{mo10, peebles93}. The first galaxies form when primordial gas (\THI and \THeI)
condenses within dark matter potential wells which leads to radiative cooling driven mostly by the Lyman~$\alpha$ line transition \cite{dijkstra06, haiman00, latif11, latif11b, latif11c, partridge67} and, probably, by H$_2$ cooling. To date Lyman~$\alpha$ emission 
has been observed in many  high redshift galaxies \cite{kashikawa06, murayam07, ouchi09, ouchi10}.
This gas condenses further to form the first stars and black holes which in turn produce radiation that starts ionizing the 
Universe. The efficiency with which these objects produce ionizing radiation is subject to many different physical processes and
assumptions (see e.g., \cite{ciardi05}).
Since this book's topic is the first galaxies, the reader is referred to the other chapters in this volume
for detailed discussion of how the first radiation emitting objects form and how efficient are they in producing
ionizing radiation.

 An important unknown in these galaxies is the so called escape fraction, namely the fraction
 of ionizing radiation that escapes the galaxy into the IGM. It is these ionizing photons that are relevant 
 to the Universe's reionization. Determining the escape fraction of ionizing radiation observationally is very difficult
 especially at high redshifts where the available information is very limited. Nevertheless, such 
 observations have been carried out by a number of authors
 \cite{giallongo02, inoue05, iwata09, shapely06, steidel01} where the measured fraction is found to be between 0.1-0.5. Theoretical
 prediction of the escape fraction is also difficult.  Early studies have assumed idealized smooth galaxies \cite{dove94, dove00, ricotti00, wood00} but later studies have simulated more realistic galaxies (see e.g., \cite{ciardi02}). Each of these studies have considered 
different set up and different sources but all conclude that the escape fraction of radiation is roughly in the range of 0.1-0.5.

The most accepted picture of how reionization unfolds is simple. The
first radiation-emitting objects ionize their immediate surroundings,
forming bubbles that expand until their ionizing photons are consume by the neutral IGM. As the number of radiating
sources increases, so do the number and size of the ionization
bubbles, which eventually spread to fill the whole Universe. However,
most of the details of this scenario are yet to be clarified.  For
example: what controls the formation of the first objects and how much
ionizing radiation do they produce? How do the bubbles expand into the
intergalactic medium and what do they ionize first, high-density or
low density regions? The answer to these important questions and many
others touch upon many fundamental questions in cosmology, galaxy
formation, quasars activity and the physical properties of very metal poor stars
\cite{barkana01, bromm04, ciardi05, choudhury06, furlanetto06a, morales10}.

 To ionize hydrogen one needs photons with energy of 13.6 eV or higher meaning
the reionization of the Universe requires ultraviolet photons. A
crucial question is which sources in the Universe provide the UV photons
needed to ionize the Universe and maintain it in that state. Obvious
candidates are the first stars (so called Population III stars), second generation stars (Population II stars) and
(mini)quasars which are objects powered by intermediate mass black
holes ($10^{3-6} M_\odot$). There are other candidate sources of reionization,
like decaying or self-annihilating dark matter particles or decaying
cosmic strings. However, the constraints on such objects make it
unlikely that they could reionize the Universe by themselves~\cite{chen04, 
kasuya04,mapelli05, mapelli06, natarajan10, padmanabhan05, ripamonti07, zhang06}.

 Massive black holes powering quasars convert mass to radiation
extremely efficiently. They produce a large amount of UV and X-ray
radiation above the ionization threshold. In fact, one of the main
discoveries of the last decade is that quasars, powered by very large black
holes with masses in excess of $10^9 M_\odot$, already existed at redshift
above 7 (e.g., QSO ULAS J1120+0641~\cite{mortlock11} from the UKIDSS survey~\cite{lauwrence07}) . How these black
holes managed to accumulate so much mass in such a short time is a
puzzle in its own right \cite{mortlock11, bolton11}. However, the mass distribution of the massive
black holes in the early Universe is unknown, rendering the role
played by quasars during reionization very uncertain.

Population III stars formed from the primordial mix of the elements and
thus only contain hydrogen and helium. This composition makes them
very different from present-day stars. In order for a star to form,
the initial proto-star has to radiate some of the energy gained by
gravitational contraction, or the collapse will rapidly halt as the
cloud reaches hydrostatic equilibrium. Population III stars are poor
radiators until the cloud from which they form reaches high
temperatures. This causes them to be very massive, and hence, they are very
efficient and abundant sources of UV photons, yet are very short
lived. Theoretically, these objects could have reionized the Universe
but our knowledge of them, including the question of whether they
existed in sufficient numbers, is very uncertain.

The first stars' early metal enrichment was likely the dominant effect
that brought about the transition from Population III to Population II
star formation. Recent numerical simulations of collapsing primordial
objects with masses of $\approx 10^6 M_\odot$, have shown that
the gas has to be enriched with heavy elements to a minimum level of
$Z_{crit} \approx 10^{-4}Z_\odot$, in order to have any effect on the
dynamics and fragmentation properties of the system. Normal, low-mass
(Population II) stars are hypothesized to form only out of gas with
metallicity $Z \ge Z_{crit}$. Thus, the characteristic mass scale for
star formation is expected to be a function of metallicity, with a
discontinuity at $Z_{crit}$ where the mass scale changes by about two
orders of magnitude. The redshift where this transition occurs has
important implications for the early growth of cosmic structure, and
the resulting observational signature includes the extended nature of
reionization (see the review by Ciardi and Ferrara~\cite{ciardi05}).

Most studies of reionization have focused on stars as being the
primary source \cite{mellema06, abel00, abel02, bromm02, yoshida03}. Due
to the deficiency of hard photons in the spectral energy distributions
(SEDs) of these ``first stars'', heating due to these objects is limited in
extent \cite{thomas08}. On the other hand, miniquasars (miniqsos),
characterized by central black hole masses $<10^6~M_\odot$,
have also been considered as an important contributor to reionization
\cite{madau97, ricotti04a, ricotti04b, nusser05, furlanetto02, furlanetto04, wyithe04a,
thomas08}. Ionization aspects of the miniquasar radiation
have been explored by several authors
\cite{madau97, ricotti04a, ricotti04b, thomas08, thomas09, zaroubi05}.
Thomas \& Zaroubi \cite{thomas08} have shown that although the ionization  pattern around
miniqsos is similar to that of stellar-type sources, the heating due to 
the presence of hard photons in miniqsos is very different. The reason being is that 
stars produce thermal radiation that is mostly in the UV range, which is very efficient in ionization,
but once it is absorbed by \THI, the energy left will be too small to be converted to heat effectively.
On the other hand black hole powered sources have hard x-ray photons as their spectral energy distribution 
(SED)
follows a power law (typically assumed to be $-1$). Such x-ray photons have lower
bound-free cross section relative to UV photons but once they are absorbed, their leftover energy
is very large and can easily be converted to heat. Also, x-ray photons penetrate much deeper into the IGM
and can heat it up much further from the source than UV radiation.

\begin{figure} \centering
\includegraphics[width=0.9 \textwidth, height =0.82 \textwidth]{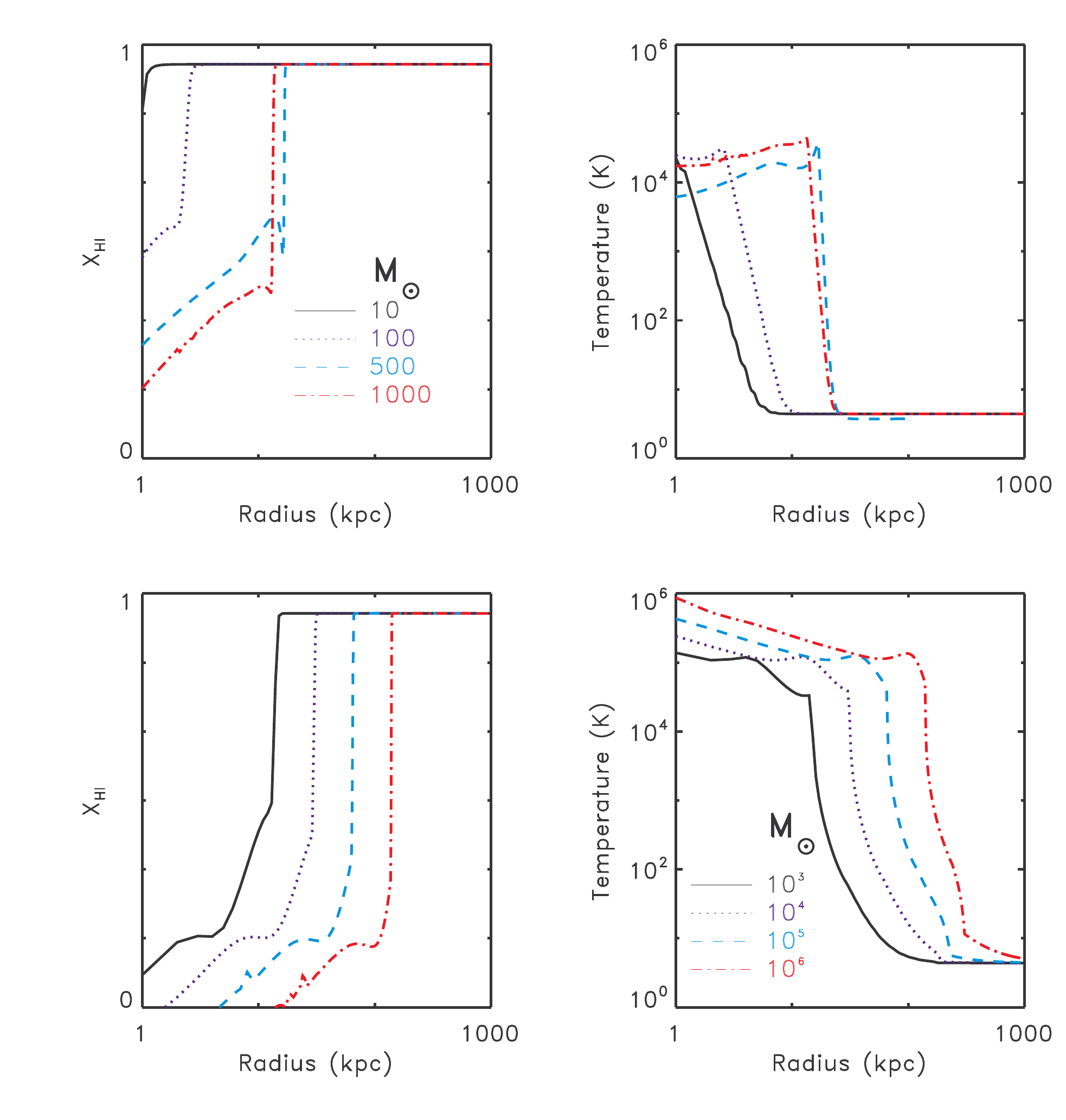}
\caption{  This figure shows the ionization and heating profile around a single star and black hole forming
in the IGM assuming spherical symmetry. The upper panels show the stars case whereas the lower panels 
show the black holes case. The left hand panels show the neutral fraction of \THI as a function of distance from
the star and the right hand panels show the gas temperature as a function of distance from the source~\cite{thomas08}.
\vspace{0.2 cm}  }
\label{fig:HeatingBHStar}
\end{figure}

Miniqsos heat the surrounding IGM well beyond
their ionization front \cite{thomas08,chuzhoy06}. Several authors (e.g., \cite{madau97, nusser05, zaroubi07} have shown
the importance of heating the IGM with respect to the observability of
the redshifted 21 cm radiation in either emission or absorption. 
Figure~\ref{fig:HeatingBHStar} shows the ionization and heating patterns 
around a number of stars (upper panels) and  miniqsos (lower panel). 
The mass of the stars and black-holes are indicated next to the lines, and their SEDs are assumed to be thermal or
to have a power law  dependence on the photon energy, $\propto E^{-1}$ , respectively. The calculation
here is spherically symmetric and assumes a single object forming in the IGM~\cite{zaroubi07, thomas08}.
The ionization pattern around stars and black holes are very similar, they both show a very abrupt 
increase in \THI  with a clear ionization front (see e.g.~\cite{kramer09, thomas08, zaroubi07}). Of course the radius at which such
front is seen depends on the mass of the star 
or the black hole but the pattern is the same (see the left hand side panels of Figure~\ref{fig:HeatingBHStar}).  
The heating profile around the two types of sources, on the 
other hand, is different since in power law sources (miniqsos) the radiation can penetrate the neutral gas and reach large
distances from the sources (see right hand panels of Figure~\ref{fig:HeatingBHStar}). This high energy radiation is efficient in heating the IGM gas through secondary electrons ~\cite{shull85} (see discussion later) whereas UV radiation is efficient in ionizing the gas but has little energy left to heat too much and can not penetrate the neutral gas as far as x-ray radiation does.

We have seen that unlike stars, x-ray source a (e.g., miniquasars) have an additional property of
heating the IGM to a large extent and through secondary Lyman~$\alpha$
radiation making the neutral IGM visible to a 21-cm
experiment. However, some authors (e.g., \cite{dijkstra04, salvaterra05b}) argue
that miniquasars alone can not reionize the Universe as they will
produce far more soft X-ray background radiation than currently
observed \cite{moretti03, soltan03} while simultaneously satisfying
the \textsc{WMAP3} polarisation results \cite{page07,spergel07}. It
should be noted, however, that these calculations have been carried out assuming
specific models for the evolution of black hole mass density and
spectral energy distributions of UV/X-ray radiation of the miniquasars,
whereas one can easily construct other models in which the discrepancy is not so severe
\cite{zaroubi07,ripamonti08}. 

Some authors~\cite{kashlinsky05} have claimed a detection of excess IR background
radiation and argued that it provides evidence for stars
being the primary source of reionization. This too has been subject to
controversy because of the sensitivity of the result to the
subtraction of the contaminants, e.g., Zodiacal light, within the same waveband
\cite{cooray07}. 

Although uncertainty looms about the sources that resided during the
\emph{dark ages}, it is conceivable from observations of our Universe up to
redshifts of 6.5, that sources of reionization could have been a
mixture of both stellar and miniquasar sources. 
Implementing radiative transfer that includes both ionizing and
hard X-ray photons has been difficult and, as a result, most 3-D
radiative transfer schemes restrict themselves to ionization due to
stars \cite{benson06, ciardi01, gnedin01, mellema06, mesinger07,
nakamoto01, pawlik08, ritzerveld03, razoumov05, susa06,
whalen06, zahn07}.
In \cite{ricotti04a}, a ``semi'' hybrid model of stars and quasars
like the one hinted above used, albeit in sequential order
instead of a simultaneous implementation. That is, pre-ionization due
to quasars has been invoked between $7 \leq z \leq 20$, after which
stars reionize the Universe at redshift 7. 

Given the numerical cost of the full 3D radiative transfer
schemes, exploring a large parameter space for models of reionization, 
 is not feasible.
Such an exploration is needed in order to understand the various physical effects introduced 
by each such parameter. It is also needed to help interpret the available data. A number of
authors have been pursuing ``quick-and-dirty" methods to simulate the reionization process.
These schemes can include very rough methods that use the initial density
field to produce a reionization cube without the need for cosmological
N-body and hydro simulations, such as 21cmFAST (~\cite{mesinger10, zahn07, zahn11}) and SimFast21~\cite{santos10}.
They also include more accurate (yet still fast) methods like BEARS~\cite{thomas08, thomas09, thomas11}
that use
N-body and hydro simulations but reduces the numerical cost by restricting
the ionization bubbles around the radiation sources to be spherical.

\section{The redshifted 21 cm as a probe of the EoR}
\label{sec:21cmProbe}

In recent years it has become clear that the 21~cm line can be used to
probe the neutral IGM prior to and during the reionization process.  This
hyperfine transition line of atomic hydrogen (in the ground state)
arises due to the interaction between the electron and proton spins
\cite{hogan79, scott90, madau97}.  The excited triplet state is a
state in which the spins are parallel whereas the spins at the lower
(singlet) state are antiparallel. The 21~cm line is a forbidden line
for which the probability for a spontaneous $1\rightarrow 0$ transition is given by the
Einstein $A$ coefficient that has the value of $A_{10} = 2.85 \times
10^{-15} sec^{-1}$. Such an extremely small value for Einstein-$A$
corresponds to a lifetime of the triplet state of $1.1\times10^7$
years for spontaneous emission. Despite its low decay rate, the 21~cm
transition line is one of the most important astrophysical probes,
simply due to the vast amounts of hydrogen in the Universe
\cite{ewen51, hulst45, muller51} as well as the  efficiency of  collisions and Lyman-$\alpha$ radiation in pumping the line
and establishing
the population of the triplet state \cite{wouthuysen52, field58}.   In this chapter I will describe the
basic physics behind this transition, especially what decides its
intensity.

\subsection{The 21~cm Spin and Brightness Temperatures}

The intensity of the 21~cm radiation is controlled by one
parameter, the so called spin temperature, $T_{spin}$. This temperature is 
defined through
the equation, 
\begin{equation}
\frac{n_1}{ n_0} = 3 \exp (- T_{\ast} / T_{spin}),
\end{equation} 
where $n_1$ and $n_0$ are the number densities of electrons in the triplet
and singlet states of the hyperfine level respectively, and $T_{\ast} = 0.0681\
\mathrm{K}$ is the temperature corresponding to the 21~cm wavelength.
The spin temperature is therefore, merely a shorthand for the ratio between the occupation
number of the two hyperfine levels. This ratio establishes the intensity of the radiation
emerging from a cloud of neutral hydrogen. Of course, in the measurement of such
radiation one has to take into account the level of background being transmitted through a given cloud 
as well as the amount of absorption and emission within the cloud. Namely, one has to use the equation of radiative transfer.

In the following derivation I follow the description in Rybicki and Lightman (\cite{rybicki86}).
 The radiative transfer equation is normally written in terms of 
 the {\it brightness} (or {\it specific intensity}) of the radiation $I_\nu$. 
This quantity is defined as the intensity per differential frequency element in the form,
$I_\nu = \frac{dI}{d\nu},$
where $\nu$
is the frequency. The intensity has the dimensions of ergs s$^{-1}$
cm$^{-2}$ sr$^{-1}$ Hz$^{-1}$, namely, it quantifies  the energy carried
by radiation traveling along a given direction, per unit area, frequency,
solid angle, and time. The radiative transfer equation 
for thermally emitting material at temperature $T$ can be written 
in terms of the optical depth for absorption as,
\begin{equation}
\frac{dI_\nu}{d\tau_\nu} = -I_\nu + B_\nu(T),
\label{eq:radtrans}
\end{equation}
where $\tau_\nu$ is the optical depth for absorption  through the cloud at a given frequency and $B_\nu$
is the Planck function.

\begin{figure} \centering
\includegraphics[width=0.9 \textwidth]{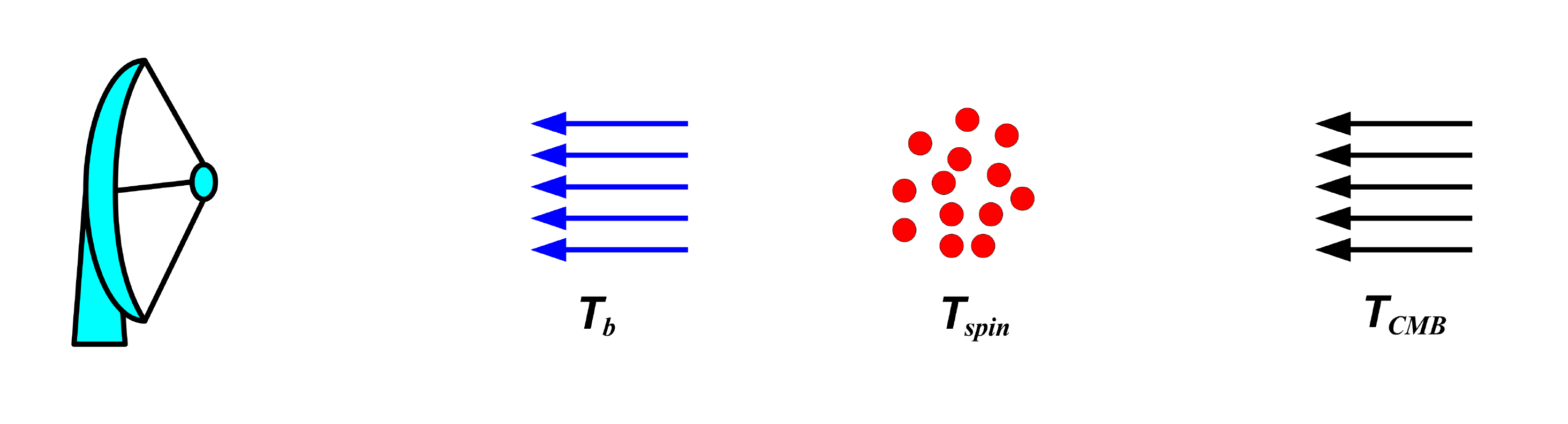}
\caption{A cartoon that shows that set up of the various components relevant to radiative transfer problem at hand starting from the background (CMB) radiation, going through a certain cloud with temperature $T_{spin} $ and emerging with a temperature $T_b$ that is 
measured by our telescopes}
\label{fig:RadTrans}
\end{figure}

 In radio astronomy the
intensity $I_{\nu}$ is often expressed by its
equivalent {\it brightness temperature}, $T_b(\nu)$. This is convenient because at
 the Rayleigh-Jeans low energy limit, the relation between
 the brightness temperature and specific intensity is given  by,
  \begin{equation}
  T_b(\nu)\approx I_{\nu} \, c^2/2k_B{\nu}^2,
  \end{equation}
where $c$ is the speed of light and $k_B$ is the Boltzmann's constant.
Expressing the radiative transfer equation~\ref{eq:radtrans} in terms of the 
brightness temperature gives it a particularly simple form,
\begin{equation}
\frac{dT_b}{d\tau_\nu} = -T_b + T_{CMB},
\label{eq:radtransT}
\end{equation}
where I substituted the CMB temperature for the background temperature.
Solving equation~\ref{eq:radtransT}  yields the temperature of the emergent radiation at frequency $\nu$,
\begin{equation} 
T_b(\nu) = T_{spin}(1-e^{-\tau_{\nu}})+T_{CMB}(\nu)e^{-\tau_{\nu}},
\label{eq:rad_trans}
\end{equation} 
where  $T_{spin}=T_b(0)$ is the brightness temperature in the absorbing cloud (see Figure~\ref{fig:RadTrans}).
Notice that  for the background radiation the
factor $\exp(-\tau_\nu)$ gives the transmission probability of the background
radiation whereas the $1-\exp(-\tau_\nu)$ factor gives the emission probability
of 21~cm photons from within the cloud. Therefore, in order to determine the brightness 
temperature, one needs to know the optical depth for absorption, $\tau_\nu$, and the spin temperature,
$T_{spin}$, in the optically thin regime relevant to our case. Notice that in the case in which $T_{spin}=T_{CMB}$
the brightness temperature gives exactly the CMB temperature. This is simply because in such a case there is a prefect balance 
between the absorption and emission at every frequency. Therefore, the measurement in  such a case does not reveal anything 
interesting about the intervening cloud, the subject we are interested in here.

I will first start with calculating the 21~cm optical depth.
The hyperfine transition of atomic hydrogen is an ideal transition to be
described by Einstein coefficients and their relations. The 21 cm radiation incident on the atom can cause $0 \rightarrow 1$
transitions (absorptions) and $1 \rightarrow 0$ transitions (induced
emissions) corresponding to Einstein coefficient $B_{01}$ and $B_{10}$ respectively. The probabilities are given by,
\begin{equation} I_\nu B_{01} = \frac{g_1}{g_0} B_{10} I_\nu,
\end{equation} and
\begin{equation} I_\nu B_{10}= A_{10}\frac{\lambda^2
I_\nu}{2h\nu_{10}},
\end{equation} respectively \cite{rybicki86}. Here $\nu_{10}=1420.4~\mathrm{MHz}$ is the frequency of 
the 21 cm transition. 

The 21 cm line absorption cross section is given by
\begin{equation} \sigma_{\nu} \equiv \sigma_{01} \phi(\nu) = \frac{3 c^2 A_{10}}{8\pi\nu^2}\phi(\nu),
\label{eq:sigma01}
\end{equation} 
where
$\phi(\nu)$ is the line profile defined so that $\int \mathrm{d} \nu
\, \phi(\nu) = 1$ and has units of time.

The optical depth of a cloud of hydrogen is then:
\begin{eqnarray} \tau_{\nu} & = & \int \mathrm{d} \ell \, \sigma_{01} \,
(1-e^{-E_{10}/k_B T_{spin}}) \, \phi(\nu) \, n_0 \label{eq:optdepth1} \\ &
\approx & \sigma_{01} \, \left(\frac{h\nu}{k_B T_{spin}}\right)
\left(\frac{N_{\HI}} {4}\right) \, \phi(\nu),
\label{eq:optdepth2}
\end{eqnarray}
where $N_{\HI}$ is
the column density of \THI and $\mathrm{d}\ell$ is a line element within the cloud. The factor of 4 connecting $n_0$ and \THI accounts for the
fraction of atoms in the hyperfine singlet state.  The second factor in equation~(\ref{eq:optdepth1})
with $E_{10}$ accounts for stimulated emission.  The approximate form in
equation~(\ref{eq:optdepth2}) assumes uniformity throughout the
cloud.
 
We now substitute for $\phi(\nu)$ and $N_\MHI$ using cosmological quantities.
In general, the line shape $\phi(\nu)$ includes natural, thermal, turbulent and
velocity broadening, as well as bulk motion (which increases the
effective Doppler spread).  Velocity broadening is the most important effect in the 
IGM. Hubble expansion of the gas results in velocity
broadening of a region of linear dimension $\ell$ will be $\Delta v\sim \ell H(z)$ so that 
$\phi(\nu)\sim c/(\ell H(z)  \nu)$. The column density along
such a segment depends on the neutral fraction ${\rm x}_{\MHI}$ of
hydrogen, so $N_{\HI} = \ell {\rm x}_{\HI} n_H(z)$ \cite{furlanetto06a}.  A more exact
solution of equation~(\ref{eq:optdepth1}) yields an expression
for the 21 cm optical depth of the diffuse IGM,
 \begin{eqnarray} \tau_{\nu_0} & = & \frac{3}{32 \pi} \, \frac{h c^3
A_{10}}{k_B T_{spin} \nu_0^2} \, \frac{\mathrm{x}_{H I} n_{H}}{(1+z) \,
(\mathrm{d} v_\parallel/\mathrm{d} r_\parallel)}
 \label{eq:optdepthcosmo} \\ & \approx & 0.0092 \, (1+\delta) \,
(1+z)^{3/2}\, \frac{\mathrm{x}_{H \, I}}{T_{spin}} \, \left[
\frac{H(z)/(1+z)}{\mathrm{d} v_\parallel/\mathrm{d} r_\parallel}
\right],
\end{eqnarray} 
where in the second relation, $T_{spin}$ is in degrees
Kelvin.  Here the factor $(1+\delta)$ is the fractional overdensity of
baryons and $\mathrm{d} v_\parallel/\mathrm{d} r_\parallel$ is the
gradient of the proper velocity along the line of sight, including
both the Hubble expansion and the peculiar velocity~\cite{kaiser87}.  In the second
line, we have substituted the velocity $H(z)/(1+z)$ appropriate for
the uniform Hubble expansion at high redshifts.

Next we need to calculate the spin temperature and substitute in Eq.~\ref{eq:rad_trans}.
In his seminal papers, George Field \cite{field58, field59b}, used the
quasi-static approximation to calculate the spin temperature,
$T_{spin},$ as a weighted average of the CMB temperature, $T_{CMB}$,
the gas kinetic temperature, $T_{kin}$, and the temperature related to
the existence of ambient Lyman-$\alpha$ photons, $T_{\alpha}$
\cite{wouthuysen52, field59b}. For almost all interesting cases, one can
safely assume that $T_{kin}=T_{\alpha}$ \cite{field58, furlanetto06a, 
madau97, morales10}.

Three competing processes determine $T_{spin}$: (1) absorption of CMB
photons (as well as stimulated emission); (2) collisions with other
hydrogen atoms, free electrons, and protons; and (3) scattering of Lyman~$\alpha$
photons through excitation and deexcitation.  Hence, the spin temperature could be recast as \cite{field58}:
\begin{equation} T_{spin}= \frac{T_{CMB} + y_{kin}
T_{kin} + y_{\alpha} T_{kin}}{1 + y_{kin} + y_{\alpha}},
\label{eq:tspin}
\end{equation} where $y_{kin}$ and $y_{\alpha}$ are the kinetic and
Lyman-$\alpha$ coupling terms, respectively.   It is
important to note that for the 21~cm radiation to be observed, it has
to attain a different temperature than that of the CMB background
\cite{field58, field59a, field59b, hogan79, wouthuysen52}. The form I use here
for Eq.~\ref{eq:tspin} is the original form used in the George Field's 1958 paper \cite{field58}, whereas some 
authors use a form that relates the inverse of the various temperatures. 
Both ways are of course equivalent  but one needs to be careful with the definitions of the
coupling coefficients in each case.

The kinetic coupling term $y_{kin}$
is due to collisional excitations of the 21~cm transitions. The
Lyman-$\alpha$ coupling term $y_\alpha$ is due to the so called Lyman-$\alpha$
pumping mechanism, also known as the Wouthyusen-Field effect,
which is produced by photo-exciting the hydrogen atoms to their Lyman
transitions \cite{field58, field59b, wouthuysen52}.
The coupling factors $y_{kin}$ and $y_{\alpha}$ depend on the rate of collisional and Lyman~$\alpha$ pumping
within the \THI cloud.  A number of authors have calculated these rates in detail 
\cite{allison69, liszt01, smith66, wild52, zygelman05}. In the case of first stars,
the  Wouthyusen-Field effect will depend on the intensity of the Lyman~$\alpha$ photons produced by these
sources. Collisions on the other hand are somewhat more complicated since it is normally done through
the so called secondary electrons which are released by the ionization of an \THI atom by an
x-ray photon. An electron with such high energy will lose it to the 
rest of the IGM through collisions. This energy will in general be divided between collisional excitation,
collisional ionization and heating \cite{furlanetto07, furlanetto10, shull85, valdes10}.

Since decoupling mechanisms can influence the spin temperature
in different ways, it is important to explore the decoupling issue for
various types of ionization sources. For instance, stars decouple the
spin temperature mainly through radiative Lyman~$\alpha$ pumping
whereas x-ray sources (e.g., mini-quasars)  decouple it through a combination of collisional
excitation and heating \cite{chuzhoy06, zaroubi07}, both
produced by the energetic secondary electrons ejected due to 
x-ray photons \cite{shull85}. The difference in the spin
temperature decoupling patterns of the two, will eventually help
disentangle the nature of the first ionization sources \cite{thomas08, pritchard07}.

Collisions could also be induced by Compton scattering of 
 the CMB photons off the residual free electrons in the IGM gas.
 This process is dominant at high redshifts $z\gsim 200$ and keeps the
gas temperature equal to that of the CMB.
However, it is not efficient enough at lower redshifts to heat the gas, it  is still
sufficient to couple the spin temperature to the gas down to $z\approx 100$.
In fact, one can show that the global spin temperature evolves in an intricate fashion
bouncing back and forth between the gas (kinetic) temperature  and the CMB temperature
based on which heating/excitations mechanism is dominant.

\begin{figure} \centering
   \includegraphics[width=0.9\textwidth,clip]{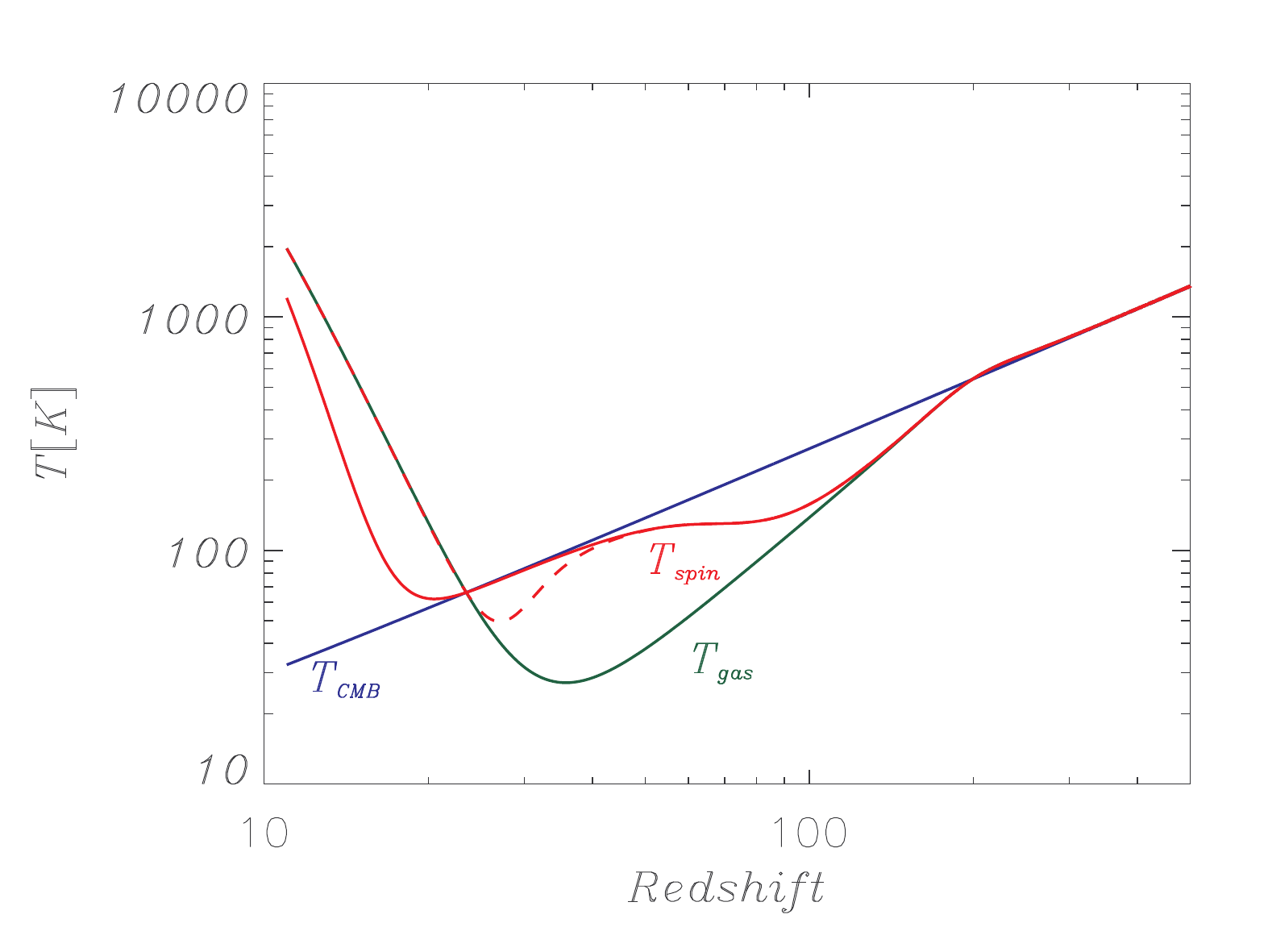}
   \caption{ The global evolution of the CMB (blue line),
gas (green line) and spin (red solid line and red dashed line)
temperatures as a function of redshift. The CMB temperature evolves
steadily as $1+z$ whereas the gas and spin temperatures evolve in a more
complicated manner (see text for detail).  \vspace{0.5 cm}}
            \label{fig:TspinHist}
           
    \end{figure}

Figure~\ref{fig:TspinHist} shows the expected global evolution of the
spin temperature as a function of redshift.  The blue solid line
represents $T_{CMB}$, which drops as $1+z$. The green line shows the
gas temperature as a function of redshift. At $z\gsim 200$, the gas
temperature is still coupled to the CMB due to Compton scattering 
of the background photons off
residual electrons leftover from the recombination era. At redshift
$\sim 200$, however, the gas decouples from the CMB radiation and starts
adiabatically cooling as a function of the redshift squared, $(1+z)^2$, until the first
objects start forming and heating up the gas at redshift below 30.
The spin temperature (shown by the red lines) has a somewhat more
complicated behavior. At $z\gsim100$ it is coupled to the gas temperature
due to collisional coupling caused by residual electrons leftover
from recombination. At $z\approx100$ the efficiency of collisional
coupling to the gas drops due to the Hubble expansion. At this stage,
the spin temperature starts veering towards $T_{CMB}$ until it is
completely dominated by it.  At lower redshifts the first
astrophysical objects that heat and ionize the IGM couple $T_{spin}$
to the gas. Here, broadly speaking, there are two possible histories,
one in which $T_{spin}$ couples to the gas as it heats up once it
obtains a temperature greater than $T_{CMB}$ (red solid line). In the
other possible evolution the spin temperature couples to the gas much
before the kinetic temperature exceeds that of the CMB (red dashed
line) \cite{baek09, pritchard08, thomas11}.  In the former case the 21~cm
radiation, after decoupling from the CMB at $z\lsim 30$, is seen only
in emission, whereas in the latter case it is seen initially in
absorption and only at later stages in emission.
    
 Currently all attempts to measure the redshifted 21~cm emission from the IGM are focused 
 on the redshift range $6 \lsim z \lsim 12$. This is due to a number of reasons that
 are related to the limitations posed by the ionosphere and the background noise 
 (see section~\ref{sec:21cmObs} for more detail). In this range  of redshifts  the spin temperature 
 is expected to be set by the astrophysics of the first objects in the Universe, namely, gas physics, feedback, etc.,
 which often involve very complicated and poorly understood processes. 
 However, observing the spin temperature of the Universe within the redshift window around $z\approx 50-100$
 will mostly probe the cosmological density field \cite{loeb04}. Such a measurement could provide a vast amount of 
 information about the pristine Universe that, given the span of its redshift coverage, could potentially  exceed that of the CMB data.
 Unfortunately however, the ionosphere at such frequencies $\nu \lsim 30~\mathrm{MHz}$ poses insurmountable
 hurdles that render such attempts futile. This has led some authors to propose setting up
 radio telescopes at these very low frequencies on the moon (see e.g., \cite{lazio09}).  

\subsection{The Differential Brightness Temperature}

%\textbf{4.2 \textit{The Brightness Temperature}}

As we mentioned above the measured quantity in radio astronomy is the brightness temperature, or more accurately 
the so called differential brightness temperature $\delta T_b \equiv T_b -T_{CMB}$ which reflects the
fact the only meaningful brightness temperature measurement insofar as the IGM is concerned is when it deviates from $T_{CMB}$. 
In order to get this quantity one should substitute the various components into Equation~\ref{eq:rad_trans}. 
Such a substitution and rearrangement
yields, \cite{field58, field59b, madau97, ciardi03a},
\begin{equation} 
\delta T_b = 28 \mathrm{{mK}}  \left( 1 + \delta
\right) x_{\HI} \left( 1 - \frac{T_{CMB}}{T_{spin}} \right) \left( \frac{\Omega_b h^2}{0.0223} \right)
\sqrt{\left( \frac{1 + z}{10} \right) \left( \frac{0.24}{\Omega_m}
\right)} \, \left[ \frac{H(z)/(1+z)}{\mathrm{d}
v_\parallel/\mathrm{d}r_\parallel} \right],
\label{eq:dTb}
\end{equation}
where $h$ is the Hubble constant in units of $100~
\mathrm{{km} \, s^{- 1} {Mpc}^{- 1}}$, $\delta$ is the mass density
contrast, $x_{\HI}$ is the neutral fraction, and $\Omega_m$ and
$\Omega_b$ are the mass and baryon densities in units of the critical
density. Note that the three quantities, $\delta$, $x_{\HI}$ and
$T_{spin}$, are all functions of 3D position. The term $(T_{spin}-T_{CMB})/T_{spin}$ can obtain a maximum 
of +1 for $T_{spin} \gg T_{CMB}$, i.e., in the emission case. It has no such bound for the case of 
$T_{spin} \ll T_{CMB}$ and can be very negative in the absorption case.

Equation~\ref{eq:dTb} shows that the differential brightness temperature is composed of a mixture of cosmology dependent 
and astrophysics dependent terms. This makes the equation a complex yet also a very 
informative one. This is simply because at different stages in the evolution of this field
$\delta T_b$ is dominated by different contributions. For example, at high redshifts and before
significant ionization takes place, i.e. $x_\MHI \approx 1$, everywhere  the brightness temperature 
is proportional to the density fluctuations making its measurement an excellent probe of cosmology.
However, at low redshifts ($z\lsim 7$) a significant fraction of the Universe is expected to be ionized and the measurement
is dominated by the contrast between the neutral and ionized regions, hence, probing the 
astrophysical source of ionization (see e.g., \cite{iliev08, thomas09}). Here I assumed that $T_{spin}\gg T_{CMB}$
at all redshifts.
Figure~\ref{fig:EoRslice}, which have discussed before, shows a typical distribution of the
differential brightness temperature. The figureis taken from the
simulations of Thomas et al. ~ \cite{thomas09}.

Most radiative transfer simulations assume that the spin temperature 
is much larger than the CMB temperature, namely the term
$\left(1-{T_{CMB}}/{T_{spin}}\right)$ in eq.~\ref{eq:dTb} is unity.  As
figure~\ref{fig:TspinHist} shows, this is a good assumption at the
later stages of reionization, however, it is probably not valid at the early
stages. Modeling this effect is somewhat complex and requires
radiative transfer codes that capture the Lyman-$\alpha$ line
formation and multifrequency effects, especially those coming from
energetic photons \cite{baek09, mesinger10, pritchard07, thomas11}.

Here we show the evolution of the brightness temperature for three reionization histories:
(1) With reionization, excitation and heating dominated by power law sources (miniqsos with x-rays);
(2) dominated by thermal (stellar) sources; (3) dominated by a mixture of the aforementioned two types of sources. 
To create a contiguous observational cube or
``frequency cube'' (right ascension (RA) $\times$ declination (DEC)
$\times$ redshift), the RA and DEC
slices, taken from individual snapshots at different redshifts (or
frequency), are stacked and interpolated smoothly to create a reionization
history. This datacube is then convolved with the point spread
function of the LOFAR telescope to simulate the mock data cube of the
redshifted 21-cm signal as seen by LOFAR. For further details on
creating this cube, refer to \cite{thomas09, thomas11}.

\begin{figure} \centering \hspace{0cm}
\includegraphics[width=0.99 \textwidth]{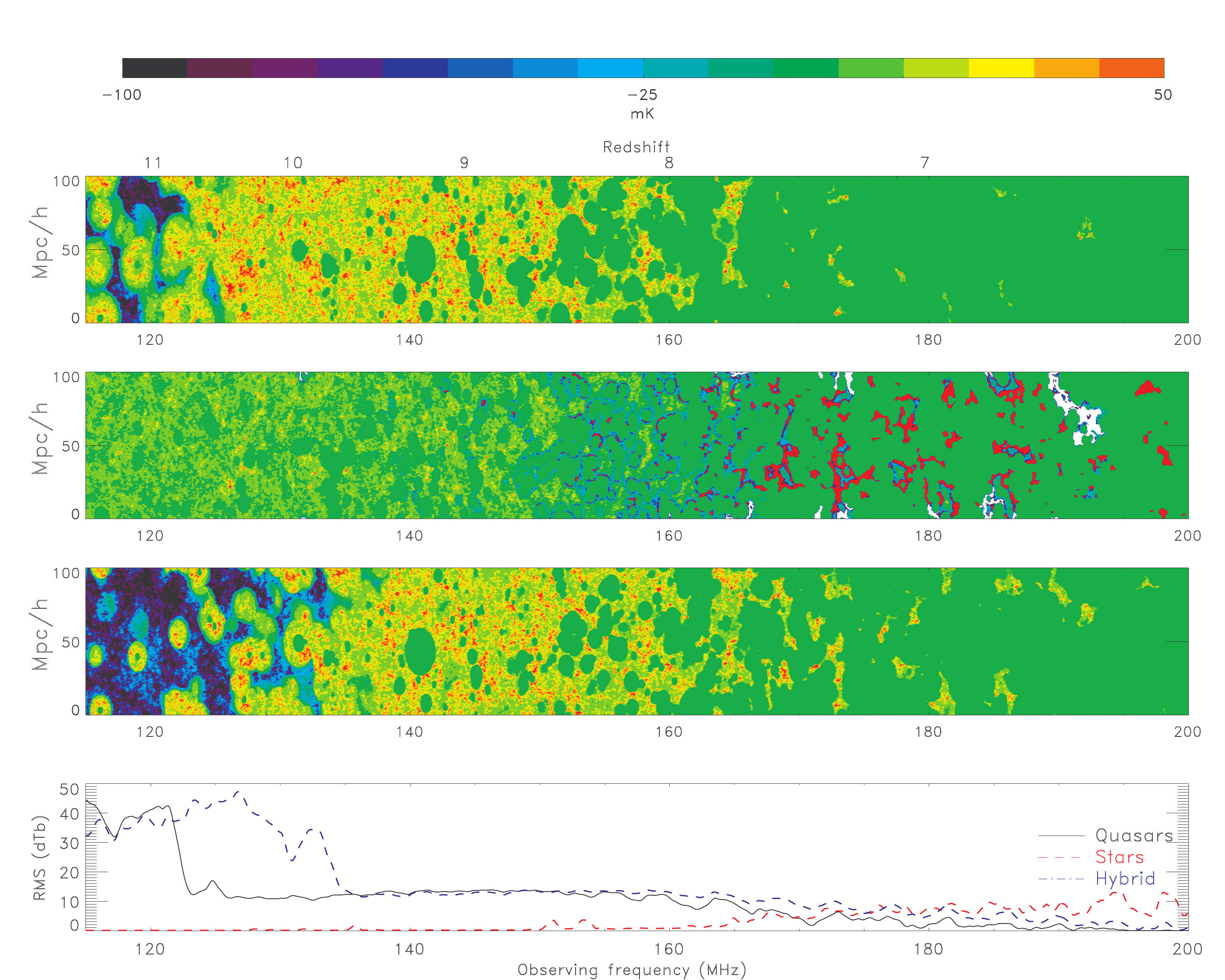}
\vspace{-.2cm}
\caption{\emph{Contrasting reionization histories}: From the top,
reionization histories ($\delta \mathrm{T_b}$ in mK as a function of
frequency or redshift) are plotted for miniqso, stellar and hybrid
sources, respectively. The bottom panel plots the r.m.s of $\delta
\mathrm{T_b}$ as a function of redshift/frequency for all the three
cases. This figure is taken from \cite{thomas11}.}
\label{fig:histcomp}
\end{figure}

As expected, the signatures (both visually and in terms of the r.m.s) of the
three scenarios (Fig.~\ref{fig:histcomp}) are markedly different. In
the miniqso-only scenario, reionization proceeds extremely quickly and
the Universe is almost completely ($\mathrm{x_{\HII}}> = 0.95$)
reionized by around redshift 7. The case in which stars are the only
source sees reionization end at a redshift of 6. Also in this case,
compared to the previous one, reionization proceeds in a rather
gradual manner. The hybrid model, as explained previously,
is in between the previous two scenarios.

In the models shown here, the transition from the absorption dominated brightness 
temperature to the emission dominated one occurs at relatively low redshifts. 
The transition redshift depends sensitively on the assumptions made in each case.
Other authors have explored such effects and conclude that the transition occurs at much higher
redshifts (see e.g., the models in \cite{mesinger10, santos10}).

The $\delta \mathrm{T_b}$ in Fig.~\ref{fig:histcomp} is calculated
based on the effectiveness of the radiation flux, produced by the source,
in decoupling the CMB temperature ($\mathrm{T_{CMB}}$) from the spin
temperature ($\mathrm{T_s}$). This flux, both in spatial extent and
amplitude, is obviously much larger in the case of miniqsos compared
to that of stars, resulting in a markedly higher brightness
temperature in both the miniqso-only and hybrid models when compared
to that of the stars. However, we know that stars themselves produce
Lyman~$\alpha$ radiation in their spectrum. Apart from providing
sufficient Lyman~$\alpha$ flux to their immediate surroundings, this
radiation builds up as the Universe evolves into a strong background
\cite{ciardi03a}, potentially filling the Universe with enough
Lyman~$\alpha$ photons to couple the spin temperature to the
kinetic temperature everywhere. 
It has to be noted that the results we are discussing here are
extremely model dependent and any changes to the parameters can
influence the results significantly.

\subsection{\textbf{The 21 cm forest at high z}} 
Finally, I will conclude this section by discussing a very different aspect of the redshifted
21~cm radiation, and that is the case of the 21~cm forest.
Very bright radio
sources might exist at high redshifts. In such a case, the emission from these sources 
is expected
to be resonantly absorbed by the neutral IGM and form a system of
absorption features just like the Lyman~$\alpha$ forest seen in the
spectra of distant quasars. Such absorption features are called the
21 cm forest and they were first investigated by Carilli et al.~\cite{carilli02} 
and subsequently by 
other authors~\cite{carilli04b, furlanetto02, furlanetto06b, mack11, xu09}. 
The discovery of such systems will provide very valuable information about 
the reionization process and the IGM's
physical properties during the EoR which will be largely independent of calibration errors 
(see section~\ref{sec:21cmObs}). Currently, we know of no very
bright high redshift sources, but with the imminent availability of
highly sensitive radio telescopes like LOFAR and SKA, the prospects  for
detecting such sources are very promising.

\begin{figure} \centering
\vspace{0.2 cm}
\includegraphics[width=0.49\textwidth]{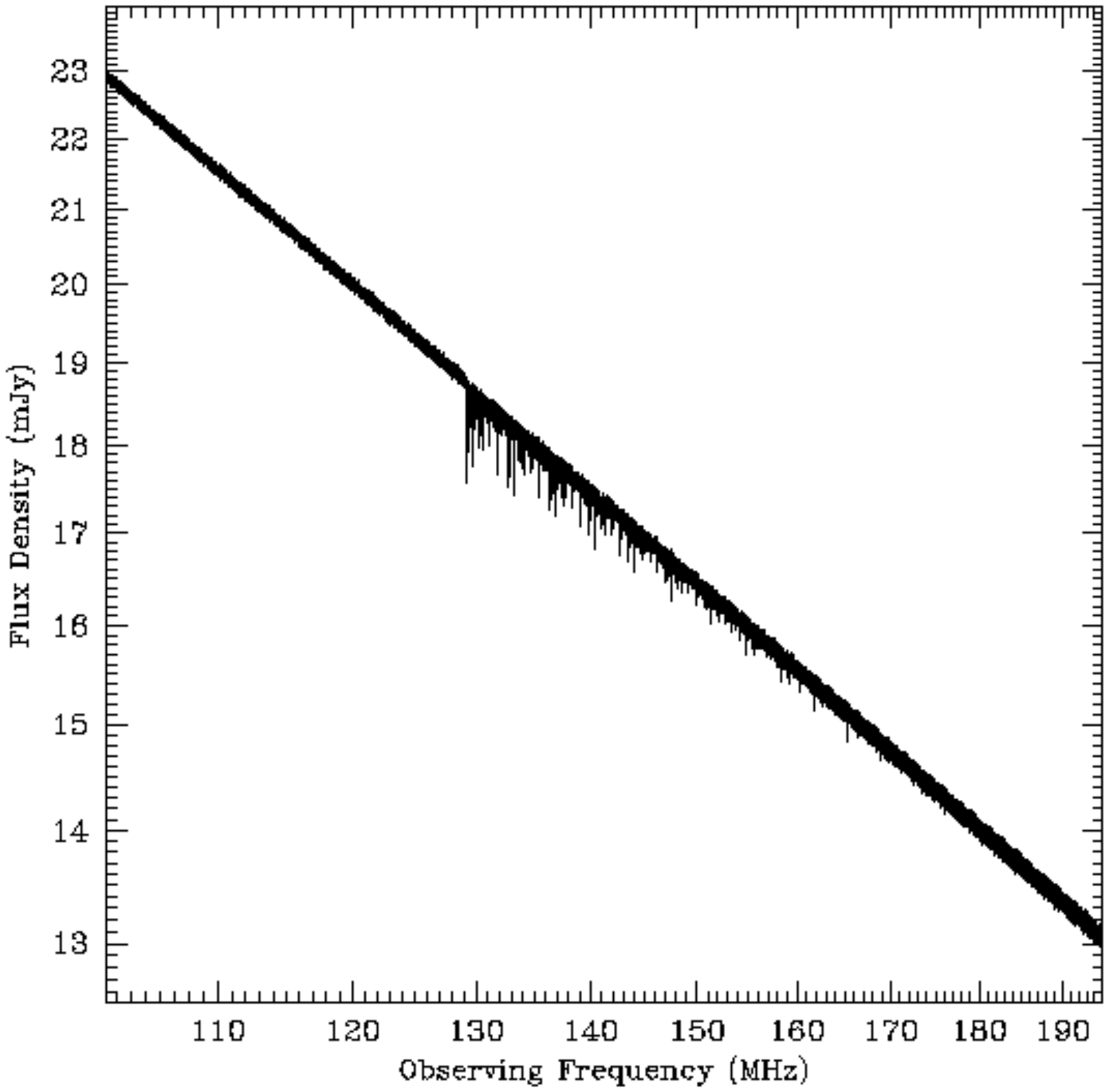}
\includegraphics[width=0.49\textwidth]{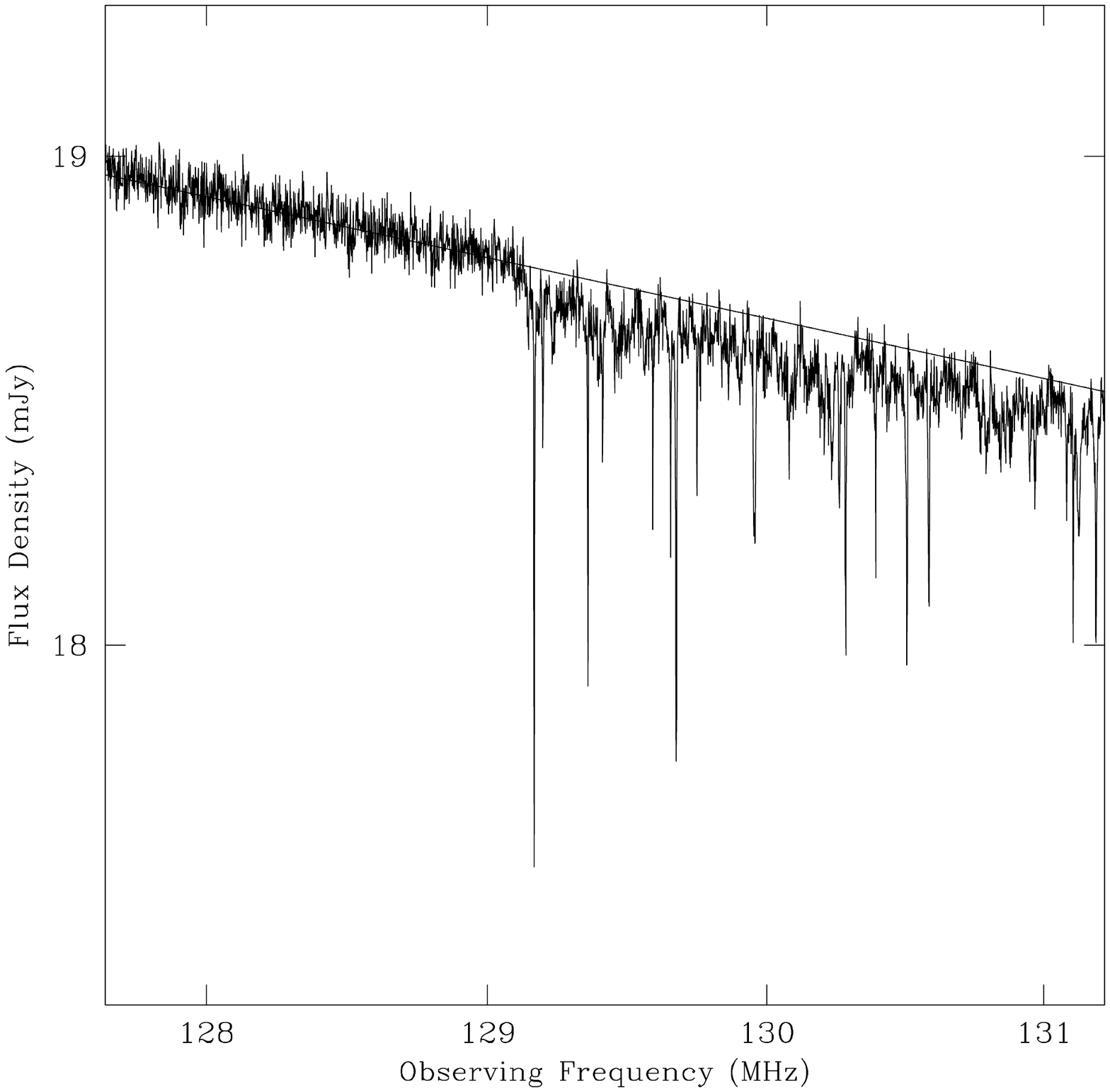}
\vspace{-0.8 cm}
\caption{ \textbf{Left hand panel:} A simulated spectrum
from 100 MHz to 200 MHz of a source with S$_{120}$ = 20 mJy at $z =
10$ using the Cygnus A spectral model and assuming \THI 21cm absorption
by the IGM.  Thermal noise has been added using the specifications of
the SKA and assuming 10 days integration with 1 kHz wide spectral
channels.  \textbf{Right hand panel:}The same as the left panel, but showing an
expanded view of the spectral region around the frequency
corresponding to the redshift \THI 21cm line at the source redshift (129
MHz).  The solid line is the Cygnus A model spectrum without noise or
absorption.  Figure taken from \cite{carilli02} }.
\label{fig:21cmForest}
\end{figure}

Figure~\ref{fig:21cmForest}  shows a simulated spectrum at 1 kHz resolution
of a $z = 10$ radio source
with a flux density of 20 mJy at an observing frequency of
120 MHz  (S$_{120}$). The implied  luminosity density at a rest frame
frequency  of 151 MHz is then $P_{151} = 2.5\times10^{35}$ erg s$^{-1}$
Hz$^{-1}$.  The left hand panel of Figure~\ref{fig:21cmForest} shows a spectrum covering a large
frequency range (100 MHz to 200 MHz, or HI 21cm redshifts
from 13 to 6), whereas the right hand panel shows an expanded view of the frequency 
range corresponding to the HI 21cm line at the source redshift
(129 MHz). At 129 MHz the spectrum shows a  1$\%$ drop 
 due to the diffuse neutral IGM. See reference \cite{carilli02} for detail.

% sec 6

\section{The redshifted 21~cm Observation}
\label{sec:21cmObs}

 In section~\ref{sec:21cmProbe} we discussed the cosmological 21~cm
signal and showed that it is expected to be on the order of $\approx
10$~mK. However, the detectable signal in the frequency range that
corresponds to the epoch of reionization is composed of a number of
components each with its own physical origin and statistical
properties.  These components are: (1) the 21~cm signal coming from
the high redshift Universe. (2) galactic and extra-galactic
foregrounds, (3) ionospheric influences, (4) telescope response effects
(5) Radio frequency interference (RFI)~\cite{offringa10a, offringa10b} and (6) thermal noise (see Figure~\ref{fig:EoRHist}).  Obviously, the
challenge of the experiments in the low frequency regime is to distill the cosmological
signal out of this complicated mixture of influences. This will depend
crucially on the ability to calibrate the data very accurately
so as to correct for the ever changing ionospheric effects and
variation of the instrument response with time. 

Currently, there are two types of redshifted 21~cm experiments that are attempting to observe 
the EoR. The first type are experiments that measure the global (mean) radio signal at the frequency range of 
 $\nu=[100-200]~\mathrm{MHz}$ averaged over the whole sky (hemisphere) as a function of frequency. In this radio
 signal the 21~cm radiation from the EoR is hidden. The expected measurement should show an increase of the 
intensity at higher redshifts due to the increase in the neutral fraction of \THI. In particular, if the reionization process occurred rapidly
such a measurement should exhibit a step-like jump in the mean brightness temperature at the redshift of reionization 
$z_i$ (in this case $z_i$ is well defined).
This type of measurement is cheap and relatively easy to perform.  However, given the amount of foreground contamination,
especially from our Galaxy, radio frequency interference (RFI), noise and calibration errors  as well as the limited amount of information contained in the data (mean intensity as a function of redshift), such experiments are in reality very hard to perform. One 
of those experiments, EDGES~\cite{bowman08}, has recently reported a lower limit on the duration of the reionization 
process to be $\Delta z> 0.06$, thus providing a very weak constraint on reionization as most realistic simulations predict that 
this process occurs over a much larger span of redshift~\cite{bowman10}.

The second type of experiment is interferometric experiments carried out in the frequency range of $\approx 100-200$~MHz, corresponding to
a redshift range of $\approx 6-12$. This type of experiment is considered more promising. The main reason
for this is that these experiments allow better control of what is being measured and contain a huge amount of information so as
to allow a much more accurate calibration of the instrument.  Furthermore, radio interferometers are more  diverse instruments 
that can be used to study many scientific topics besides the EoR, which makes them appealing for a much wider community. 
Having said that, however, one should note that 
the cost involved in building and running  such facilities is much higher than for the global signal experiments.  

\begin{figure}
\centering
\includegraphics[width=0.48\textwidth]{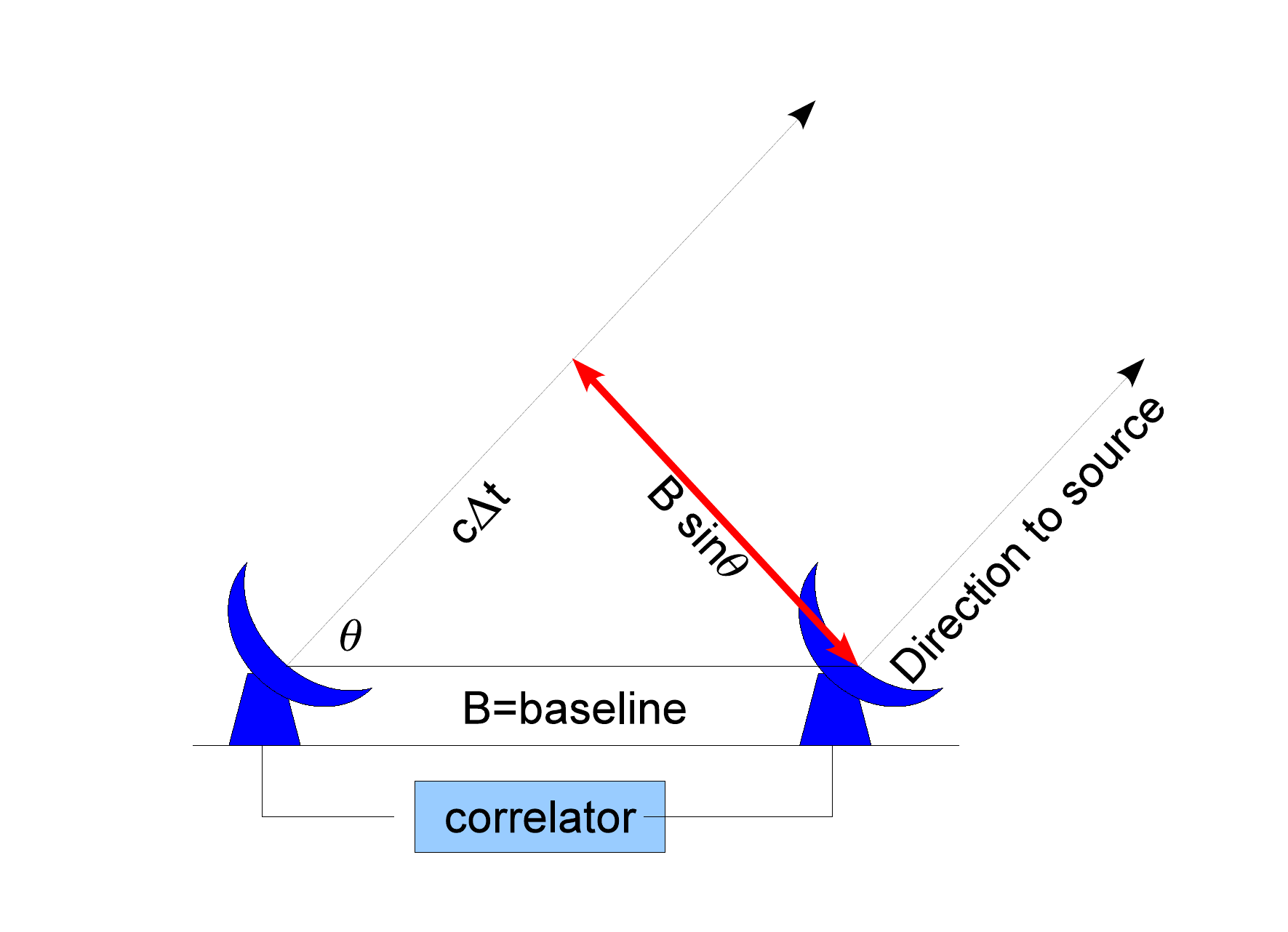}
\hspace{0.01\textwidth}
\includegraphics[width=0.48\textwidth]{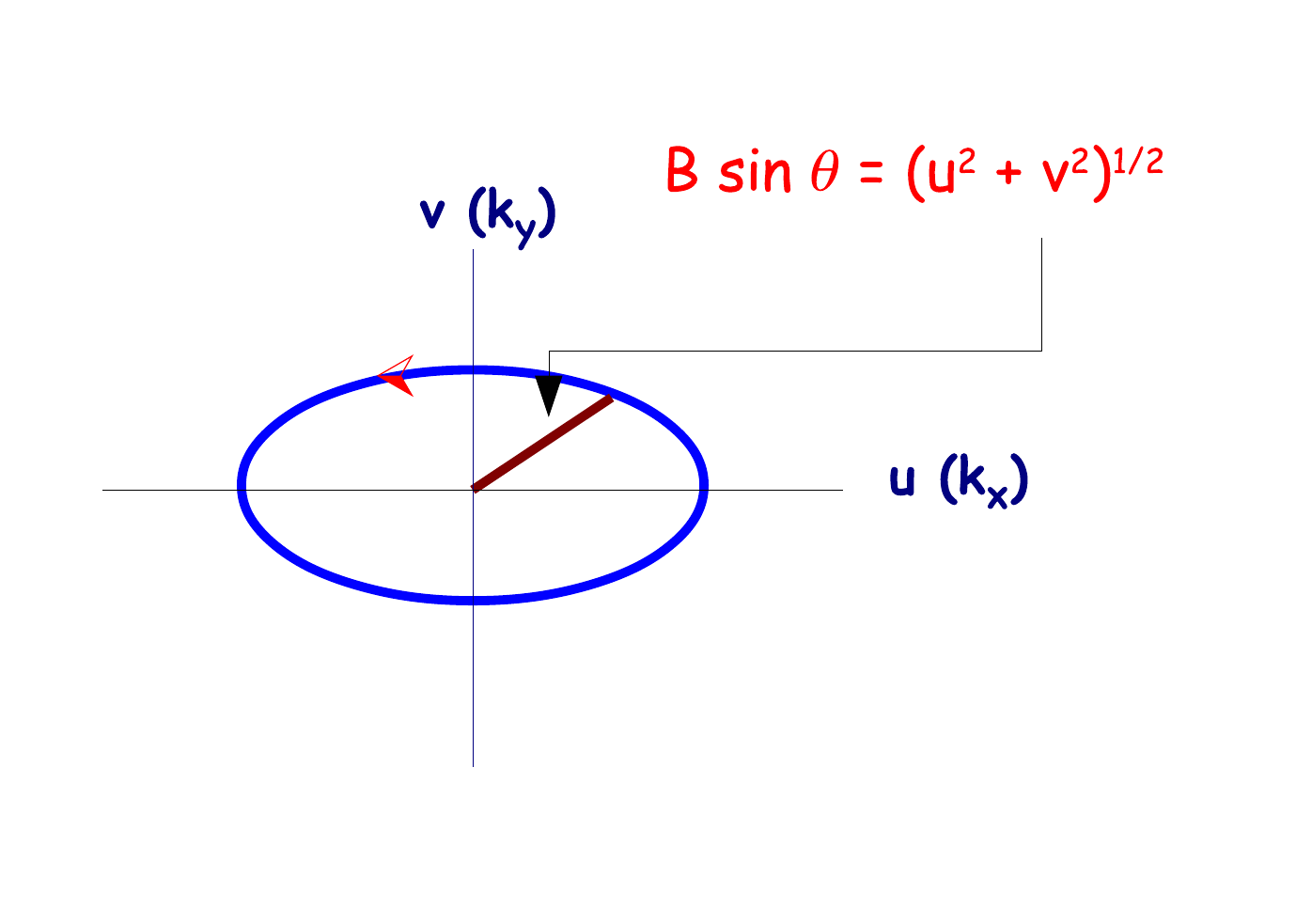}
\caption{Left hand panel: A sketch showing the basic principle of radio interferometry, the delay time
$\Delta t$ between the two antennas is set by the direction of the observed object on the sky.
Right hand panel: The projection of the baseline on the sky gives the uv point measured at time, t.
The rotation of earth produces a track in the uv-plane which completes half of the drawn track after 12 hours.
The other half is produced due to the fact that the intensity is a real
function, so its Fourier transform has a complex symmetry. The width of the uv-track is set by the size of 
the station, i.e., the larger the station the thicker its tracks. In this case we assumed an east-west baseline, other baselines will produce upper half track and lower half track that are not part of the same ellipse. 
}
\label{fig:interferometer}
\end{figure}

\subsection{Radio Interferometry and the Calibration Problem}

Interferometers measure the spatial correlation of the electric field   vector emanating from a distant source in the sky, 
$\mathbf{E}(\mathbf{R},t)$, located at position $\mathbf{R}$ and measured at time $t$.
The sketch presented in the left hand panel of Figure~\ref{fig:interferometer} shows the basic principle of interferometry. The two stations (dishes) 
receive a wavefront from a distant source and the receivers are timed to account for the difference in the
pathway to the two stations which obviously depends on the source location on the sky. The signals measured
at the two stations, taken with the appropriate time difference $\Delta t$, are then cross-correlated (see for example~\cite{taylor99, thompson01}).

The measured spatial correlation of the electric field between two
interferometric elements (stations) $i$ and $j$ is called the ``visibility" and is approximately given
by \cite{taylor99, thompson01}:
\begin{equation}\label{vis}
  V_{\nu}(\mathrm{u,v})=\int A(l,m; \nu)I_\nu(l,m)e^{i(\mathrm{u}l+\mathrm{v}m)}dldm,
\end{equation}
where $A$ is the normalized station response pattern and $I_\nu$ is the  observed intensity at frequency $\nu$.
The coordinates $l$ and $m$ are the projections (direction cosines) of the source in terms of  the baseline 
in units of wavelength. As a side note,  here we ignore the effect
of the Earth's curvature, the so called w-projection. From this equation it is clear that the observed visibility
is basically the Fourier transform of the intensity measured at the coordinates u and v. Notice that coordinates u and v
depend on the baseline and its direction relative to the source position (see the right hand panel of Figure~\ref{fig:interferometer}). 
Therefore, the coordinates u and v produced by a given baseline vary with time due to Earth's 
rotation and will create an arc in the uv plane that completes half of the drawn track after 12 hours
as seen in the right hand panel of Figure~\ref{fig:interferometer}.
The other half is produced due the fact that the intensity is a real function. The width of the uv-track is set by the size of 
the station, such that large stations produce thick tracks. I will discuss the issue of uv coverage in more detail below.
One also should note that the coordinates u and v are a function
of wavelength, namely their value will change as a function of frequency, which one has to take into account
when combining or comparing results from different frequencies.

In the interferometric visibilities there always exist errors introduced
by the sky, the atmosphere (e.g. troposphere and ionosphere),
the instrument (e.g. beam-shape, frequency response, receiver
gains etc.) and by Radio Frequency Interference (RFI). The
process of estimating and reducing the errors in these measurements
is called ÒcalibrationÓ and is an essential step before understanding the measured data.
Calibration normally involves knowing very well the position and intensities of the bright
sources within and without the field of view of the radio telescope and using them to correct
for the ionospheric and instrumental effects introduced into the data~\cite{hamaker96, kazemi11, pearson84, yatawatta09}. 
This is similar to the adaptive optics techniques used in the optical regime except that here one needs to account for 
the variations in polarization of the radiation as well as in its total intensity. 

Since most current instruments are composed of simple dipoles as their fundamental elements which
have a polarized response (preferred x and y direction), the main danger in insufficient calibration lies 
in the possible leakage of 
polarized components into total intensity, thereby severely polluting the signal (see e.g., \cite{jelic10b}).
That is to say, since the cosmological signal is not expected to be polarized, if  the polarized response
of the instrument is not very well understood and taken into account it will mix some of the polarization
that exists in the Galactic foregrounds (see subsection~\ref{sec:foregrounds}) with the cosmological signal and create a spurious
signal that can not be distinguished from the cosmological signal. Hence, a very accurate calibration of
these instruments is absolutely needed. Another issue one needs to deal with is that of the Radio 
Frequency interference, but we 
will not discuss it here and refer the reader instead
to the papers by Offringa et. al.~\cite{offringa10a, offringa10b}.

\subsection{Current and Future EoR Experiments}

Currently, there are
a number of new generation radio telescopes,
GMRT\footnote{Giant
Metrewave Telescope, http://gmrt.ncra.tifr.res.in}, LOFAR\footnote{Low
Frequency Array, http://www.lofar.org}, MWA\footnote{Murchinson
Widefield Array, http://www.mwatelescope.org/}, 21CMA\footnote{21
Centimeter Array, http://21cma.bao.ac.cn/} and PAPER\footnote{Precision
Array to Probe EoR, http://astro.berkeley.edu/$^\sim$dbacker/eor}, that plan to capture the lower
redshift part of the $\delta T_b$ evolution ($z\lsim 12$).  Unfortunately, however, none of these experiments
has enough signal-to-noise to provide images of the EoR as it evolves with redshift. Instead,
they are all designed to detect the signal statistically. In what follows I will focus on LOFAR more 
than the other telescopes, simply because this is the instrument I know best, but the general points I will
make are applicable to the other telescopes as well.

\begin{figure*}
\centering
\hspace{0cm}
\includegraphics[width=0.44\textwidth]{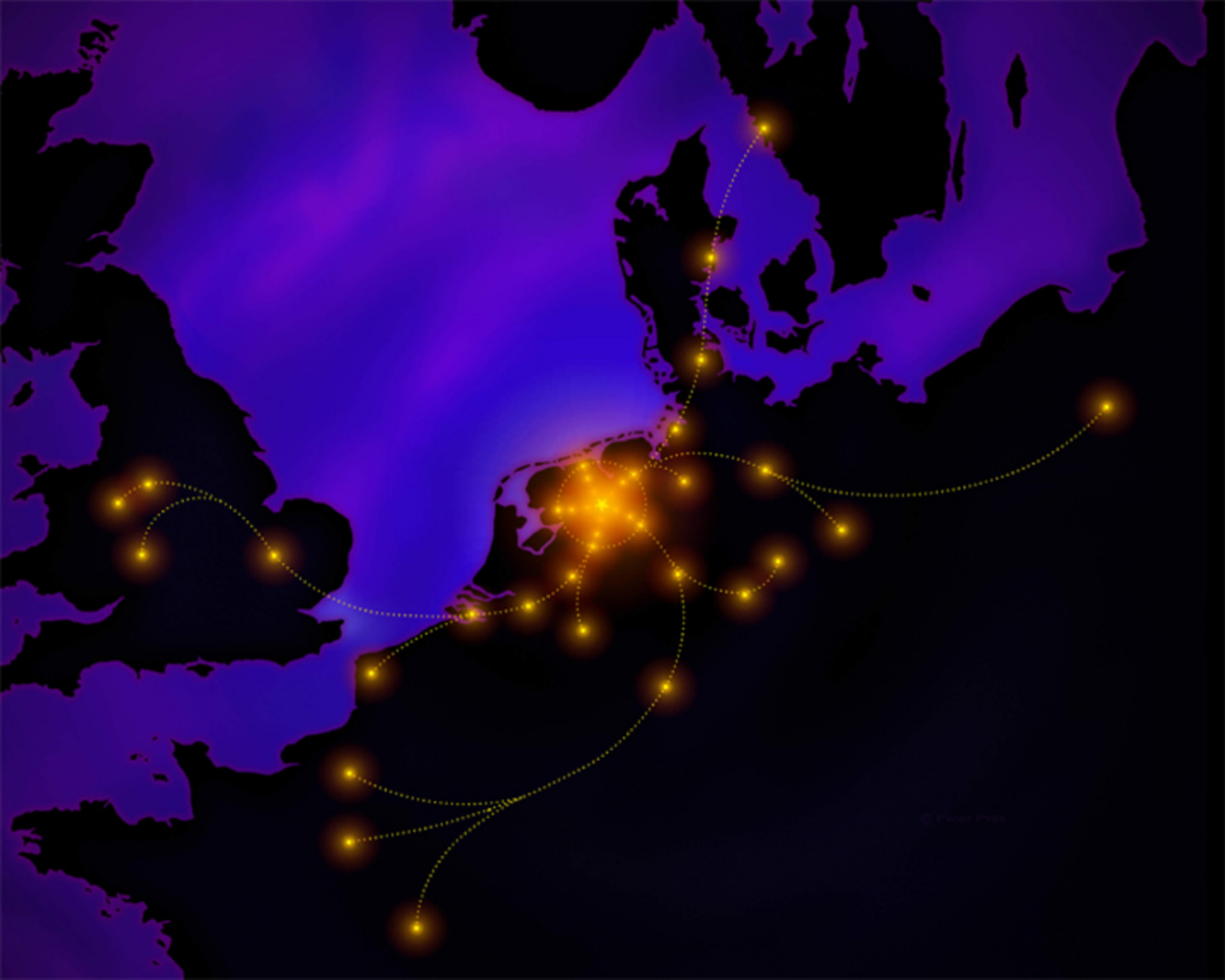}
\hspace{0.02\textwidth}
\includegraphics[width=0.47\textwidth]{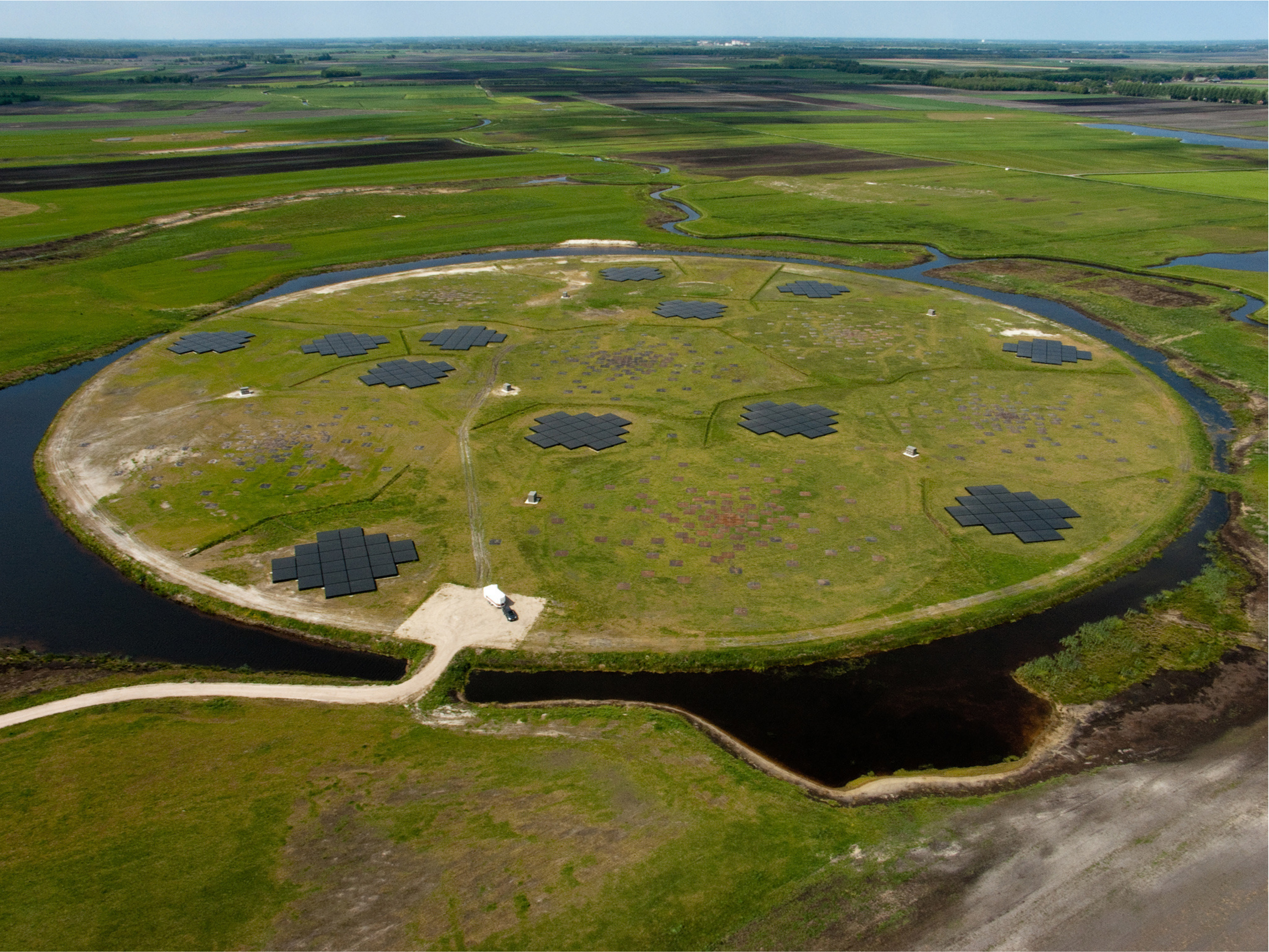}
\caption{Left hand panel: An artistÕs impression of the layout of the LOFAR telescope over Western Europe. For the
EoR, only the central part of the telescope is relevant. (courtesy of Peter Prijs). Right hand panel: The very central area of 
LOFAR. This circular area is know as the superterp and is the heart of the LOFAR core. The high-band array stations (covered in black blastic sheets) are clearly seen in this picture. In between one can also see the Low-Band Array antennas.}
\label{fig:LOFAR}
\end{figure*}

The LOw Frequency ARray (LOFAR) is a European telescope built mostly in the Netherlands and
has two observational bands, a low band and a high band covering the frequency range of 30-85 MHz and 115-230 MHz, respectively. 
The high
band array is expected to be sensitive enough to measure the
redshifted 21~cm radiation coming from the neutral IGM within the
redshift range of z=11.4 (115 MHz) to z=6 (203 MHz), with a resolution
of 3-4 arcminutes and a typical field of view of $\sim 120$ square
degrees (with 5 beams) and a sensitivity on the order of 80 mK per
resolution element for a 1 MHz frequency bandwidth.  At frequencies
below the FM band, probed by the low band array, the LOFAR sensitivity
drops significantly and the sky noise increases so dramatically (roughly like $\approx \nu^{-2.6}$) that
detection of \THI signals at these frequencies is beyond the reach of
LOFAR \cite{harker10,jelic08,panos09} and all other current generation telescopes for that  matter.  Figure~\ref{fig:LOFAR} shows
an artistic impression of the LOFAR telescope and its spread over Europe (left hand panel).
The right hand panel shows the inner most center of the core located in the north of the Netherlands and
 shows the two types of stations used in the array.

 In the future, SKA\footnote{Square Kilometer Array, http://www.skatelescope.org/}~\cite{carilli04} 
 will significantly improve on the current
instruments in two major ways.  Firstly, it will have at least an
order of magnitude higher signal-to-noise which will allow us to actually image
the reionization process. It will also  give us access to the Universe's
\emph{dark ages} up to redshifts as high as $z\approx 30$ (assuming lowest frequency of about 50~MHz), 
hence, providing crucial information about cosmology
which none of the current telescopes is able to probe. Thirdly,  SKA will
have a resolution better by a factor of few, at least, relative to the
current telescopes~\cite{zaroubi10}. These three advantages -- sensitivity, resolution and frequency coverage --  will not only improve on the
understanding we gain with current telescopes but give the opportunity
to address a host of fundamental issues that current telescopes will
not be able to address at all. Here, I give a few examples: (1) due to
the limited resolution and poor signal-to-noise of current telescopes, the nature of the
ionizing sources is expected to remain poorly constrained; (2) the
mixing between the astrophysical effects and the cosmological
evolution is severe during the EoR but much less so during the
\emph{dark ages}, an epoch beyond the reach of the current generation
of telescopes, but within SKA's reach; (3) at redshifts larger than 30, the
21~cm could potentially provide very strong constraints, potentially much more so
than the CMB, on the primordial non-gaussianity of the cosmological
density field, which is essential in order to distinguish between
theories of the very early Universe (e.g., between
different inflationary models).

\subsection{Station configuration and uv coverage}

In principle, Fourier space measurement and real space measurement are equivalent. However, this is only true
if one has a perfect coverage of both spaces. In reality, each baseline will cover a certain line in the so called uv plane which needs 
to be convolved with the width of the track (see right hand panel of Figure~\ref{fig:interferometer}).
The combination of all the tracks of the array produces the uv coverage of the interferometer.
The low frequency arrays must be configured so that they have a very
good uv coverage. This is crucial to the calibration effort of the
data where a filled uv plane is important for obtaining precise Local
\cite{nijboer06} and Global \cite{smirnov04} Sky models (LSM/GSM; i.e.
catalogues of the brightest, mostly compact, sources in and outside of
the beam, i.e. local versus global).  It is also crucial for the
ability to accurately fit for the foregrounds \cite{harker09b,
jelic08} and to the measurement of the EoR signal power spectrum
\cite{bowman06, harker10, hobson02, santos05}.

 The uv coverage of an interferometric array depends on the layout of
the stations (interferometric elements), their number and size as well
as on the integration time, especially, when the number of stations is
not large enough to have a good instantaneous uv coverage.

 For a given total collecting area one can achieve a better uv
coverage by having smaller elements (stations). For example LOFAR has
chosen to have large stations resulting in about $\approx 10^3$
baselines in the core area.  Such a small number of baselines needs
about 5-6 hours of integration time per field in order to fill the uv
plane (using the Earth's rotation). In comparison, MWA which has
roughly 1/3 of the total collecting area of LOFAR but chose to have smaller
stations with about $\approx 10^5$ baselines resulting in an almost
instantaneous full uv coverage.

 The decision on which strategy to follow has to do with a number of
considerations that include the ability to store the raw visibilities,
hence, allowing for a better calibration and an acceptable noise level
for both the foreground extraction needs as well the power spectrum
measurement (see the following sections; sec.~\ref{sec:noise} and
sec.~\ref{sec:foregrounds}).  A compromise between these issues as
well the use of the telescopes for scientific projects other than the EoR
is what drives the decision on the specific layout of the antennas.

\subsection{Noise Issues}
\label{sec:noise}

  \begin{figure} \centering
    \includegraphics[width=0.9\textwidth]{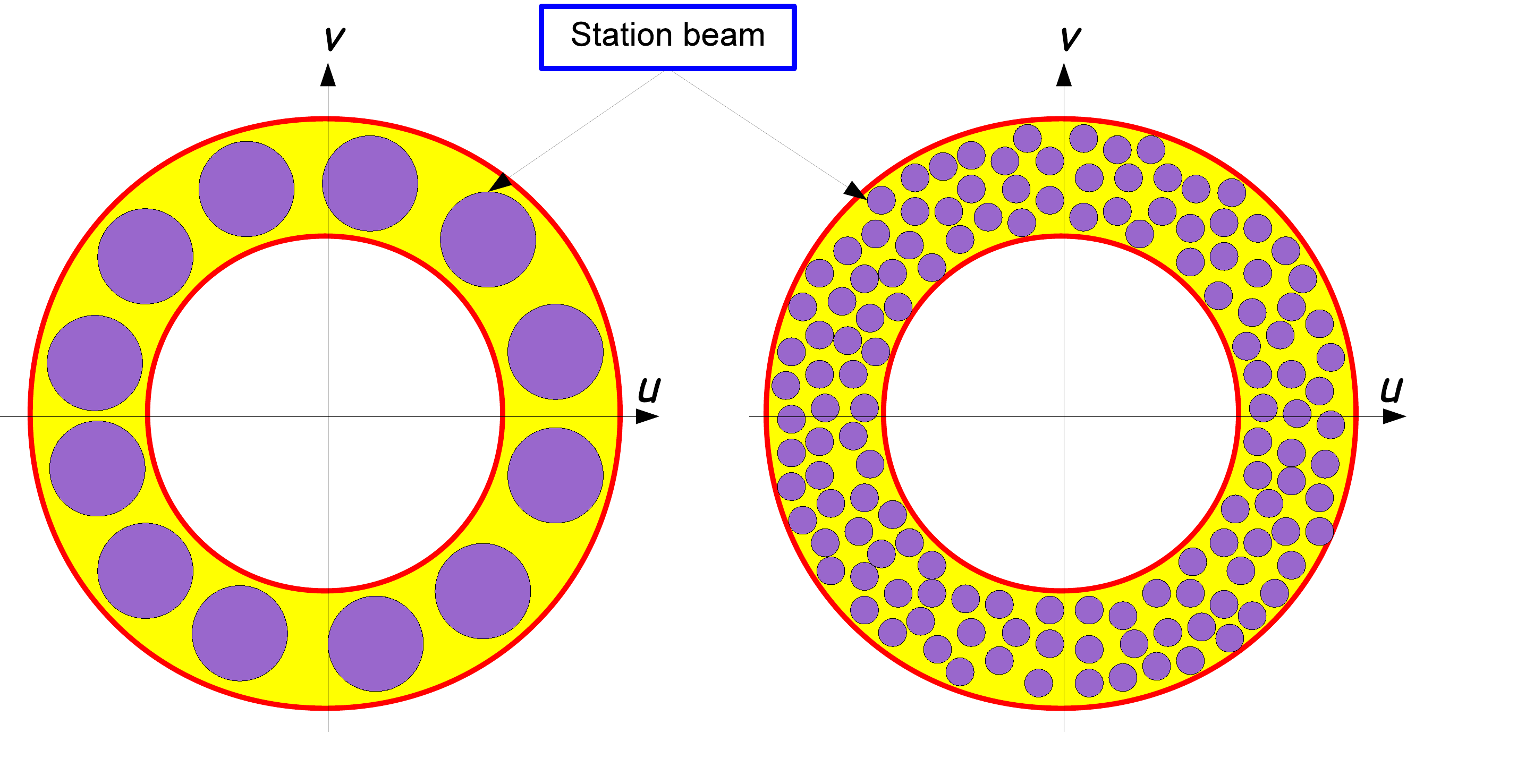}
      \caption{ This figure shows how two different
experiments might sample an annulus in the uv plane. The size of uv point is
given by the station (interferometric element) size where a larger station
(left hand panel) has a larger footprint relative to the smaller station
case (right hand panel) in the uv plane; the footprint is shown by the purple
circles. Even though the sampled area in the two cases might be the
same, the fact that smaller stations sample the annulus more results
in an increased accuracy in their estimation of the power spectrum.  
\vspace{0.5 cm}}
          \label{fig:UVannulus}
   \end{figure}

In the low frequency regime the random component of the noise, i.e., the thermal noise, is
set by two effects: the sky noise and the receiver noise. At frequencies $\nu$ below $\approx 160$MHz
the sky is so bright that the dominant source of noise is the sky itself, whereas at higher frequencies
the receiver noise starts to be more important. The combination of the two effects is normally written 
in terms of the so called system temperature, $T_{sys}$. One can show that the thermal noise level for a given visibility, i.e., uv point, is,

\begin{equation} {\Delta V}(u,v) \approx \frac{2 k_{B}
T_{sys}}{\epsilon dA\sqrt{B t}}\, ,
\label{eq:vnoise}
\end{equation} where  $\epsilon$
is the efficiency of the system, $dA$ is the station area, $B$ is the bandwidth and
$t$ is the observation time (see e.g., \cite{morales05}). This
expression is simple to understand in that the more one observes -- either in
terms of integration time, frequency bandwidth or station collecting
area -- the less uncertainty one has. Obviously, if the signal we are
after is well localized in either time, space or frequency the
relevant noise calculation should take that into account.

 In order to calculate the noise in the 3D power spectrum, the main
quantity we are after, one should remember that the frequency
direction in the observed datacube can be mapped one-to-one with the redshift,
which in turn can be easily translated to distance, whereas the u and v
coordinates are in effect Fourier space coordinates. Therefore, to
calculate the power spectrum one should first Fourier transform the
data cube along the frequency direction. Following Morales' work \cite{morales05} I
will call the new Fourier space coordinate $\eta$ (with $d\eta$
resolution), which together with u and v defines the Fourier space
vector $\vec{u}=\{u,v,\eta\}$.  From this, one can calculate the noise
contribution to the power spectrum at a given $\vert\vec{u}\vert$,

\begin{equation} P_{noise}(\vert \mathbf{u}\vert) \approx 2
N_{beam}^{-1} N_{cell}^{-1/2} \left(\frac{2 k_{B} T_{sys}}{\epsilon dA
d\eta}\right)^2 \frac{1}{B n(\vert\vec{u}\vert) t} \, ,
\label{eq:PSnoise}
\end{equation} where $N_{beam}$ is the number of simultaneous beams
that could be measured, $N_{cell}$ is the number of independent
Fourier samplings per annulus and $n(\vert\vec{u}\vert)$ is the number
of baselines covering this annulus \cite{morales05}. Note that
$n(\vert\vec{u}\vert)$ is proportional to the square of the number of stations,
hence, $n(\vert\vec{u}\vert) dA^2$ is proportional to the square of
the total collecting area of the array regardless of the station
size. This means that in rough terms the noise power spectrum measurement does not
depend only on the total collecting area, bandwidth and integration
time, it also depends the number of stations per annulus. This is easy
to understand as follows. The power in a certain Fourier space annulus
is given by the variance of the measured visibilities in the annulus
which carries uncertainty proportional to the inverse square root of
the number of points. This point is demonstrated in
Figure~\ref{fig:UVannulus}~\cite{morales05, morales04, zaroubi10}.

\subsection{The Foregrounds}
\label{sec:foregrounds}

   \begin{figure} \centering
    \includegraphics[width=0.99\textwidth,clip]{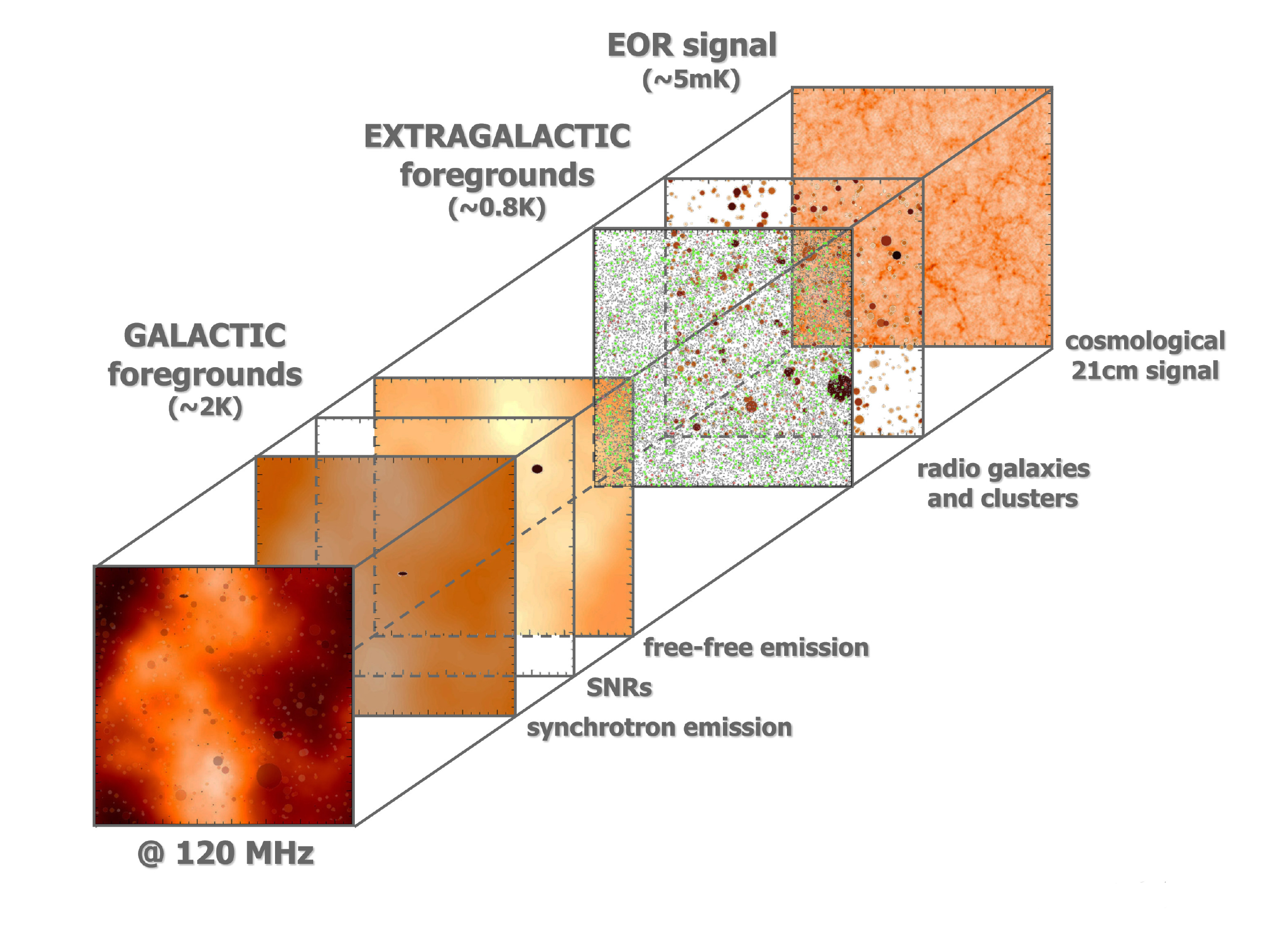}
      \caption{ A figure showing the various cosmological
and galactic components that contribute to the measured signal at a
given frequency. The slices are color coded with different color tables owing
to the vast difference in the range of brightness temperature in
each component. The figure also shows the rms of the galactic
foregrounds, extra galactic foregrounds and cosmological signal.
\vspace{0.2 cm}
}
 \label{fig:FGslice}
  \end{figure}

 The foregrounds in the frequency regime ($40-200$MHz) are very bright
and dominate the sky. In fact the amplitude of the foreground
contribution, $T_{sky}$, at $150 \mathrm{MHz}$ is about 4 orders of
magnitude larger than that of the expected signal. However, since we
are considering radio interferometers the important part of the
foregrounds is that of the fluctuations and not the mean signal, which reduces the ratio
between them and the cosmological signal to about 2-3 orders of
magnitude, which is still a formidable obstacle to surmount.

  The most prominent foreground is the synchrotron emission from
relativistic electrons in the Galaxy: this source of contamination
contributes about 75\% of the foregrounds. Other sources that
contribute to the foregrounds are radio galaxies, galaxy clusters,
resolved supernovae remnants and free-free emission, which together provide 25\% of the
foreground contribution (see e.g., \cite{shaver99}).
Figure~\ref{fig:FGslice} shows the simulated foreground contribution at
120 MHz taking into account all the foreground sources mentioned.

Observationally, the regime of frequencies relevant to the EoR is obviously
not very well explored. 
There are several all-sky maps of the total Galactic diffuse radio
emission at different frequencies and angular resolutions. 
The $150~{\rm MHz}$ map of Landecker \& Wielebinski \cite{landecker70} is the only all-sky map in the frequency range
($100-200~{\rm MHz}$) relevant for the EoR experiments, but has only
$5^\circ$ resolution.
 
In addition to current all-sky maps, a number of recent dedicated 
observations have given estimates of Galactic foregrounds in small selected areas.For example, 
Ali et al. \cite{ali08} have used 153 MHz observations with GMRT
to characterize the visibility correlation function of the
foregrounds. Rogers  and Bowman (\cite{rogers08}) have measured the spectral index of the
diffuse radio background between 100 and 200~{\rm MHz}. Pen et al. (\cite{pen09})
have set an upper limit to the diffuse polarized Galactic emission.

Recently, a comprehensive program was initiated by the LOFAR-EoR
collaboration to directly measure the properties of the Galactic radio
emission in the frequency range relevant for the EoR experiments. The
observations were carried out using the Low Frequency Front Ends
(LFFE) on the WSRT radio telescope. Three different fields were
observed. The first field was a highly polarized region known as the Fan
region in the 2nd Galactic quadrant  at a low Galactic latitude of $\sim10^\circ$
\cite{bernardi09}. This field is not ideal for measuring the EoR but it is a good
field to learn from about calibration issues and about the influence of strong polarization.

\begin{figure} \centering
    \includegraphics[width=0.43\textwidth]{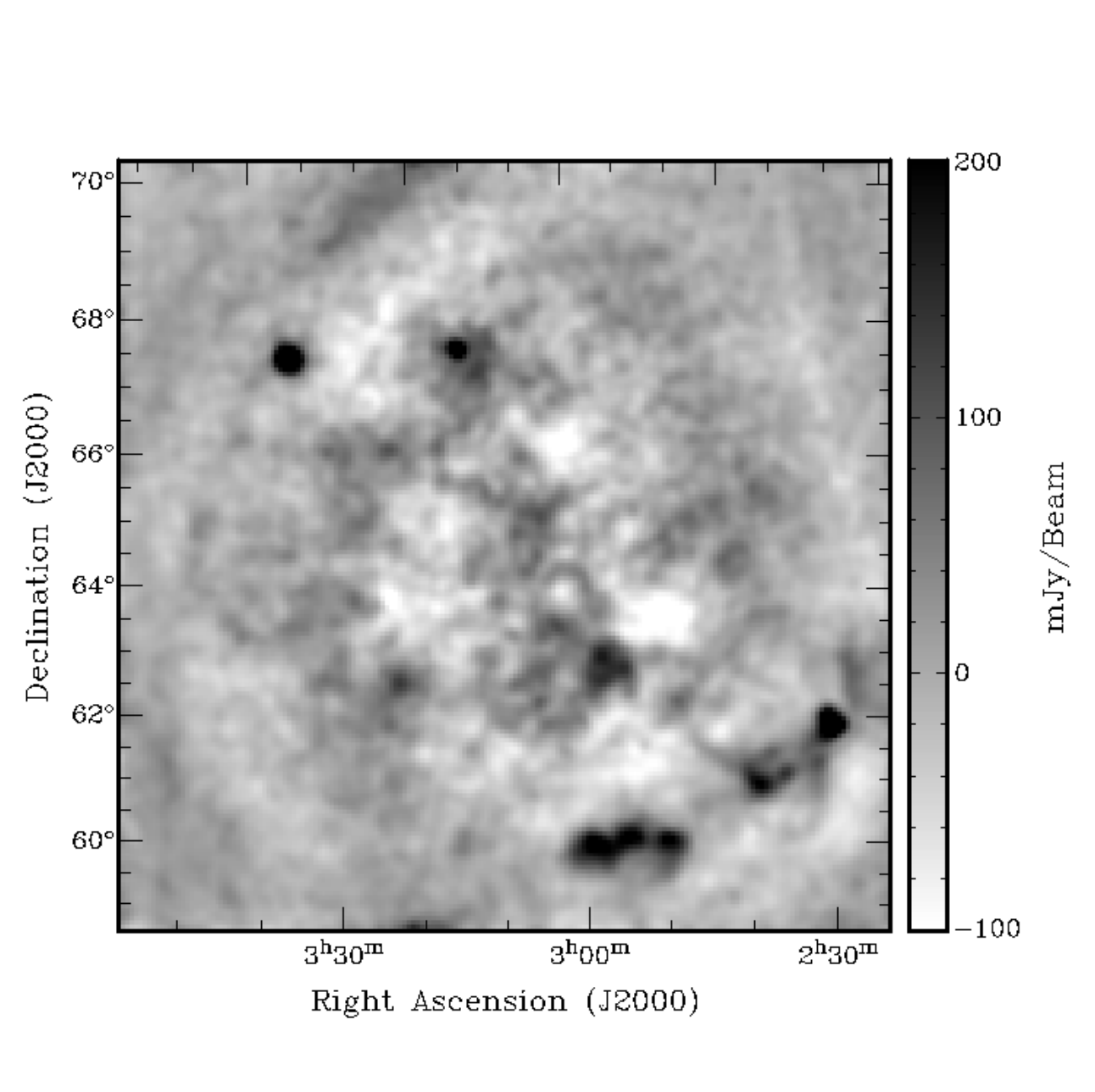}
    \includegraphics[width=0.53\textwidth]{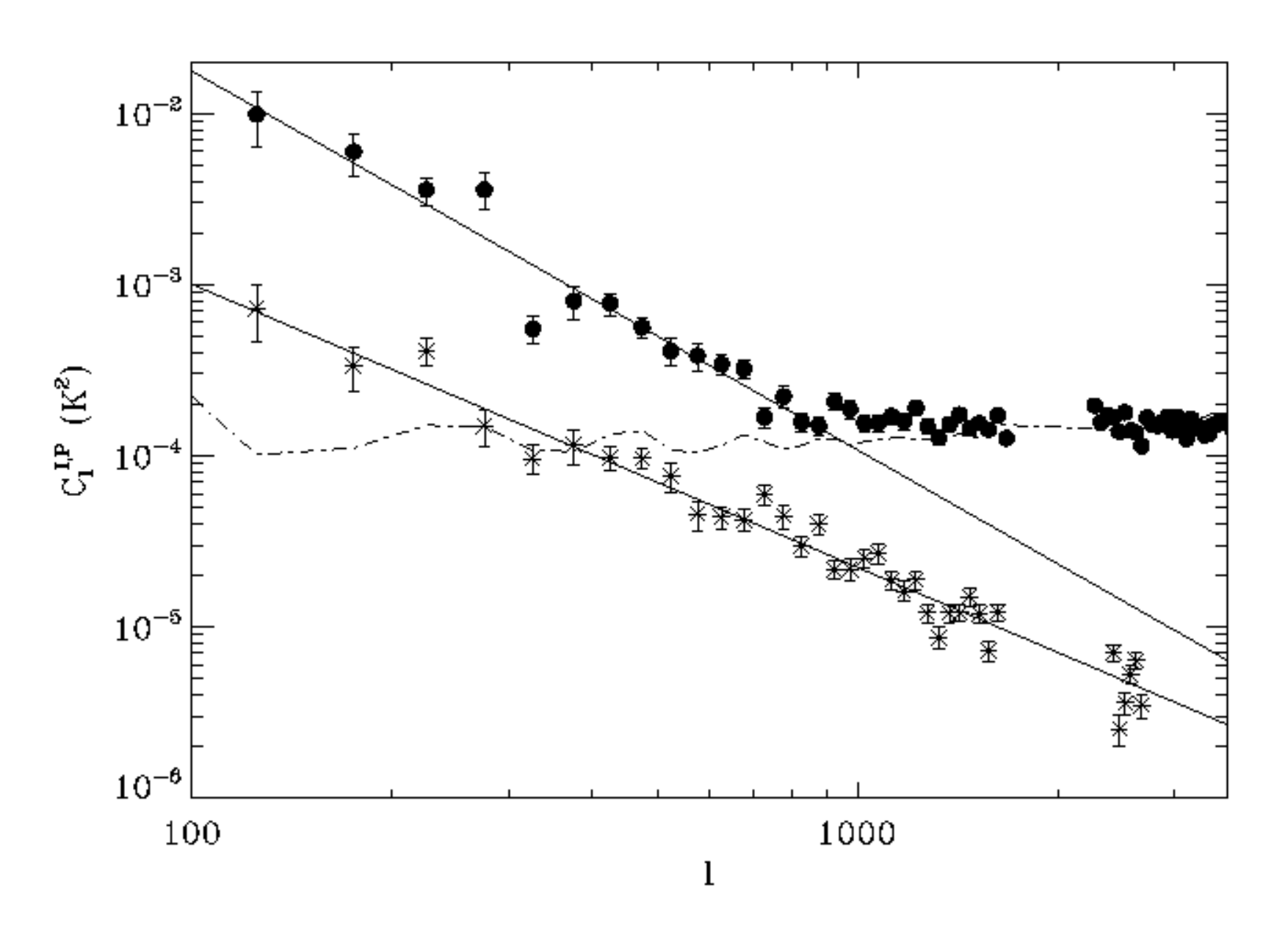}
      \caption{Left hand panel: Stokes I map of the Galactic emission in the soÐcalled Fan
region, at Galactic coordinates $l = 137^\circ$ and $b = +8^\circ$ in
the $2^\mathrm{nd}$ Galactic quadrant \cite{brow76, haverkorn03}. The conversion factor is
from flux (Jansky) to temperature is $1~\mathrm{Jy~beam}^{-1} = 105.6~\mathrm{K}$. Right hand panel: power spectrum (filled circles: total intensity; asterisks: polarized
intensity) of the Galactic emission in Fan region with the best power-law fit. The plotted $1~\sigma$ error bars only account for the statistical
errors. Power spectra are computed in the inner $6^\circ\times 6^\circ$ square
of the map. This figure is taken from \cite{bernardi09}.}
          \label{fig:fanregion}
   \end{figure}

The second field 
was a very cold region in the Galactic halo ($l\sim170^\circ$) around the bright radio quasar
 3C196, and the third was a region around the North Celestial Pole (NCP) \cite{bernardi10}. 
In the Fan region fluctuations of the Galactic diffuse emission were 
detected at $150~{\rm MHz}$ for the first time (see Fig.~\ref{fig:fanregion}). 
The fluctuations were detected 
both in total and polarized intensity, with an $rms$ of $14~{\rm K}$ 
($13~{\rm arcmin}$ resolution) and $7.2~{\rm K}$ ($4~{\rm arcmin}$ resolution)
respectively \cite{bernardi09}. Their spatial structure appeared to
have a power law behavior with a slope of $-2.2\pm0.3$ in total intensity
and $-1.65\pm0.15$ in polarized intensity  (see Fig.~\ref{fig:fanregion}). 
Note that, due to its strong 
polarized emission, the ``Fan region'' is not a representative part of the
 high Galactic latitude sky.

The foregrounds in the context of the EoR measurements have been
studied theoretically by various authors \cite{shaver99, dimatteo02, dimatteo04, cooray04,
santos05, jelic08, gleser08, wilman08, angelica08, sun08, waelkens09, sun09, bowman09}.
The first comprehensive simulation of the EoR foregrounds was carried out by Jeli\'c et al. (\cite{jelic08})
focusing on  the LOFAR-EoR project. The Jeli\'c et al.
model takes into account the Galactic diffuse synchrotron \& free-free
emission, synchrotron emission from Galactic supernova remnants and
extragalactic emission from radio galaxies and clusters, both in total intensity and polarization. The simulated 
foreground maps, in their angular and
frequency characteristics, are similar to the maps expected from the LOFAR-EoR experiment (see
Fig.~\ref{fig:FGslice}).

One major problem faced when considering the LOFAR-EoR data  is  disentangling 
the desired cosmological signal from the foreground signals.
Even though  the foregrounds are very prominent they are very
smooth along the frequency direction~\cite{shaver99, jelic08, jelic10b, bernardi09, bernardi10}, 
as opposed to the cosmological signal that fluctuates along the same direction. Hence,
the separation of the two is,  at a first glance, very simple. One would fit a smooth function to the data
along the frequency direction and subtract it to obtain the desired signal.
In reality, however, things are much more complicated as the existence of thermal noise and systematic
errors due to calibration imperfections make the extraction much harder. In addition, the foregrounds are 
partially polarized, with a complicated structure along
the frequency direction. The confluence of this with the ionospheric distortions and the polarized
instrumental response makes it imperative to calibrate the data very accurately over a very wide field in order to obtain
a very high dynamic range of  observations.
These factors make the fitting non-trivial, that might result in either
under-fitting or over-fitting the signal.  In the former case the deduced signal retains a large contribution of the foregrounds and
produce a spurious ``signal".  Whereas in the  over-fitting case one  fits out the foregrounds and some of the signal
resulting in an underestimation of the cosmological signal. 

The simplest method for foreground removal in total intensity as a function of frequency is a polynomial fitting 
performed on the log-log scale which reduces to a power law in the first order case ~\cite{jelic08}. However, one should be careful in choosing the order of the polynomial to
perform the fitting. If the order of the polynomial is too small, the
foregrounds will be under-fitted and the EoR signal could be dominated
and corrupted by the fitting residuals, while if the order of the
polynomial is too big, the EoR signal could be fitted out. Arguably,  it would be better to fit the foregrounds 
non-parametrically, i.e.,  allowing the data to determine their shape 
rather than selecting some functional form in advance and then fitting its 
parameters (see \cite{harker09a}).

\begin{figure} 
\centering
    \includegraphics[width=0.45 \textwidth]{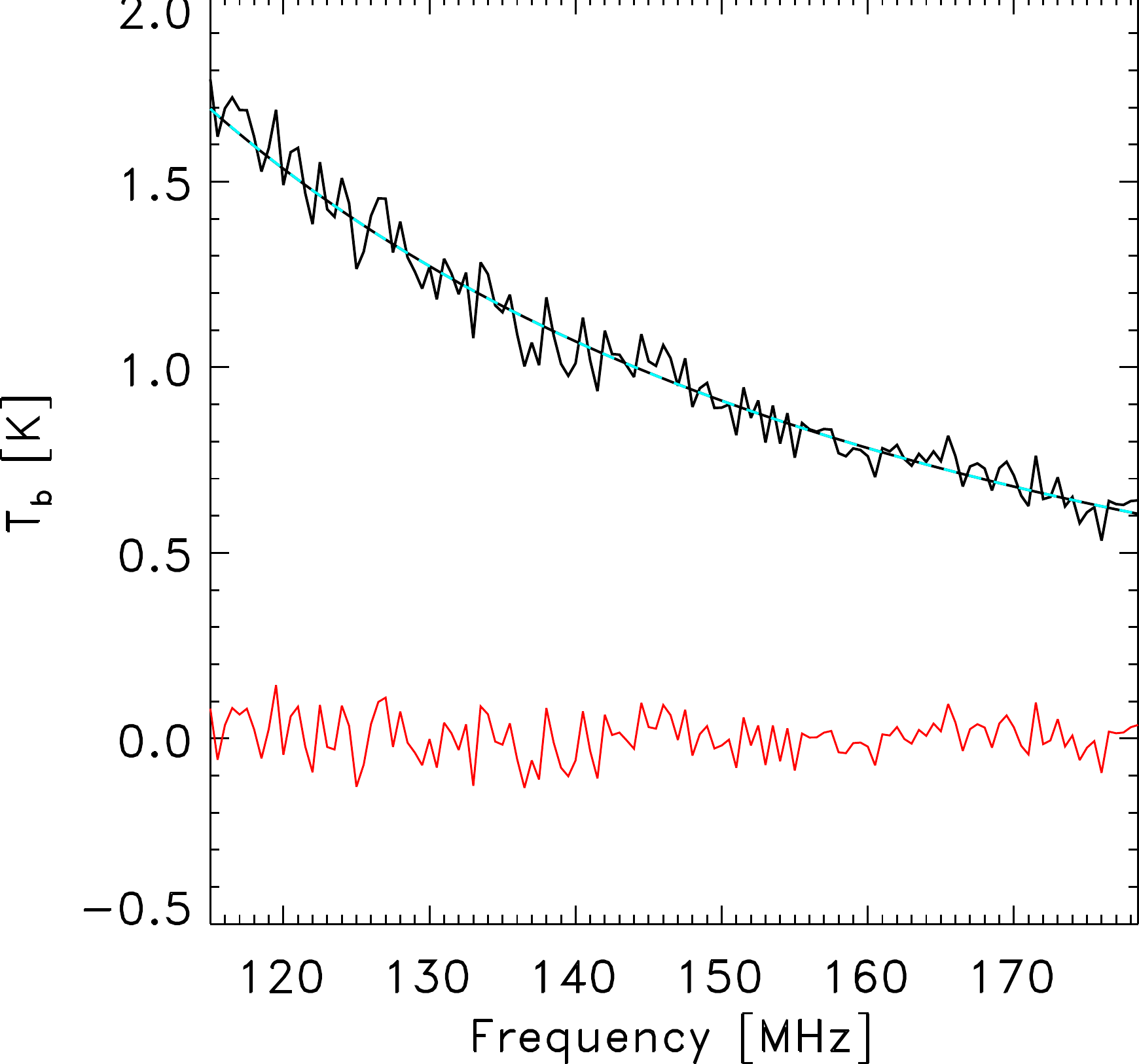}
    \hspace{0.02 \textwidth}
    \includegraphics[width=0.45\textwidth]{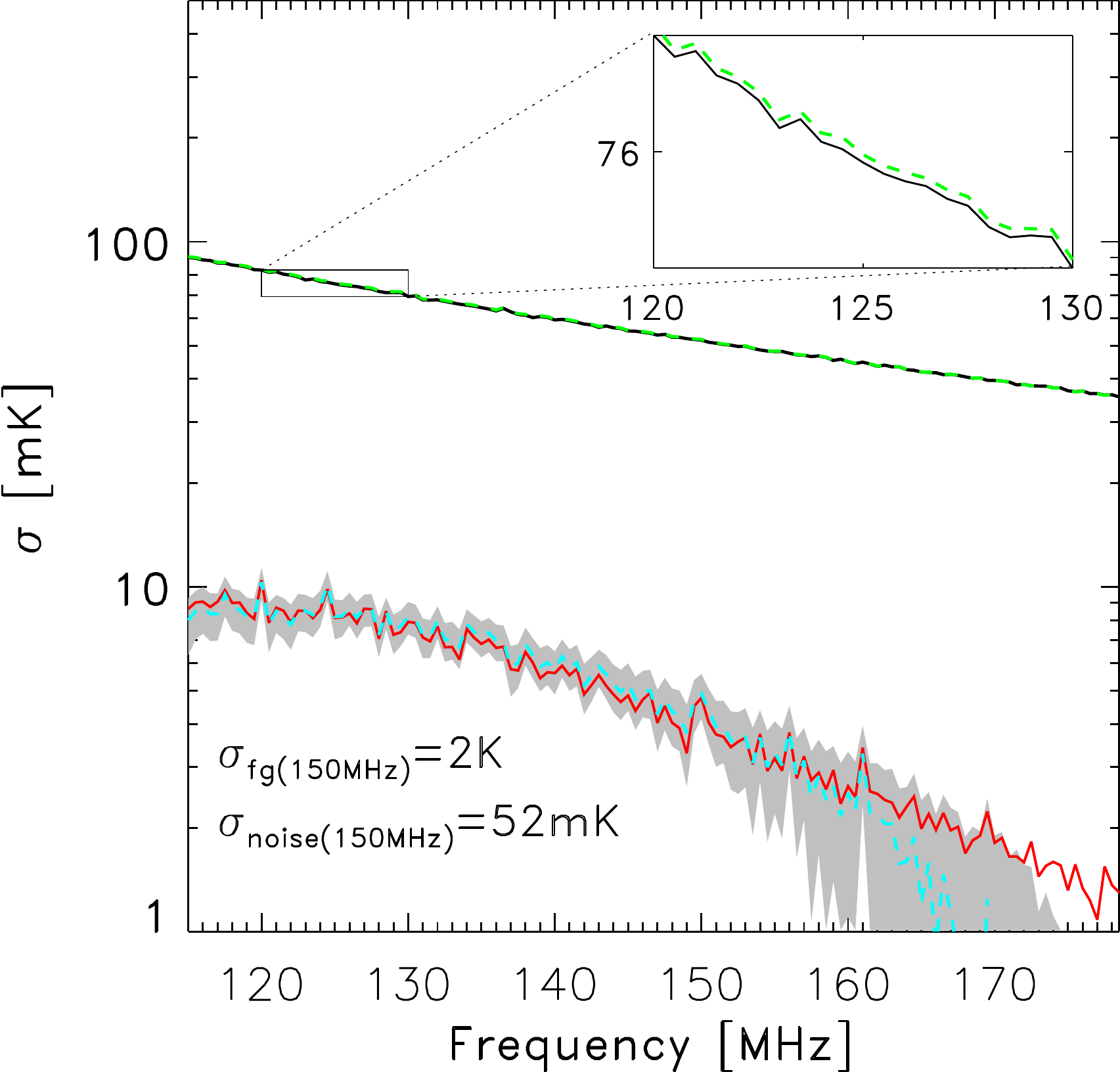}
\caption{This figure shows the ability to statistically extract the EoR signal
from the foregrounds. Please  notice the difference in the vertical axis units 
between the two panels.
Left hand panel: One line of sight (one pixel along frequency) of the
LOFAR-EoR data maps (black solid line), smooth component of the
foregrounds (dashed black line), fitted foregrounds (dashed cyan line)
and residuals (red solid line) after taking out of the foregrounds. 
Right hand panel: Statistical detection of the EoR signal
from the LOFAR-EoR data maps that include diffuse components of the
foregrounds and realistic instrumental noise ($\sigma_{noise}(150~{\rm
MHz})=52~{\rm mK}$). Black dashed line represents standard deviation
($\sigma$) of the noise as a function of frequency, cyan dashed line
is the $\sigma$ of the residuals after taking out smooth 
component due to the foregrounds and the red solid line the $\sigma$ of the original EoR signal. The
grey shaded surface represents the $90\%$ of detected EoR signals from
1000 independent realisations of the noise, while the cyan dashed line is
the mean of the detected EoR signal. Note that the y-axis is in logarithmic
scale \cite{jelic08}.
}
\label{fig:FGfit}
\end{figure}

After foreground subtraction from the EoR observations, 
the residuals will be dominated by instrumental
noise, i.e., the level of the noise is expected to be an order of magnitude 
larger than the EoR signal (assuming 300 hours of observation with LOFAR). Thus, 
general statistical properties of the noise should be determined and
used to statistically detect the cosmological 21 cm signal, e.g., the variance of 
the EoR signal over the image,  $\sigma_{\rm EoR}^{2}$, as a 
function of frequency (redshift) obtained by
subtracting the variance of the noise, $\sigma_{\rm noise}^{2}$, 
from that of the residuals, $\sigma_{\rm residuals}^{2}$. It has been 
shown the such statistical detection of the EoR signal using the
fiducial model of the LOFAR-EoR experiment is possible \cite{jelic08} (see Figure~\ref{fig:FGfit}). 
Similar results by using different statistics are
the skewness of the one-point distribution of brightness temperature of the
EoR signal, measured as a function of observed frequency \cite{harker09b},
and the power spectrum of variations in the intensity of redshifted 21 cm 
radiation from the EoR \cite{harker10}.

\section{The Statistics of the observed cosmological signal}
\label{sec:statistics}
\subsection{The 21~cm Power Spectrum}

\begin{figure} \centering
    \includegraphics[width=0.95\textwidth]{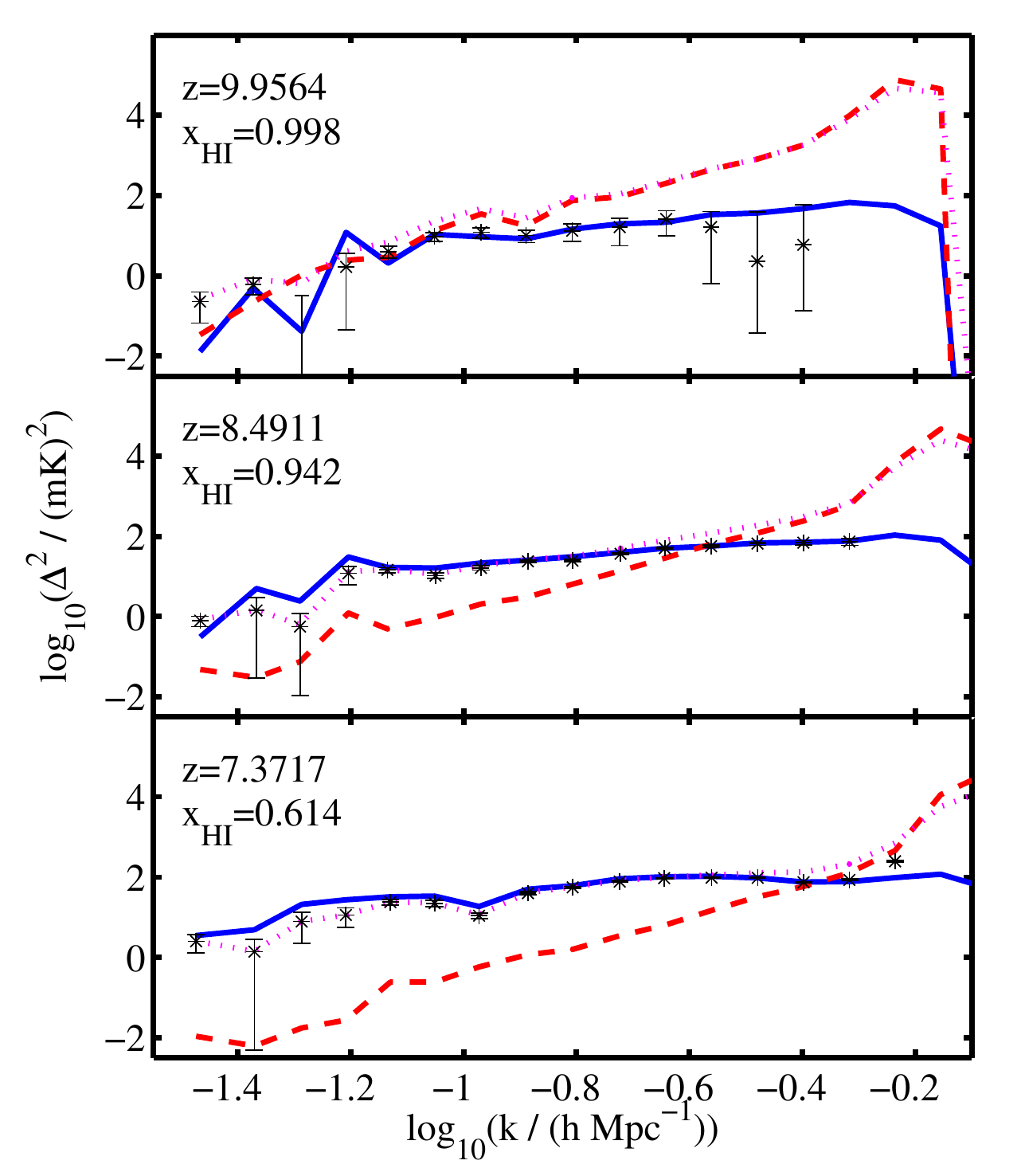}
      \caption{ Power spectra of the cosmic signal (blue
solid line), the noise (red dashed line), the residuals (magenta
dotted line) and the extracted signal (black points with error bars)
at three different redshifts. Here the assumption is that, like in the real experiment, the uv coverage is
frequency-dependent. Furthermore, the field is assumed to have been observed for 300~hours per frequency
channel with a single station beam and the foreground fitting is done
using the so called Wp method but performed in Fourier space \cite{harker10}.
\vspace{-0.2cm} }
          \label{fig:PS}
   \end{figure}

 One of the main goals of the EoR projects is to measure the power
spectrum of variations in the intensity of redshifted 21 cm radiation
from the EoR~\cite{barkana05, bharadwaj04, morales04, morales05, zaldarriaga04}.  
As shown in Equation~\ref{eq:dTb} the power spectrum
depends on a number of astrophysical and cosmological quantities. The
sensitivity with which this power spectrum can be estimated depends on
the level of thermal noise (Eq.~\ref{eq:PSnoise}),  sample variance
and  systematic errors arising from the extraction process,
in particular from the subtraction of foreground contamination. In the
LOFAR case~\cite{harker09b, harker10, jelic08, jelic10b, panos09}, for example, the extraction process is modeled using
realistic simulations of the cosmological signal, the foregrounds and
the noise. In doing so we estimate the sensitivity of the LOFAR EoR
experiment to the redshifted 21~cm power spectrum. Detection of
emission from the EoR should be possible within 300 hours of
observation with a single station beam. Integrating for longer, and
synthesizing multiple station beams within the primary (tile) beam, will
then enable us to extract progressively more accurate estimates of
the power at a greater range of scales and redshifts (see
Figure~\ref{fig:PS} taken from \cite{harker10}).

\subsection{High order statistics}

\begin{figure} \centering
\vspace{0.5 cm}
    \includegraphics[height=0.45\textwidth]{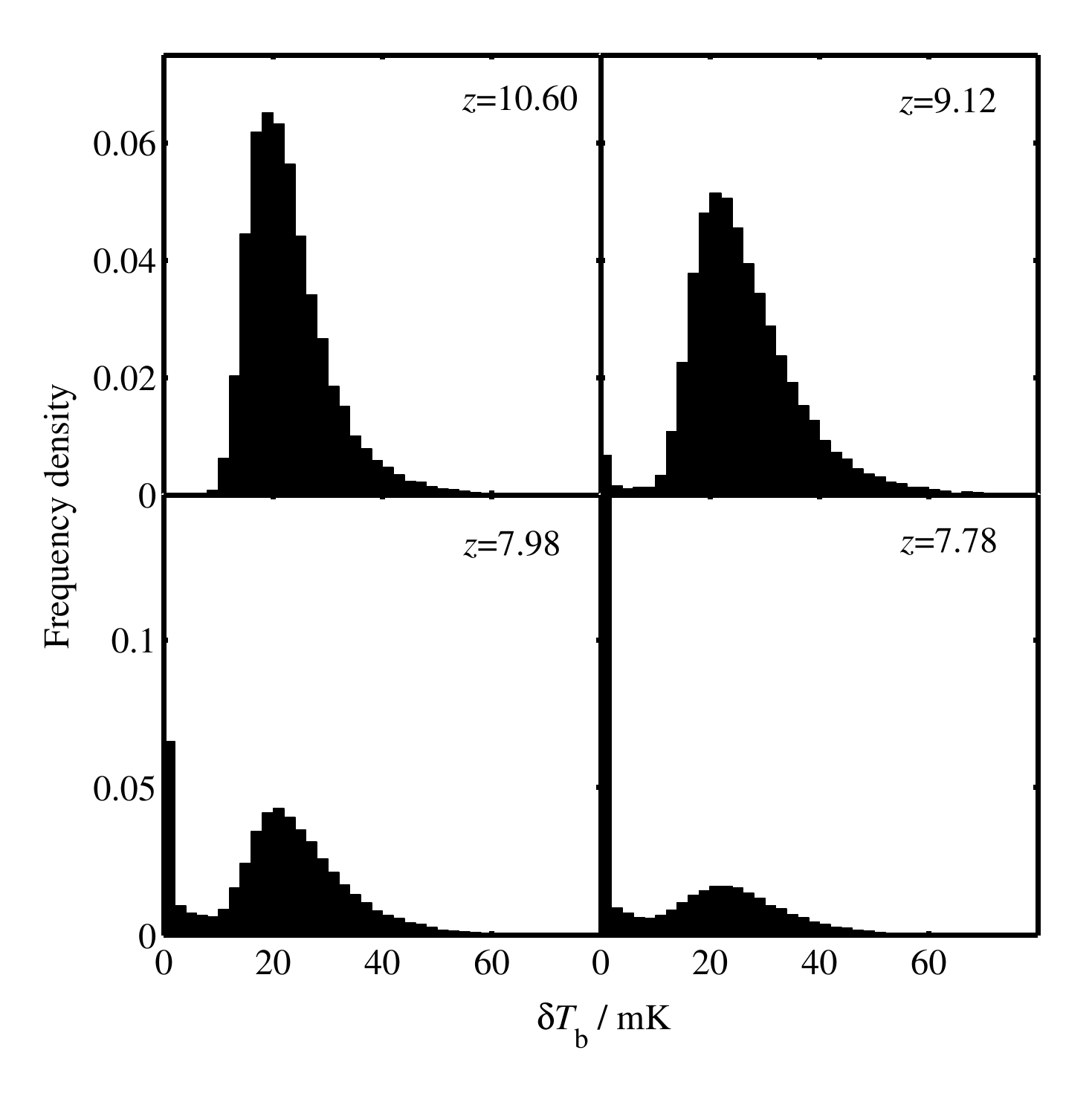}
        \includegraphics[height=0.45\textwidth]{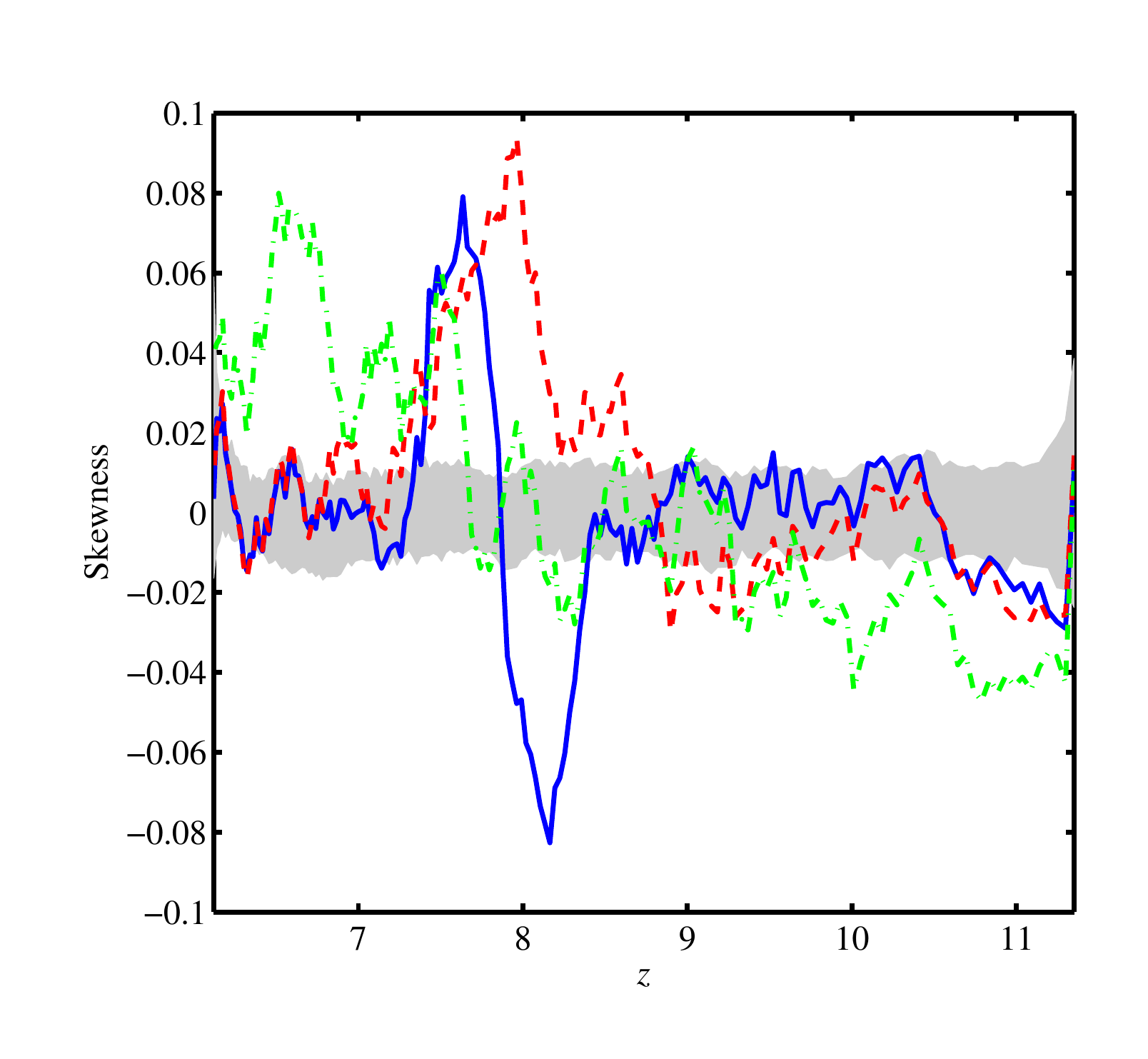}
\caption{ Left hand panel: The distribution of $\delta T_\mathrm{b}$ in a
certain cosmological simulation of reionization \cite{iliev08} at four
different redshifts, showing how the PDF evolves as reionization
proceeds. Note that the y-axis scale in the top two panels is
different from that in the bottom two panels. The delta-function at
$\delta T_\mathrm{b}=0$ grows throughout this period while the rest of
the distribution retains a similar shape. The bar for the first bin in
the bottom-right hand panel has been cut off; approximately 58 per cent of
points are in the first bin at $z=7.78$ \cite{harker09a}.  
Right hand panel: Skewness of the fitting residuals from
data cubes with uncorrelated noise, but in which the residual image
has been denoised by smoothing at each frequency before calculating
the skewness. The three lines correspond to results from three
different simulations \cite{thomas09, iliev08}. Each line has been
smoothed with a moving average (boxcar) filter with a span of nine
points. The grey, shaded area shows the errors, estimated using 100
realizations of the noise (see \cite{harker09a}).}
          \label{fig:nongauss}
   \end{figure}

Given the nature of the reionization process the expected signal is non-Gaussian, hence using high order statistics
to characterize the data can reveal information that the power spectrum does not include.
 The left hand panel of Figure~\ref{fig:nongauss} shows the Probability Density Function (PDF)  of the brightness
temperature at four different redshifts; the PDF is clearly
non Gaussian in all four cases. Therefore, high order moments, like the skewness,
as a function of redshift could be a useful tool for signal extraction
in the presence of realistic overall levels of foregrounds and noise.
Harker et al~\cite{harker09a}, (see also~\cite{gleser06, ichikawa10}) has shown that the cosmological signal, under
generic assumptions, has a very characteristic pattern in the
skewness as a function of redshift (the right hand panel of Figure~\ref{fig:nongauss}). At
sufficiently high redshifts the signal is controlled by the
cosmological density fluctuations which, in the linear regime, are
Gaussian. At lower redshifts, and as nonlinearity becomes important, the signal
starts getting a slightly positive skewness.  As the ionization
bubbles begin to show up the skewness starts veering towards 0 until
it crosses to the negative side when the weight of the ionized
bubbles becomes more important than the high density outliers --note
high density outliers are likely to ionize first-- but the
distribution is still dominated by the density fluctuations. At lower
redshifts the bubbles dominate the PDF and the neutral areas become the
``new" outliers giving rise to a sharp positive peak to the skewness. At
redshift around 6 the instrument noise, assumed to be Gaussian,
dominates driving the skewness again towards zero.  Exploiting this
characteristic behavior might allow us to pick up the cosmological
signal with this high order statistic.

\subsection{Cross-correlating the LOFAR-EoR data with other data sets}

  Given the challenges and uncertainties involved in measuring the redshifted 21cm signal from the EoR, it is vital to corroborate this result with other probes of the EoR.  Namely, other
 astrophysical data that probe the EoR
 signal. The list of such data is long but here we focus on the most promising
 two such probes: CMB maps and high
 redshift galaxy catalogs.
 
\begin{figure}
 \centering \includegraphics[width=0.95\textwidth]{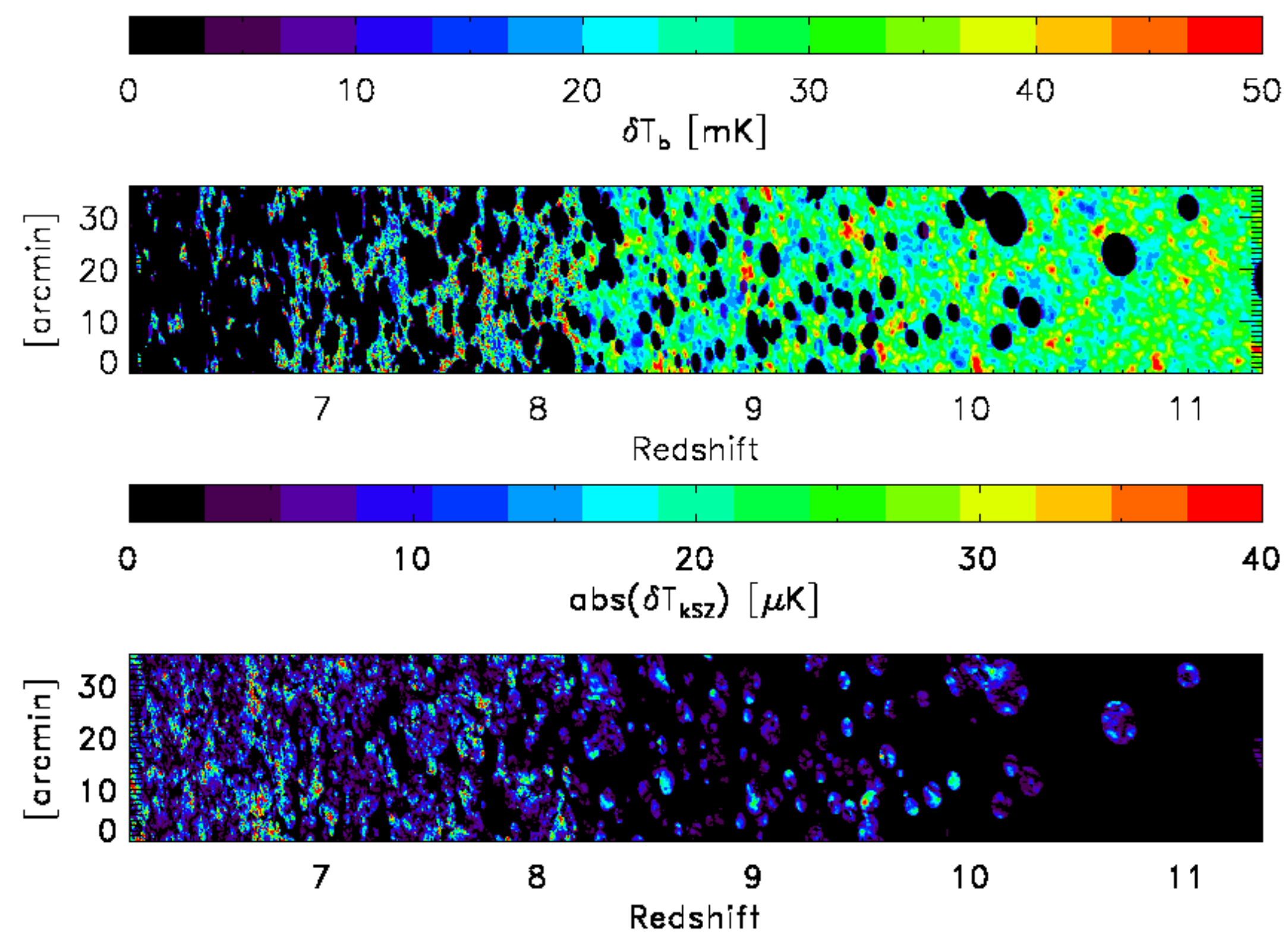}
 \caption{\footnotesize{A slice through a simulated reionization history of the
 the cosmological 21cm signal (top panel) and the so called kinematic Sunyaev-Zeldovich effect
 which reflects the effect of reionization on the
 CMB photons (bottom panel) as time evolves~\cite{ostriker86, sunyaev72, sunyaev80, sunyaev81}. The expected anti-correlation between the two
 phenomena is patently clear.
 Unfortunately however, the CMB data stores no redshift (time) information and the actual map that 
 one gets is the integral of the lower panel over redshift (time), which will make the anti-correlation more
challenging to detect. This Figure is taken from~\cite{jelic10a}.}
\vspace{0.3 cm}
}
 \label{fig:EoRkSZs}
 \end{figure}

   The CMB photons are scattered by the free electrons released during the
   reionization process. This scattering produces anti-correlation
   between the CMB signal and the EoR through a number of physical
   processes. This has been studied in recent years by a number of
   groups \cite{alvarez06, adshead08, cooray04, jelic08, jelic10a, salvaterra05a, tashiro08, tashiro10, tashiro11}. 
   Figure~\ref{fig:EoRkSZs} shows a slice through reionization 
   history of the 21cm signal (top panel) and the so called kinematic
   Sunyaev-Zeldovich effect which reflects the effect of reionization on the
   CMB photons (bottom panel) as time evolves~~\cite{ostriker86, sunyaev72, sunyaev80, sunyaev81}. 
   Notice the clear anti-correlation  between the two maps.  However, whereas the 21cm data store the 
   redshift information in them the actual
   CMB data do not, instead they are sensitive to an integral of the map in the bottom panel over time
   (redshift). In both cases many instrumental, foreground and background effects 
   might hamper the anti-correlation analysis.
   
   Of special interest here are the CMB data expected to be
   collected by the Planck satellite, which will have a resolution comparable
   to the LOFAR core and, since it is an all-sky survey, will probe the same
   regions as the LOFAR-EoR project.

   The other interesting data will come from high $z$ galaxies and
   quasars. These objects harbor the sources of ionization and are located 
   at the center of the ionization bubbles. Hence, they are
   expected to anti-correlate with the EoR signal. Currently, a large
   amount of effort is being put into gathering possible galaxy candidates at
   high redshifts, e.g., Lyman-$\alpha$ emitters and Z, Y and J-bands
   drop outs, etc. Recent studies have shown that the anti-correlation
   signal will be detectable provided the number of galaxies in the
   regions with EoR observations is significant (e.g. see~\cite{lidz09}).

%ch7.tex
\section{Summary}\label{sec:summary}

The EoR, which starts about 400 million years after the Big-Bang, represents a major phase
transition for hydrogen. Due to the formation of the first astrophysical sources, hydrogen in this
epoch transforms from fully neutral to fully ionized. The EoR could be traced in space and time
using relic radio emission that will be observed by the LOFAR radio telescope starting from the
end of this year.

The EoR is determined by how and when the Universe started forming astrophysical objects and
how the ionizing radiation from these objects permeates and fills the intergalactic medium. The
EoR is related to many fundamental questions in cosmology, properties of the first (mini-)quasars,
formation of very metal-poor stars and a slew of other important research topics in astrophysics.
Hence uncovering it will have far reaching implications on the study of structure formation in
the early Universe.

Currently, there are only few  observational constraints on the epoch of reionization. The CMB temperature and polarization data obtained by the WMAP satellite allow measurement of the total Thomson scattering of the primordial CMB photons off intervening free electrons produced by the epoch of reionization along the line of sight. They show that the CMB intensity has only been damped by $\sim 9\%$, indicating that the Universe was mostly neutral for 400 million years and then ionized. However, the Thomson scattering measurement is an integral constraint telling us little about the sources of reionization, its duration or how it propagated to fill the whole Universe.

Another constraint comes from specific features in the spectra of distant quasars, known as the Lyman~$\alpha$ forest. These features, which are due to neutral hydrogen, indicate two important facts about reionization. First, hydrogen in the recent Universe is highly ionized, only 1 part in 10000 being neutral. Second, the neutral fraction of hydrogen in the distant Universe suddenly increases at redshift 6.5, i.e., about 900 million years after the Big Bang, demarcating the end of the reionization process. Despite these data providing strong constraints on the ionization state of the Universe at redshifts below 6.5, they say very little about the reionization process itself. 
Another couple of constraints  come also from the Lyman~$\alpha$ forest systems, IGM temperature and the number of ionizing photons per
baryon, suggesting the bulk of the reionization process occurs at late redshifts $z\approx [6-9]$. 

 A whole slew of possible constraints currently discussed in the literature are either very controversial, very weak or, as is often the case, both. Most are very interesting and exciting, but can be investigated reliably only with a new generation of instruments such as the James Webb Space Telescope, replacing the Hubble Space Telescope in the next decade.
 
The imminent availability of observations of redshifted 21~cm
radiation from the Universe's \emph{dark ages} and the EoR will be one
of the most exciting developments in the study of cosmology and galaxy
and structure formation in recent years. Currently, there are a number
of instruments that are designed to measure this radiation. In this
contribution I have argued that despite the many difficulties that face
such measurements they will provide a major breakthrough in our
understanding of this crucial epoch. In particular current radio
telescopes, such as LOFAR, will be able to provide us with the global
history of the EoR progression, the fluctuations power spectrum during
the EoR, etc., up to $z\approx 11$. These measurements will usher the
study of the high redshift Universe into a new era which will bridge,
at least in part, the large gap that currently exists in
observation between the very high redshift Universe ($z\approx 1100$)
as probed by the CMB and the low redshift Universe ($z\lsim 6$).

Although the current generation of telescopes have a great promise 
they will also have limitations.  For example they have neither the
resolution, the sensitivity nor the frequency coverage to address many
fundamental issues, like the nature of the first sources. Crucially,
they will not provide a lot of information about the \emph{dark
ages} which is only accessible through very low frequencies in the
range of $40-120$ ($z\approx 35-11$).
Fortunately, in the future SKA can improve dramatically on the
current instruments in terms of sensitivity, redshift coverage and
resolution.
 
The next decade will be extremely exciting for studying the high
redshift Universe, especially as these radio telescopes gradually come
online, starting with LOFAR, GMRT and MWA. They promise to resolve
many of the puzzles we have today pertaining to the formation and
evolution of the first objects cosmology, and the physical processes in
the high redshift intergalactic medium.

\begin{acknowledgement} 
I would like to thank Geraint Harker, Stephen Rafter and Rajat M. Thomas for careful reading of the manuscript. 
Many of the results shown here have been obtained
in collaboration with the members of the LOFAR EoR project
whose contribution I would like to acknowledge. I would like also thank the editors of this book for giving
me the opportunity to write this chapter.
\end{acknowledgement}
%
% \section*{Appendix} 
% \addcontentsline{toc}{section}{Appendix}
%
%

% input{referenc}

\begin{thebibliography}{100}

\bibitem{abel00}
T.~{Abel}, G.~L. {Bryan}, and M.~L. {Norman}.
\newblock {The Formation and Fragmentation of Primordial Molecular Clouds}.
\newblock {\em \apj}, 540:39--44, September 2000.

\bibitem{abel02}
T.~{Abel}, G.~L. {Bryan}, and M.~L. {Norman}.
\newblock {The Formation of the First Star in the Universe}.
\newblock {\em Science}, 295:93--98, January 2002.

\bibitem{adshead08}
P.~J. {Adshead} and S.~R. {Furlanetto}.
\newblock {Reionization and the large-scale 21-cm cosmic microwave background
  cross-correlation}.
\newblock {\em \mnras}, 384:291--304, February 2008.

\bibitem{aghanim08}
N.~{Aghanim}, S.~{Majumdar}, and J.~{Silk}.
\newblock {Secondary anisotropies of the CMB}.
\newblock {\em Reports on Progress in Physics}, 71(6):066902--+, June 2008.

\bibitem{ali08}
S.~S. {Ali}, S.~{Bharadwaj}, and J.~N. {Chengalur}.
\newblock {Foregrounds for redshifted 21-cm studies of reionization: Giant
  Meter Wave Radio Telescope 153-MHz observations}.
\newblock {\em \mnras}, 385:2166--2174, April 2008.

\bibitem{allison69}
A.~C. {Allison} and A.~{Dalgarno}.
\newblock {Spin Change in Collisions of Hydrogen Atoms}.
\newblock {\em \apj}, 158:423--+, October 1969.

\bibitem{alvarez06}
M.~A. {Alvarez}, E.~{Komatsu}, O.~{Dor{\'e}}, and P.~R. {Shapiro}.
\newblock {The Cosmic Reionization History as Revealed by the Cosmic Microwave
  Background Doppler-21 cm Correlation}.
\newblock {\em \apj}, 647:840--852, August 2006.

\bibitem{baek09}
S.~{Baek}, P.~{Di Matteo}, B.~{Semelin}, F.~{Combes}, and Y.~{Revaz}.
\newblock {The simulated 21 cm signal during the epoch of reionization: full
  modeling of the Ly-{$\alpha$} pumping}.
\newblock {\em \aap}, 495:389--405, February 2009.

\bibitem{barkana01}
R.~{Barkana} and A.~{Loeb}.
\newblock {In the beginning: the first sources of light and the reionization of
  the universe}.
\newblock {\em \physrep}, 349:125--238, July 2001.

\bibitem{barkana05}
R.~{Barkana} and A.~{Loeb}.
\newblock {A Method for Separating the Physics from the Astrophysics of
  High-Redshift 21 Centimeter Fluctuations}.
\newblock {\em \apjl}, 624:L65--L68, May 2005.

\bibitem{bechtold03}
J.~{Bechtold}.
\newblock {Quasar absorption lines}.
\newblock In {I.~P{\'e}rez-Fournon, M.~Balcells, F.~Moreno-Insertis, \&
  F.~S{\'a}nchez}, editor, {\em Galaxies at High Redshift}, pages 131--184,
  2003.

\bibitem{benson06}
A.~J. {Benson}, N.~{Sugiyama}, A.~{Nusser}, and C.~G. {Lacey}.
\newblock {The epoch of reionization}.
\newblock {\em \mnras}, 369:1055--1080, July 2006.

\bibitem{bernardi09}
G.~{Bernardi}, A.~G. {de Bruyn}, M.~A. {Brentjens}, B.~{Ciardi}, G.~{Harker},
  V.~{Jeli{\'c}}, L.~V.~E. {Koopmans}, P.~{Labropoulos}, A.~{Offringa}, V.~N.
  {Pandey}, J.~{Schaye}, R.~M. {Thomas}, S.~{Yatawatta}, and S.~{Zaroubi}.
\newblock {Foregrounds for observations of the cosmological 21 cm line. I.
  First Westerbork measurements of Galactic emission at 150 MHz in a low
  latitude field}.
\newblock {\em \aap}, 500:965--979, June 2009.

\bibitem{bernardi10}
G.~{Bernardi}, A.~G. {de Bruyn}, G.~{Harker}, M.~A. {Brentjens}, B.~{Ciardi},
  V.~{Jeli{\'c}}, L.~V.~E. {Koopmans}, P.~{Labropoulos}, A.~{Offringa}, V.~N.
  {Pandey}, J.~{Schaye}, R.~M. {Thomas}, S.~{Yatawatta}, and S.~{Zaroubi}.
\newblock {Foregrounds for observations of the cosmological 21 cm line. II.
  Westerbork observations of the fields around 3C 196 and the North Celestial
  Pole}.
\newblock {\em \aap}, 522:A67, November 2010.

\bibitem{bharadwaj04}
S.~{Bharadwaj} and S.~S. {Ali}.
\newblock {The cosmic microwave background radiation fluctuations from HI
  perturbations prior to reionization}.
\newblock {\em \mnras}, 352:142--146, July 2004.

\bibitem{bi92}
H.~G. {Bi}, G.~{Boerner}, and Y.~{Chu}.
\newblock {An alternative model for the Ly-alpha absorption forest}.
\newblock {\em \aap}, 266:1--5, December 1992.

\bibitem{bolton10}
J.~S. {Bolton}, G.~D. {Becker}, J.~S.~B. {Wyithe}, M.~G. {Haehnelt}, and
  W.~L.~W. {Sargent}.
\newblock {A first direct measurement of the intergalactic medium temperature
  around a quasar at z = 6}.
\newblock {\em \mnras}, page 771, May 2010.

\bibitem{bolton07}
J.~S. {Bolton} and M.~G. {Haehnelt}.
\newblock {The observed ionization rate of the intergalactic medium and the
  ionizing emissivity at z {>}= 5: evidence for a photon-starved and extended
  epoch of reionization}.
\newblock {\em \mnras}, 382:325--341, November 2007.

\bibitem{bolton05}
J.~S. {Bolton}, M.~G. {Haehnelt}, M.~{Viel}, and V.~{Springel}.
\newblock {The Lyman {$\alpha$} forest opacity and the metagalactic hydrogen
  ionization rate at z\~{} 2-4}.
\newblock {\em \mnras}, 357:1178--1188, March 2005.

\bibitem{bolton11}
J.~S. {Bolton}, M.~G. {Haehnelt}, S.~J. {Warren}, P.~C. {Hewett}, D.~J.
  {Mortlock}, B.~P. {Venemans}, R.~G. {McMahon}, and C.~{Simpson}.
\newblock {How neutral is the intergalactic medium surrounding the redshift
  z=7.085 quasar ULAS J1120+0641?}
\newblock {\em ArXiv e-prints}, June 2011.

\bibitem{bond84}
J.~R. {Bond} and G.~{Efstathiou}.
\newblock {Cosmic background radiation anisotropies in universes dominated by
  nonbaryonic dark matter}.
\newblock {\em \apjl}, 285:L45--L48, October 1984.

\bibitem{bouwens11}
R.~J. {Bouwens}, G.~D. {Illingworth}, I.~{Labbe}, P.~A. {Oesch}, M.~{Trenti},
  C.~M. {Carollo}, P.~G. {van Dokkum}, M.~{Franx}, M.~{Stiavelli},
  V.~{Gonz{\'a}lez}, D.~{Magee}, and L.~{Bradley}.
\newblock {A candidate redshift z\~{}10 galaxy and rapid changes in that
  population at an age of 500Myr}.
\newblock {\em \nat}, 469:504--507, January 2011.

\bibitem{bouwens10}
R.~J. {Bouwens}, G.~D. {Illingworth}, P.~A. {Oesch}, M.~{Stiavelli}, P.~{van
  Dokkum}, M.~{Trenti}, D.~{Magee}, I.~{Labb{\'e}}, M.~{Franx}, C.~M.
  {Carollo}, and V.~{Gonzalez}.
\newblock {Discovery of z \~{} 8 Galaxies in the Hubble Ultra Deep Field from
  Ultra-Deep WFC3/IR Observations}.
\newblock {\em \apjl}, 709:L133--L137, February 2010.

\bibitem{bouwens05}
R.~J. {Bouwens}, G.~D. {Illingworth}, R.~I. {Thompson}, and M.~{Franx}.
\newblock {Constraints on z\~{}10 Galaxies from the Deepest Hubble Space
  Telescope NICMOS Fields}.
\newblock {\em \apjl}, 624:L5--L8, May 2005.

\bibitem{bowman06}
J.~D. {Bowman}, M.~F. {Morales}, and J.~N. {Hewitt}.
\newblock {The Sensitivity of First-Generation Epoch of Reionization
  Observatories and Their Potential for Differentiating Theoretical Power
  Spectra}.
\newblock {\em \apj}, 638:20--26, February 2006.

\bibitem{bowman09}
J.~D. {Bowman}, M.~F. {Morales}, and J.~N. {Hewitt}.
\newblock {Foreground Contamination in Interferometric Measurements of the
  Redshifted 21 cm Power Spectrum}.
\newblock {\em \apj}, 695:183--199, April 2009.

\bibitem{bowman10}
J.~D. {Bowman} and A.~E.~E. {Rogers}.
\newblock {A lower limit of {$\Delta z > 0.06$} for the duration of the
  reionization epoch}.
\newblock {\em \nat}, 468:796--798, December 2010.

\bibitem{bowman08}
J.~D. {Bowman}, A.~E.~E. {Rogers}, and J.~N. {Hewitt}.
\newblock {Toward Empirical Constraints on the Global Redshifted 21 cm
  Brightness Temperature During the Epoch of Reionization}.
\newblock {\em \apj}, 676:1--9, March 2008.

\bibitem{bromm02}
V.~{Bromm}, P.~S. {Coppi}, and R.~B. {Larson}.
\newblock {The Formation of the First Stars. I. The Primordial Star-forming
  Cloud}.
\newblock {\em \apj}, 564:23--51, January 2002.

\bibitem{bromm04}
V.~{Bromm} and R.~B. {Larson}.
\newblock {The First Stars}.
\newblock {\em \araa}, 42:79--118, September 2004.

\bibitem{bromm06}
V.~{Bromm} and A.~{Loeb}.
\newblock {High-Redshift Gamma-Ray Bursts from Population III Progenitors}.
\newblock {\em \apj}, 642:382--388, May 2006.

\bibitem{brow76}
W.~N. {Brouw} and T.~A.~T. {Spoelstra}.
\newblock {Linear polarization of the galactic background at frequencies
  between 408 and 1411 MHz. Reductions.}
\newblock {\em \aaps}, 26:129--+, October 1976.

\bibitem{bunker10}
A.~J. {Bunker}, S.~{Wilkins}, R.~S. {Ellis}, D.~P. {Stark}, S.~{Lorenzoni},
  K.~{Chiu}, M.~{Lacy}, M.~J. {Jarvis}, and S.~{Hickey}.
\newblock {The contribution of high-redshift galaxies to cosmic reionization:
  new results from deep WFC3 imaging of the Hubble Ultra Deep Field}.
\newblock {\em \mnras}, 409:855--866, December 2010.

\bibitem{calverley11}
A.~P. {Calverley}, G.~D. {Becker}, M.~G. {Haehnelt}, and J.~S. {Bolton}.
\newblock {Measurements of the ultraviolet background at 4.6 < z < 6.4 using
  the quasar proximity effect}.
\newblock {\em \mnras}, 412:2543--2562, April 2011.

\bibitem{carilli04}
C.~L. {Carilli}, S.~{Furlanetto}, F.~{Briggs}, M.~{Jarvis}, S.~{Rawlings}, and
  H.~{Falcke}.
\newblock {Probing the dark ages with the Square Kilometer Array}.
\newblock {\em New Astronomy Reviews}, 48:1029--1038, December 2004.

\bibitem{carilli04b}
C.~L. {Carilli}, N.~{Gnedin}, S.~{Furlanetto}, and F.~{Owen}.
\newblock {Observations of HI 21-cm absorption by the neutral IGM during the
  epoch of re-ionization with the Square Kilometer Array}.
\newblock {\em \nar}, 48:1053--1061, December 2004.

\bibitem{carilli02}
C.~L. {Carilli}, N.~Y. {Gnedin}, and F.~{Owen}.
\newblock {H I 21 Centimeter Absorption beyond the Epoch of Reionization}.
\newblock {\em \apj}, 577:22--30, September 2002.

\bibitem{cen02}
R.~{Cen} and P.~{McDonald}.
\newblock {Evolution of the Ionizing Radiation Background and Star Formation in
  the Aftermath of Cosmological Reionization}.
\newblock {\em \apj}, 570:457--462, May 2002.

\bibitem{cen94}
R.~{Cen}, J.~{Miralda-Escud{\'e}}, J.~P. {Ostriker}, and M.~{Rauch}.
\newblock {Gravitational collapse of small-scale structure as the origin of the
  Lyman-alpha forest}.
\newblock {\em \apjl}, 437:L9--L12, December 1994.

\bibitem{chen04}
X.~{Chen} and M.~{Kamionkowski}.
\newblock {Particle decays during the cosmic dark ages}.
\newblock {\em \prd}, 70(4):043502--+, August 2004.

\bibitem{choudhury06}
T.~R. {Choudhury} and A.~{Ferrara}.
\newblock {Physics of Cosmic Reionization}.
\newblock {\em ArXiv Astrophysics e-prints}, March 2006.

\bibitem{chuzhoy06}
L.~{Chuzhoy}, M.~A. {Alvarez}, and P.~R. {Shapiro}.
\newblock {Recognizing the First Radiation Sources through Their 21 cm
  Signature}.
\newblock {\em \apjl}, 648:L1--L4, September 2006.

\bibitem{ciardi02}
B.~{Ciardi}, S.~{Bianchi}, and A.~{Ferrara}.
\newblock {Lyman continuum escape from an inhomogeneous interstellar medium}.
\newblock {\em \mnras}, 331:463--473, March 2002.

\bibitem{ciardi05}
B.~{Ciardi} and A.~{Ferrara}.
\newblock {The First Cosmic Structures and Their Effects}.
\newblock {\em Space Science Reviews}, 116:625--705, February 2005.

\bibitem{ciardi01}
B.~{Ciardi}, A.~{Ferrara}, S.~{Marri}, and G.~{Raimondo}.
\newblock {Cosmological reionization around the first stars: Monte Carlo
  radiative transfer}.
\newblock {\em \mnras}, 324:381--388, June 2001.

\bibitem{ciardi03a}
B.~{Ciardi} and P.~{Madau}.
\newblock {Probing beyond the Epoch of Hydrogen Reionization with 21 Centimeter
  Radiation}.
\newblock {\em \apj}, 596:1--8, October 2003.

\bibitem{cooray04}
A.~{Cooray}.
\newblock {Cross-correlation studies between CMB temperature anisotropies and
  21cm fluctuations}.
\newblock {\em \prd}, 70(6):063509--+, September 2004.

\bibitem{cooray07}
A.~{Cooray}, I.~{Sullivan}, R.-R. {Chary}, J.~J. {Bock}, M.~{Dickinson}, H.~C.
  {Ferguson}, B.~{Keating}, A.~{Lange}, and E.~L. {Wright}.
\newblock {IR Background Anisotropies in Spitzer GOODS Images and Constraints
  on First Galaxies}.
\newblock {\em \apjl}, 659:L91--L94, April 2007.

\bibitem{angelica08}
A.~{de Oliveira-Costa}, M.~{Tegmark}, B.~M. {Gaensler}, J.~{Jonas}, T.~L.
  {Landecker}, and P.~{Reich}.
\newblock {A model of diffuse Galactic radio emission from 10 MHz to 100 GHz}.
\newblock {\em \mnras}, 388:247--260, July 2008.

\bibitem{dimatteo04}
T.~{Di Matteo}, B.~{Ciardi}, and F.~{Miniati}.
\newblock {The 21-cm emission from the reionization epoch: extended and point
  source foregrounds}.
\newblock {\em \mnras}, 355:1053--1065, December 2004.

\bibitem{dimatteo02}
T.~{Di Matteo}, R.~{Perna}, T.~{Abel}, and M.~J. {Rees}.
\newblock {Radio Foregrounds for the 21 Centimeter Tomography of the Neutral
  Intergalactic Medium at High Redshifts}.
\newblock {\em \apj}, 564:576--580, January 2002.

\bibitem{dijkstra04}
M.~{Dijkstra}, Z.~{Haiman}, and A.~{Loeb}.
\newblock {A Limit from the X-Ray Background on the Contribution of Quasars to
  Reionization}.
\newblock {\em \apj}, 613:646--654, October 2004.

\bibitem{dijkstra06}
M.~{Dijkstra}, Z.~{Haiman}, and M.~{Spaans}.
\newblock {Ly{$\alpha$} Radiation from Collapsing Protogalaxies. I.
  Characteristics of the Emergent Spectrum}.
\newblock {\em \apj}, 649:14--36, September 2006.

\bibitem{dore07}
O.~{Dor{\'e}}, G.~{Holder}, M.~{Alvarez}, I.~T. {Iliev}, G.~{Mellema}, U.-L.
  {Pen}, and P.~R. {Shapiro}.
\newblock {Signature of patchy reionization in the polarization anisotropy of
  the CMB}.
\newblock {\em \prd}, 76(4):043002--+, August 2007.

\bibitem{dove94}
J.~B. {Dove} and J.~M. {Shull}.
\newblock {Photoionization of the diffuse interstellar medium and galactic halo
  by OB associations}.
\newblock {\em \apj}, 430:222--235, July 1994.

\bibitem{dove00}
J.~B. {Dove}, J.~M. {Shull}, and A.~{Ferrara}.
\newblock {The Escape of Ionizing Photons from OB Associations in Disk
  Galaxies: Radiation Transfer through Superbubbles}.
\newblock {\em \apj}, 531:846--860, March 2000.

\bibitem{dunkley09}
J.~{Dunkley}, E.~{Komatsu}, M.~R. {Nolta}, D.~N. {Spergel}, D.~{Larson},
  G.~{Hinshaw}, L.~{Page}, C.~L. {Bennett}, B.~{Gold}, N.~{Jarosik}, J.~L.
  {Weiland}, M.~{Halpern}, R.~S. {Hill}, A.~{Kogut}, M.~{Limon}, S.~S. {Meyer},
  G.~S. {Tucker}, E.~{Wollack}, and E.~L. {Wright}.
\newblock {Five-Year Wilkinson Microwave Anisotropy Probe Observations:
  Likelihoods and Parameters from the WMAP Data}.
\newblock {\em \apjs}, 180:306--329, February 2009.

\bibitem{ewen51}
H.~I. {Ewen} and E.~M. {Purcell}.
\newblock {Observation of a Line in the Galactic Radio Spectrum: Radiation from
  Galactic Hydrogen at 1,420 Mc./sec.}
\newblock {\em \nat}, 168:356--+, September 1951.

\bibitem{fan03}
X.~{Fan,~\textit{et al.}}
\newblock {A Survey of z{$>$}5.7 Quasars in the Sloan Digital Sky Survey. II.
  Discovery of Three Additional Quasars at z{$>$}6}.
\newblock {\em \aj}, 125:1649--1659, April 2003.

\bibitem{fan06}
X.~{Fan,~\textit{et al.}}
\newblock {A Survey of z{$>$}5.7 Quasars in the Sloan Digital Sky Survey. IV.
  Discovery of Seven Additional Quasars}.
\newblock {\em \aj}, 131:1203--1209, March 2006.

\bibitem{field58}
G.~B. {Field}.
\newblock {Excitation of the Hydrogen 21-cm Line,}.
\newblock {\em Proc. IRE}, 46:240--+, May 1958.

\bibitem{field59b}
G.~B. {Field}.
\newblock {The Spin Temperature of Intergalactic Neutral Hydrogen.}
\newblock {\em \apj}, 129:536--+, May 1959.

\bibitem{field59a}
G.~B. {Field}.
\newblock {The Time Relaxation of a Resonance-Line Profile.}
\newblock {\em \apj}, 129:551--+, May 1959.

\bibitem{furlanetto06b}
S.~R. {Furlanetto}.
\newblock {The 21-cm forest}.
\newblock {\em \mnras}, 370:1867--1875, August 2006.

\bibitem{furlanetto07}
S.~R. {Furlanetto} and M.~R. {Furlanetto}.
\newblock {Spin exchange rates in proton-hydrogen collisions}.
\newblock {\em \mnras}, 379:130--134, July 2007.

\bibitem{furlanetto02}
S.~R. {Furlanetto} and A.~{Loeb}.
\newblock {The 21 Centimeter Forest: Radio Absorption Spectra as Probes of
  Minihalos before Reionization}.
\newblock {\em \apj}, 579:1--9, November 2002.

\bibitem{furlanetto06a}
S.~R. {Furlanetto}, S.~P. {Oh}, and F.~H. {Briggs}.
\newblock {Cosmology at low frequencies: The 21 cm transition and the
  high-redshift Universe}.
\newblock {\em \physrep}, 433:181--301, October 2006.

\bibitem{furlanetto10}
S.~R. {Furlanetto} and S.~J. {Stoever}.
\newblock {Secondary ionization and heating by fast electrons}.
\newblock {\em \mnras}, 404:1869--1878, June 2010.

\bibitem{furlanetto04}
S.~R. {Furlanetto}, M.~{Zaldarriaga}, and L.~{Hernquist}.
\newblock {The Growth of H II Regions During Reionization}.
\newblock {\em \apj}, 613:1--15, September 2004.

\bibitem{giallongo02}
E.~{Giallongo}, S.~{Cristiani}, S.~{D'Odorico}, and A.~{Fontana}.
\newblock {A Low Upper Limit to the Lyman Continuum Emission of Two Galaxies at
  z\~{}=3}.
\newblock {\em \apjl}, 568:L9--L12, March 2002.

\bibitem{gleser08}
L.~{Gleser}, A.~{Nusser}, and A.~J. {Benson}.
\newblock {Decontamination of cosmological 21-cm maps}.
\newblock {\em \mnras}, 391:383--398, November 2008.

\bibitem{gleser06}
L.~{Gleser}, A.~{Nusser}, B.~{Ciardi}, and V.~{Desjacques}.
\newblock {The morphology of cosmological reionization by means of Minkowski
  functionals}.
\newblock {\em \mnras}, 370:1329--1338, August 2006.

\bibitem{gnedin01}
N.~Y. {Gnedin} and T.~{Abel}.
\newblock {Multi-dimensional cosmological radiative transfer with a Variable
  Eddington Tensor formalism}.
\newblock {\em New Astronomy}, 6:437--455, October 2001.

\bibitem{gunn65}
J.~E. {Gunn} and B.~A. {Peterson}.
\newblock {On the Density of Neutral Hydrogen in Intergalactic Space.}
\newblock {\em \apj}, 142:1633--1641, November 1965.

\bibitem{haiman00}
Z.~{Haiman}, M.~{Spaans}, and E.~{Quataert}.
\newblock {Ly{$\alpha$} Cooling Radiation from High-Redshift Halos}.
\newblock {\em \apjl}, 537:L5--L8, July 2000.

\bibitem{hamaker96}
J.~P. {Hamaker}, J.~D. {Bregman}, and R.~J. {Sault}.
\newblock {Understanding radio polarimetry. I. Mathematical foundations.}
\newblock {\em \aaps}, 117:137--147, May 1996.

\bibitem{harker10}
G.~{Harker}, S.~{Zaroubi}, G.~{Bernardi}, M.~A. {Brentjens}, A.~G. {de Bruyn},
  B.~{Ciardi}, V.~{Jeli{\'c}}, L.~V.~E. {Koopmans}, P.~{Labropoulos},
  G.~{Mellema}, A.~{Offringa}, V.~N. {Pandey}, A.~H. {Pawlik}, J.~{Schaye},
  R.~M. {Thomas}, and S.~{Yatawatta}.
\newblock {Power spectrum extraction for redshifted 21-cm Epoch of Reionization
  experiments: the LOFAR case}.
\newblock {\em \mnras}, 405:2492--2504, July 2010.

\bibitem{harker09b}
G.~{Harker}, S.~{Zaroubi}, G.~{Bernardi}, M.~A. {Brentjens}, A.~G. {de Bruyn},
  B.~{Ciardi}, V.~{Jeli{\'c}}, L.~V.~E. {Koopmans}, P.~{Labropoulos},
  G.~{Mellema}, A.~{Offringa}, V.~N. {Pandey}, J.~{Schaye}, R.~M. {Thomas}, and
  S.~{Yatawatta}.
\newblock {Non-parametric foreground subtraction for 21-cm epoch of
  reionization experiments}.
\newblock {\em \mnras}, 397:1138--1152, August 2009.

\bibitem{harker09a}
G.~J.~A. {Harker}, S.~{Zaroubi}, R.~M. {Thomas}, V.~{Jeli{\'c}},
  P.~{Labropoulos}, G.~{Mellema}, I.~T. {Iliev}, G.~{Bernardi}, M.~A.
  {Brentjens}, A.~G. {de Bruyn}, B.~{Ciardi}, L.~V.~E. {Koopmans}, V.~N.
  {Pandey}, A.~H. {Pawlik}, J.~{Schaye}, and S.~{Yatawatta}.
\newblock {Detection and extraction of signals from the epoch of reionization
  using higher-order one-point statistics}.
\newblock {\em \mnras}, 393:1449--1458, March 2009.

\bibitem{haverkorn03}
M.~{Haverkorn}, P.~{Katgert}, and A.~G. {de Bruyn}.
\newblock {Multi-frequency polarimetry of the Galactic radio background around
  350 MHz. I. A region in Auriga around l = 161 deg, b = 16 deg}.
\newblock {\em \aap}, 403:1031--1044, June 2003.

\bibitem{hernquist96}
L.~{Hernquist}, N.~{Katz}, D.~H. {Weinberg}, and J.~{Miralda-Escud{\'e}}.
\newblock {The Lyman-Alpha Forest in the Cold Dark Matter Model}.
\newblock {\em \apjl}, 457:L51+, February 1996.

\bibitem{hobson02}
M.~P. {Hobson} and K.~{Maisinger}.
\newblock {Maximum-likelihood estimation of the cosmic microwave background
  power spectrum from interferometer observations}.
\newblock {\em \mnras}, 334:569--588, August 2002.

\bibitem{hogan79}
C.~J. {Hogan} and M.~J. {Rees}.
\newblock {Spectral appearance of non-uniform gas at high Z}.
\newblock {\em \mnras}, 188:791--798, September 1979.

\bibitem{holder03}
G.~P. {Holder}, Z.~{Haiman}, M.~{Kaplinghat}, and L.~{Knox}.
\newblock {The Reionization History at High Redshifts. II. Estimating the
  Optical Depth to Thomson Scattering from Cosmic Microwave Background
  Polarization}.
\newblock {\em \apj}, 595:13--18, September 2003.

\bibitem{hu95}
W.~{Hu}.
\newblock {Wandering in the Background: A CMB Explorer}.
\newblock {\em ArXiv Astrophysics e-prints}, August 1995.

\bibitem{hu97}
W.~{Hu} and M.~{White}.
\newblock {A CMB polarization primer}.
\newblock {\em New Astronomy}, 2:323--344, October 1997.

\bibitem{hui97}
L.~{Hui} and N.~Y. {Gnedin}.
\newblock {Equation of state of the photoionized intergalactic medium}.
\newblock {\em \mnras}, 292:27--+, November 1997.

\bibitem{hui03}
L.~{Hui} and Z.~{Haiman}.
\newblock {The Thermal Memory of Reionization History}.
\newblock {\em \apj}, 596:9--18, October 2003.

\bibitem{ichikawa10}
K.~{Ichikawa}, R.~{Barkana}, I.~T. {Iliev}, G.~{Mellema}, and P.~R. {Shapiro}.
\newblock {Measuring the history of cosmic reionization using the 21-cm
  probability distribution function from simulations}.
\newblock {\em \mnras}, 406:2521--2532, August 2010.

\bibitem{iliev08}
I.~T. {Iliev}, G.~{Mellema}, {U.-L.} {Pen}, J.~R. {Bond}, and P.~R. {Shapiro}.
\newblock {Current models of the observable consequences of cosmic reionization
  and their detectability}.
\newblock {\em \mnras}, 384:863--874, March 2008.

\bibitem{inoue05}
A.~K. {Inoue}, I.~{Iwata}, J.-M. {Deharveng}, V.~{Buat}, and D.~{Burgarella}.
\newblock {VLT narrow-band photometry in the Lyman continuum of two galaxies at
  z \~ 3. Limits to the escape of ionizing flux}.
\newblock {\em \aap}, 435:471--482, May 2005.

\bibitem{iwata09}
I.~{Iwata}, A.~K. {Inoue}, Y.~{Matsuda}, H.~{Furusawa}, T.~{Hayashino},
  K.~{Kousai}, M.~{Akiyama}, T.~{Yamada}, D.~{Burgarella}, and J.-M.
  {Deharveng}.
\newblock {Detections of Lyman Continuum from Star-Forming Galaxies at z \~{} 3
  through Subaru/Suprime-Cam Narrow-Band Imaging}.
\newblock {\em \apj}, 692:1287--1293, February 2009.

\bibitem{jelic10a}
V.~{Jeli{\'c}}, S.~{Zaroubi}, N.~{Aghanim}, M.~{Douspis}, L.~V.~E. {Koopmans},
  M.~{Langer}, G.~{Mellema}, H.~{Tashiro}, and R.~M. {Thomas}.
\newblock {A cross-correlation study between the cosmological 21 cm signal and
  the kinetic Sunyaev-Zel'dovich effect}.
\newblock {\em \mnras}, 402:2279--2290, March 2010.

\bibitem{jelic10b}
V.~{Jeli{\'c}}, S.~{Zaroubi}, P.~{Labropoulos}, G.~{Bernardi}, A.~G. {de
  Bruyn}, and L.~V.~E. {Koopmans}.
\newblock {Realistic simulations of the Galactic polarized foreground:
  consequences for 21-cm reionization detection experiments}.
\newblock {\em \mnras}, 409:1647--1659, December 2010.

\bibitem{jelic08}
V.~{Jeli{\'c}}, S.~{Zaroubi}, P.~{Labropoulos}, R.~M. {Thomas}, G.~{Bernardi},
  M.~A. {Brentjens}, A.~G. {de Bruyn}, B.~{Ciardi}, G.~{Harker}, L.~V.~E.
  {Koopmans}, V.~N. b~{Pandey}, J.~{Schaye}, and S.~{Yatawatta}.
\newblock {Foreground simulations for the LOFAR-epoch of reionization
  experiment}.
\newblock {\em \mnras}, 389:1319--1335, September 2008.

\bibitem{kaiser87}
N.~{Kaiser}.
\newblock {Clustering in real space and in redshift space}.
\newblock {\em \mnras}, 227:1--21, July 1987.

\bibitem{kamionkowski97}
M.~{Kamionkowski}, A.~{Kosowsky}, and A.~{Stebbins}.
\newblock {A Probe of Primordial Gravity Waves and Vorticity}.
\newblock {\em Physical Review Letters}, 78:2058--2061, March 1997.

\bibitem{kashikawa06}
N.~{Kashikawa}, K.~{Shimasaku}, M.~A. {Malkan}, M.~{Doi}, Y.~{Matsuda},
  M.~{Ouchi}, Y.~{Taniguchi}, C.~{Ly}, T.~{Nagao}, M.~{Iye}, K.~{Motohara},
  T.~{Murayama}, K.~{Murozono}, K.~{Nariai}, K.~{Ohta}, S.~{Okamura},
  T.~{Sasaki}, Y.~{Shioya}, and M.~{Umemura}.
\newblock {The End of the Reionization Epoch Probed by Ly{$\alpha$} Emitters at
  z = 6.5 in the Subaru Deep Field}.
\newblock {\em \apj}, 648:7--22, September 2006.

\bibitem{kashlinsky05}
A.~{Kashlinsky}, R.~G. {Arendt}, J.~{Mather}, and S.~H. {Moseley}.
\newblock {Tracing the first stars with fluctuations of the cosmic infrared
  background}.
\newblock {\em \nat}, 438:45--50, November 2005.

\bibitem{kasuya04}
S.~{Kasuya} and M.~{Kawasaki}.
\newblock {Early reionization by decaying particles and cosmic microwave
  background radiation}.
\newblock {\em \prd}, 70(10):103519--+, November 2004.

\bibitem{kazemi11}
S.~{Kazemi}, S.~{Yatawatta}, S.~{Zaroubi}, P.~{Lampropoulos}, A.~G. {de Bruyn},
  L.~V.~E. {Koopmans}, and J.~{Noordam}.
\newblock {Radio interferometric calibration using the SAGE algorithm}.
\newblock {\em \mnras}, 414:1656--1666, June 2011.

\bibitem{kramer09}
R.~H. {Kramer} and Z.~{Haiman}.
\newblock {Probing re-ionization with quasar spectra: the impact of the
  intrinsic Lyman {$\alpha$} emission line shape uncertainty}.
\newblock {\em \mnras}, 400:1493--1511, December 2009.

\bibitem{panos09}
P.~{Labropoulos}, L.~V.~E. {Koopmans}, V.~{Jelic}, S.~{Yatawatta}, R.~M.
  {Thomas}, G.~{Bernardi}, M.~{Brentjens}, G.~{de Bruyn}, B.~{Ciardi},
  G.~{Harker}, A.~{Offringa}, V.~N. {Pandey}, J.~{Schaye}, and S.~{Zaroubi}.
\newblock {The LOFAR EoR Data Model: (I) Effects of Noise and Instrumental
  Corruptions on the 21-cm Reionization Signal-Extraction Strategy}.
\newblock {\em ArXiv e-prints}, January 2009.

\bibitem{landecker70}
T.~L. {Landecker} and R.~{Wielebinski}.
\newblock {The Galactic Metre Wave Radiation: A two-frequency survey between
  declinations $+25^{o}$ and $-25^{o}$ and the preparation of a map of the
  whole sky}.
\newblock {\em Australian Journal of Physics Astrophysical Supplement},
  16:1--+, 1970.

\bibitem{latif11c}
M.~A. {Latif}, D.~R.~G. {Schleicher}, M.~{Spaans}, and S.~{Zaroubi}.
\newblock {Lyman alpha emission from the first galaxies: Implications of UV
  backgrounds and the formation of molecules}.
\newblock {\em ArXiv e-prints}, June 2011.

\bibitem{latif11b}
M.~A. {Latif}, D.~R.~G. {Schleicher}, M.~{Spaans}, and S.~{Zaroubi}.
\newblock {Lyman {$\alpha$} emission from the first galaxies: signatures of
  accretion and infall in the presence of line trapping}.
\newblock {\em \mnras}, 413:L33--L37, May 2011.

\bibitem{latif11}
M.~A. {Latif}, S.~{Zaroubi}, and M.~{Spaans}.
\newblock {The impact of Lyman {$\alpha$} trapping on the formation of
  primordial objects}.
\newblock {\em \mnras}, 411:1659--1670, March 2011.

\bibitem{lauwrence07}
A.~{Lawrence}, S.~J. {Warren}, O.~{Almaini}, A.~C. {Edge}, N.~C. {Hambly},
  R.~F. {Jameson}, P.~{Lucas}, M.~{Casali}, A.~{Adamson}, S.~{Dye}, J.~P.
  {Emerson}, S.~{Foucaud}, P.~{Hewett}, P.~{Hirst}, S.~T. {Hodgkin}, M.~J.
  {Irwin}, N.~{Lodieu}, R.~G. {McMahon}, C.~{Simpson}, I.~{Smail},
  D.~{Mortlock}, and M.~{Folger}.
\newblock {The UKIRT Infrared Deep Sky Survey (UKIDSS)}.
\newblock {\em \mnras}, 379:1599--1617, August 2007.

\bibitem{lazio09}
J.~{Lazio}, C.~{Carilli}, J.~{Hewitt}, S.~{Furlanetto}, and J.~{Burns}.
\newblock {The lunar radio array (LRA)}.
\newblock In {\em Society of Photo-Optical Instrumentation Engineers (SPIE)
  Conference Series}, volume 7436 of {\em Society of Photo-Optical
  Instrumentation Engineers (SPIE) Conference Series}, August 2009.

\bibitem{lewis06}
A.~{Lewis}, J.~{Weller}, and R.~{Battye}.
\newblock {The cosmic microwave background and the ionization history of the
  Universe}.
\newblock {\em \mnras}, 373:561--570, December 2006.

\bibitem{lidz10}
A.~{Lidz}, C.-A. {Faucher-Gigu{\`e}re}, A.~{Dall'Aglio}, M.~{McQuinn},
  C.~{Fechner}, M.~{Zaldarriaga}, L.~{Hernquist}, and S.~{Dutta}.
\newblock {A Measurement of Small-scale Structure in the 2.2 <= z <= 4.2
  Ly{$\alpha$} Forest}.
\newblock {\em \apj}, 718:199--230, July 2010.

\bibitem{lidz02}
A.~{Lidz}, L.~{Hui}, M.~{Zaldarriaga}, and R.~{Scoccimarro}.
\newblock {How Neutral Is the Intergalactic Medium at z \~{} 6?}
\newblock {\em \apj}, 579:491--499, November 2002.

\bibitem{lidz09}
A.~{Lidz}, O.~{Zahn}, S.~R. {Furlanetto}, M.~{McQuinn}, L.~{Hernquist}, and
  M.~{Zaldarriaga}.
\newblock {Probing Reionization with the 21 cm Galaxy Cross-Power Spectrum}.
\newblock {\em \apj}, 690:252--266, January 2009.

\bibitem{liszt01}
H.~{Liszt}.
\newblock {The spin temperature of warm interstellar H I}.
\newblock {\em \aap}, 371:698--707, May 2001.

\bibitem{loeb01}
A.~{Loeb} and R.~{Barkana}.
\newblock {The First Light}.
\newblock {\em \araa}, 39:19--130, November 2001.

\bibitem{loeb04}
A.~{Loeb} and M.~{Zaldarriaga}.
\newblock {Measuring the Small-Scale Power Spectrum of Cosmic Density
  Fluctuations through 21cm Tomography Prior to the Epoch of Structure
  Formation}.
\newblock {\em Physical Review Letters}, 92(21):211301--+, May 2004.

\bibitem{ma95}
C.-P. {Ma} and E.~{Bertschinger}.
\newblock {Cosmological Perturbation Theory in the Synchronous and Conformal
  Newtonian Gauges}.
\newblock {\em \apj}, 455:7--+, December 1995.

\bibitem{machacek00}
M.~E. {Machacek}, G.~L. {Bryan}, A.~{Meiksin}, P.~{Anninos}, D.~{Thayer},
  M.~{Norman}, and Y.~{Zhang}.
\newblock {Hydrodynamical Simulations of the Ly{$\alpha$} Forest: Model
  Comparisons}.
\newblock {\em \apj}, 532:118--135, March 2000.

\bibitem{mack11}
K.~J. {Mack} and J.~S.~B. {Wyithe}.
\newblock {Detecting the redshifted 21cm forest during reionization}.
\newblock {\em ArXiv e-prints}, January 2011.

\bibitem{madau97}
P.~{Madau}, A.~{Meiksin}, and M.~J. {Rees}.
\newblock {21 Centimeter Tomography of the Intergalactic Medium at High
  Redshift}.
\newblock {\em \apj}, 475:429--+, February 1997.

\bibitem{mapelli05}
M.~{Mapelli} and A.~{Ferrara}.
\newblock {Background radiation from sterile neutrino decay and reionization}.
\newblock {\em \mnras}, 364:2--12, November 2005.

\bibitem{mapelli06}
M.~{Mapelli}, A.~{Ferrara}, and E.~{Pierpaoli}.
\newblock {Impact of dark matter decays and annihilations on reionization}.
\newblock {\em \mnras}, 369:1719--1724, July 2006.

\bibitem{maselli09}
A.~{Maselli}, A.~{Ferrara}, and S.~{Gallerani}.
\newblock {Interpreting the transmission windows of distant quasars}.
\newblock {\em \mnras}, 395:1925--1933, June 2009.

\bibitem{maselli07}
A.~{Maselli}, S.~{Gallerani}, A.~{Ferrara}, and T.~R. {Choudhury}.
\newblock {On the size of HII regions around high-redshift quasars}.
\newblock {\em \mnras}, 376:L34--L38, March 2007.

\bibitem{mclure10}
R.~J. {McLure}, J.~S. {Dunlop}, M.~{Cirasuolo}, A.~M. {Koekemoer}, E.~{Sabbi},
  D.~P. {Stark}, T.~A. {Targett}, and R.~S. {Ellis}.
\newblock {Galaxies at z = 6-9 from the WFC3/IR imaging of the Hubble Ultra
  Deep Field}.
\newblock {\em \mnras}, 403:960--983, April 2010.

\bibitem{mellema06}
G.~{Mellema}, I.~T. {Iliev}, M.~A. {Alvarez}, and P.~R. {Shapiro}.
\newblock {C$^{2}$-ray: A new method for photon-conserving transport of
  ionizing radiation}.
\newblock {\em New Astronomy}, 11:374--395, March 2006.

\bibitem{mesinger07}
A.~{Mesinger} and S.~{Furlanetto}.
\newblock {Efficient Simulations of Early Structure Formation and
  Reionization}.
\newblock {\em \apj}, 669:663--675, November 2007.

\bibitem{mesinger09}
A.~{Mesinger} and S.~{Furlanetto}.
\newblock {The inhomogeneous ionizing background following reionization}.
\newblock {\em \mnras}, 400:1461--1471, December 2009.

\bibitem{mesinger10}
A.~{Mesinger}, S.~{Furlanetto}, and R.~{Cen}.
\newblock {21cmFAST: A Fast, Semi-Numerical Simulation of the High-Redshift
  21-cm Signal}.
\newblock {\em ArXiv e-prints}, March 2010.

\bibitem{mesinger04}
A.~{Mesinger} and Z.~{Haiman}.
\newblock {Evidence of a Cosmological Str{\"o}mgren Surface and of Significant
  Neutral Hydrogen Surrounding the Quasar SDSS J1030+0524}.
\newblock {\em \apjl}, 611:L69--L72, August 2004.

\bibitem{miralda-Escude96}
J.~{Miralda-Escud{\'e}}, R.~{Cen}, J.~P. {Ostriker}, and M.~{Rauch}.
\newblock {The Ly alpha Forest from Gravitational Collapse in the Cold Dark
  Matter + Lambda Model}.
\newblock {\em \apj}, 471:582--+, November 1996.

\bibitem{mo10}
H.~{Mo}, F.~C. {van den Bosch}, and S.~{White}.
\newblock {\em {Galaxy Formation and Evolution}}.
\newblock 2010.

\bibitem{morales05}
M.~F. {Morales}.
\newblock {Power Spectrum Sensitivity and the Design of Epoch of Reionization
  Observatories}.
\newblock {\em \apj}, 619:678--683, February 2005.

\bibitem{morales04}
M.~F. {Morales} and J.~{Hewitt}.
\newblock {Toward Epoch of Reionization Measurements with Wide-Field Radio
  Observations}.
\newblock {\em \apj}, 615:7--18, November 2004.

\bibitem{morales10}
M.~F. {Morales} and J.~S.~B. {Wyithe}.
\newblock {Reionization and Cosmology with 21-cm Fluctuations}.
\newblock {\em \araa}, 48:127--171, September 2010.

\bibitem{moretti03}
A.~{Moretti}, S.~{Campana}, D.~{Lazzati}, and G.~{Tagliaferri}.
\newblock {The Resolved Fraction of the Cosmic X-Ray Background}.
\newblock {\em \apj}, 588:696--703, May 2003.

\bibitem{mortlock11}
D.~J. {Mortlock}, S.~J. {Warren}, B.~P. {Venemans}, M.~{Patel}, P.~C. {Hewett},
  R.~G. {McMahon}, C.~{Simpson}, T.~{Theuns}, E.~A. {Gonz{\'a}les-Solares},
  A.~{Adamson}, S.~{Dye}, N.~C. {Hambly}, P.~{Hirst}, M.~J. {Irwin},
  E.~{Kuiper}, A.~{Lawrence}, and H.~J.~A. {R{\"o}ttgering}.
\newblock {A luminous quasar at a redshift of z = 7.085}.
\newblock {\em \nat}, 474:616--619, June 2011.

\bibitem{mortonson08}
M.~J. {Mortonson} and W.~{Hu}.
\newblock {Model-Independent Constraints on Reionization from Large-Scale
  Cosmic Microwave Background Polarization}.
\newblock {\em \apj}, 672:737--751, January 2008.

\bibitem{muller51}
C.~A. {Muller} and J.~H. {Oort}.
\newblock {Observation of a Line in the Galactic Radio Spectrum: The
  Interstellar Hydrogen Line at 1,420 Mc./sec., and an Estimate of Galactic
  Rotation}.
\newblock {\em \nat}, 168:357--358, September 1951.

\bibitem{murayam07}
T.~{Murayama}, Y.~{Taniguchi}, N.~Z. {Scoville}, M.~{Ajiki}, D.~B. {Sanders},
  B.~{Mobasher}, H.~{Aussel}, P.~{Capak}, A.~{Koekemoer}, Y.~{Shioya},
  T.~{Nagao}, C.~{Carilli}, R.~S. {Ellis}, B.~{Garilli}, M.~{Giavalisco}, M.~G.
  {Kitzbichler}, O.~{Le F{\`e}vre}, D.~{Maccagni}, E.~{Schinnerer}, V.~{Smol{\v
  c}i{\'c}}, S.~{Tribiano}, A.~{Cimatti}, Y.~{Komiyama}, S.~{Miyazaki}, S.~S.
  {Sasaki}, J.~{Koda}, and H.~{Karoji}.
\newblock {Ly{$\alpha$} Emitters at Redshift 5.7 in the COSMOS Field}.
\newblock {\em \apjs}, 172:523--544, September 2007.

\bibitem{nakamoto01}
T.~{Nakamoto}, M.~{Umemura}, and H.~{Susa}.
\newblock {The effects of radiative transfer on the reionization of an
  inhomogeneous universe}.
\newblock {\em \mnras}, 321:593--604, March 2001.

\bibitem{natarajan10}
A.~{Natarajan} and D.~J. {Schwarz}.
\newblock {Distinguishing standard reionization from dark matter models}.
\newblock {\em \prd}, 81(12):123510--+, June 2010.

\bibitem{nijboer06}
R.~J. {Nijboer}, J.~E. {Noordam}, and S.~B. {Yatawatta}.
\newblock {LOFAR Self-Calibration using a Local Sky Model}.
\newblock In {C.~Gabriel, C.~Arviset, D.~Ponz, \& S.~Enrique}, editor, {\em
  Astronomical Data Analysis Software and Systems XV}, volume 351 of {\em
  Astronomical Society of the Pacific Conference Series}, pages 291--+, July
  2006.

\bibitem{nusser05}
A.~{Nusser}.
\newblock {The spin temperature of neutral hydrogen during cosmic
  pre-reionization}.
\newblock {\em \mnras}, 359:183--190, May 2005.

\bibitem{oesch10}
P.~A. {Oesch}, R.~J. {Bouwens}, G.~D. {Illingworth}, C.~M. {Carollo},
  M.~{Franx}, I.~{Labb{\'e}}, D.~{Magee}, M.~{Stiavelli}, M.~{Trenti}, and
  P.~G. {van Dokkum}.
\newblock {z \~{} 7 Galaxies in the HUDF: First Epoch WFC3/IR Results}.
\newblock {\em \apjl}, 709:L16--L20, January 2010.

\bibitem{offringa10a}
A.~R. {Offringa}, A.~G. {de Bruyn}, M.~{Biehl}, S.~{Zaroubi}, G.~{Bernardi},
  and V.~N. {Pandey}.
\newblock {Post-correlation radio frequency interference classification
  methods}.
\newblock {\em \mnras}, 405:155--167, June 2010.

\bibitem{offringa10b}
A.~R. {Offringa}, A.~G. {de Bruyn}, S.~{Zaroubi}, and M.~{Biehl}.
\newblock {A LOFAR RFI detection pipeline and its first results}.
\newblock {\em ArXiv e-prints}, July 2010.

\bibitem{ostriker86}
J.~P. {Ostriker} and E.~T. {Vishniac}.
\newblock {Generation of microwave background fluctuations from nonlinear
  perturbations at the ERA of galaxy formation}.
\newblock {\em \apjl}, 306:L51--L54, July 1986.

\bibitem{ouchi09}
M.~{Ouchi}, Y.~{Ono}, E.~{Egami}, T.~{Saito}, M.~{Oguri}, P.~J. {McCarthy},
  D.~{Farrah}, N.~{Kashikawa}, I.~{Momcheva}, K.~{Shimasaku}, K.~{Nakanishi},
  H.~{Furusawa}, M.~{Akiyama}, J.~S. {Dunlop}, A.~M.~J. {Mortier},
  S.~{Okamura}, M.~{Hayashi}, M.~{Cirasuolo}, A.~{Dressler}, M.~{Iye}, M.~J.
  {Jarvis}, T.~{Kodama}, C.~L. {Martin}, R.~J. {McLure}, K.~{Ohta},
  T.~{Yamada}, and M.~{Yoshida}.
\newblock {Discovery of a Giant Ly{$\alpha$} Emitter Near the Reionization
  Epoch}.
\newblock {\em \apj}, 696:1164--1175, May 2009.

\bibitem{ouchi10}
M.~{Ouchi}, K.~{Shimasaku}, H.~{Furusawa}, T.~{Saito}, M.~{Yoshida},
  M.~{Akiyama}, Y.~{Ono}, T.~{Yamada}, K.~{Ota}, N.~{Kashikawa}, M.~{Iye},
  T.~{Kodama}, S.~{Okamura}, C.~{Simpson}, and M.~{Yoshida}.
\newblock {Statistics of 207 Ly{$\alpha$} Emitters at a Redshift Near 7:
  Constraints on Reionization and Galaxy Formation Models}.
\newblock {\em \apj}, 723:869--894, November 2010.

\bibitem{padmanabhan05}
N.~{Padmanabhan} and D.~P. {Finkbeiner}.
\newblock {Detecting dark matter annihilation with CMB polarization: Signatures
  and experimental prospects}.
\newblock {\em \prd}, 72(2):023508--+, July 2005.

\bibitem{page07}
L.~{Page}, G.~{Hinshaw}, E.~{Komatsu}, M.~R. {Nolta}, D.~N. {Spergel}, C.~L.
  {Bennett}, C.~{Barnes}, R.~{Bean}, O.~{Dor{\'e}}, J.~{Dunkley}, M.~{Halpern},
  R.~S. {Hill}, N.~{Jarosik}, A.~{Kogut}, M.~{Limon}, S.~S. {Meyer},
  N.~{Odegard}, H.~V. {Peiris}, G.~S. {Tucker}, L.~{Verde}, J.~L. {Weiland},
  E.~{Wollack}, and E.~L. {Wright}.
\newblock {Three-Year Wilkinson Microwave Anisotropy Probe (WMAP) Observations:
  Polarization Analysis}.
\newblock {\em \apjs}, 170:335--376, June 2007.

\bibitem{partridge67}
R.~B. {Partridge} and P.~J.~E. {Peebles}.
\newblock {Are Young Galaxies Visible?}
\newblock {\em \apj}, 147:868--+, March 1967.

\bibitem{pawlik08}
A.~H. {Pawlik} and J.~{Schaye}.
\newblock {TRAPHIC - radiative transfer for smoothed particle hydrodynamics
  simulations}.
\newblock {\em \mnras}, 389:651--677, September 2008.

\bibitem{pearson84}
T.~J. {Pearson} and A.~C.~S. {Readhead}.
\newblock {Image Formation by Self-Calibration in Radio Astronomy}.
\newblock {\em \araa}, 22:97--130, 1984.

\bibitem{peebles93}
P.~J.~E. {Peebles}.
\newblock {\em {Principles of Physical Cosmology}}.
\newblock 1993.

\bibitem{peebles70}
P.~J.~E. {Peebles} and J.~T. {Yu}.
\newblock {Primeval Adiabatic Perturbation in an Expanding Universe}.
\newblock {\em \apj}, 162:815--+, December 1970.

\bibitem{pen09}
U.-L. {Pen}, T.-C. {Chang}, C.~M. {Hirata}, J.~B. {Peterson}, J.~{Roy},
  Y.~{Gupta}, J.~{Odegova}, and K.~{Sigurdson}.
\newblock {The GMRT EoR experiment: limits on polarized sky brightness at 150
  MHz}.
\newblock {\em \mnras}, 399:181--194, October 2009.

\bibitem{pritchard07}
J.~R. {Pritchard} and S.~R. {Furlanetto}.
\newblock {21-cm fluctuations from inhomogeneous X-ray heating before
  reionization}.
\newblock {\em \mnras}, 376:1680--1694, April 2007.

\bibitem{pritchard08}
J.~R. {Pritchard} and A.~{Loeb}.
\newblock {Evolution of the 21cm signal throughout cosmic history}.
\newblock {\em \prd}, 78(10):103511--+, November 2008.

\bibitem{rasera06}
Y.~{Rasera} and R.~{Teyssier}.
\newblock {The history of the baryon budget. Cosmic logistics in a hierarchical
  universe}.
\newblock {\em \aap}, 445:1--27, January 2006.

\bibitem{rauch98}
M.~{Rauch}.
\newblock {The Lyman Alpha Forest in the Spectra of QSOs}.
\newblock {\em \araa}, 36:267--316, 1998.

\bibitem{razoumov05}
A.~O. {Razoumov} and C.~Y. {Cardall}.
\newblock {Fully threaded transport engine: new method for multi-scale
  radiative transfer}.
\newblock {\em \mnras}, 362:1413--1417, October 2005.

\bibitem{richard06}
J.~{Richard}, R.~{Pell{\'o}}, D.~{Schaerer}, J.-F. {Le Borgne}, and J.-P.
  {Kneib}.
\newblock {Constraining the population of 6 <z < 10 star-forming galaxies with
  deep near-IR images of lensing clusters}.
\newblock {\em \aap}, 456:861--880, September 2006.

\bibitem{ricotti04a}
M.~{Ricotti} and J.~P. {Ostriker}.
\newblock {Reionization, chemical enrichment and seed black holes from the
  first stars: is Population III important?}
\newblock {\em \mnras}, 350:539--551, May 2004.

\bibitem{ricotti04b}
M.~{Ricotti} and J.~P. {Ostriker}.
\newblock {X-ray pre-ionization powered by accretion on the first black holes -
  I. A model for the WMAP polarization measurement}.
\newblock {\em \mnras}, 352:547--562, August 2004.

\bibitem{ricotti00}
M.~{Ricotti} and J.~M. {Shull}.
\newblock {Feedback from Galaxy Formation: Escaping Ionizing Radiation from
  Galaxies at High Redshift}.
\newblock {\em \apj}, 542:548--558, October 2000.

\bibitem{ripamonti07}
E.~{Ripamonti}, M.~{Mapelli}, and A.~{Ferrara}.
\newblock {The impact of dark matter decays and annihilations on the formation
  of the first structures}.
\newblock {\em \mnras}, 375:1399--1408, March 2007.

\bibitem{ripamonti08}
E.~{Ripamonti}, M.~{Mapelli}, and S.~{Zaroubi}.
\newblock {Radiation from early black holes - I. Effects on the neutral
  intergalactic medium}.
\newblock {\em \mnras}, 387:158--172, June 2008.

\bibitem{ritzerveld03}
J.~{Ritzerveld}, V.~{Icke}, and {E.-J.} {Rijkhorst}.
\newblock {Triangulating Radiation: Radiative Transfer on Unstructured Grids}.
\newblock {\em ArXiv Astrophysics e-prints}, December 2003.

\bibitem{rogers08}
A.~E.~E. {Rogers} and J.~D. {Bowman}.
\newblock {Spectral Index of the Diffuse Radio Background Measured from 100 TO
  200 MHz}.
\newblock {\em \aj}, 136:641--648, August 2008.

\bibitem{rudie12}
G.~C. {Rudie}, C.~C. {Steidel}, R.~F. {Trainor}, O.~{Rakic},
  M.~{Bogosavljevic}, M.~{Pettini}, N.~{Reddy}, A.~E. {Shapley}, D.~K. {Erb},
  and D.~R. {Law}.
\newblock {The Gaseous Environment of High-z Galaxies: Precision Measurements
  of Neutral Hydrogen in the Circumgalactic Medium of z \~{} 2-3 Galaxies in
  the Keck Baryonic Structure Survey}.
\newblock {\em ArXiv e-prints}, February 2012.

\bibitem{rybicki86}
G.~B. {Rybicki} and A.~P. {Lightman}.
\newblock {\em {Radiative Processes in Astrophysics}}.
\newblock June 1986.

\bibitem{salvaterra05a}
R.~{Salvaterra}, B.~{Ciardi}, A.~{Ferrara}, and C.~{Baccigalupi}.
\newblock {Reionization history from coupled cosmic microwave background/21-cm
  line data}.
\newblock {\em \mnras}, 360:1063--1068, July 2005.

\bibitem{salvaterra05b}
R.~{Salvaterra}, F.~{Haardt}, and A.~{Ferrara}.
\newblock {Cosmic backgrounds from miniquasars}.
\newblock {\em \mnras}, 362:L50--L54, September 2005.

\bibitem{santos05}
M.~G. {Santos}, A.~{Cooray}, and L.~{Knox}.
\newblock {Multifrequency Analysis of 21 Centimeter Fluctuations from the Era
  of Reionization}.
\newblock {\em \apj}, 625:575--587, June 2005.

\bibitem{santos10}
M.~G. {Santos}, L.~{Ferramacho}, M.~B. {Silva}, A.~{Amblard}, and A.~{Cooray}.
\newblock {Fast large volume simulations of the 21-cm signal from the
  reionization and pre-reionization epochs}.
\newblock {\em \mnras}, 406:2421--2432, August 2010.

\bibitem{schaye00}
J.~{Schaye}, T.~{Theuns}, M.~{Rauch}, G.~{Efstathiou}, and W.~L.~W. {Sargent}.
\newblock {The thermal history of the intergalactic medium}.
\newblock {\em \mnras}, 318:817--826, November 2000.

\bibitem{scheuer65}
P.~A.~G. {Scheuer}.
\newblock {A Sensitive Test for the Presence of Atomic Hydrogen in
  Intergalactic Space}.
\newblock {\em \nat}, 207:963--+, August 1965.

\bibitem{scott90}
D.~{Scott} and M.~J. {Rees}.
\newblock {The 21-cm line at high redshift: a diagnostic for the origin of
  large scale structure}.
\newblock {\em \mnras}, 247:510--+, December 1990.

\bibitem{shapely06}
A.~E. {Shapley}, C.~C. {Steidel}, M.~{Pettini}, K.~L. {Adelberger}, and D.~K.
  {Erb}.
\newblock {The Direct Detection of Lyman Continuum Emission from Star-forming
  Galaxies at z\~{}3}.
\newblock {\em \apj}, 651:688--703, November 2006.

\bibitem{shaver99}
P.~A. {Shaver}, R.~A. {Windhorst}, P.~{Madau}, and A.~G. {de Bruyn}.
\newblock {Can the reionization epoch be detected as a global signature in the
  cosmic background?}
\newblock {\em \aap}, 345:380--390, May 1999.

\bibitem{shull85}
J.~M. {Shull} and M.~E. {van Steenberg}.
\newblock {X-ray secondary heating and ionization in quasar emission-line
  clouds}.
\newblock {\em \apj}, 298:268--274, November 1985.

\bibitem{smirnov04}
O.~M. {Smirnov} and J.~E. {Noordam}.
\newblock {The LOFAR Global Sky Model: Some Design Challenges}.
\newblock In {F.~Ochsenbein, M.~G.~Allen, \& D.~Egret}, editor, {\em
  Astronomical Data Analysis Software and Systems (ADASS) XIII}, volume 314 of
  {\em Astronomical Society of the Pacific Conference Series}, pages 18--+,
  July 2004.

\bibitem{smith66}
F.~J. {Smith}.
\newblock {Hydrogen atom spin-change collisions}.
\newblock {\em \planss}, 14:929--+, October 1966.

\bibitem{soltan03}
A.~M. {So{\l}tan}.
\newblock {The diffuse X-ray background}.
\newblock {\em \aap}, 408:39--42, September 2003.

\bibitem{spergel07}
D.~N. {Spergel}, R.~{Bean}, O.~{Dor{\'e}}, M.~R. {Nolta}, C.~L. {Bennett},
  J.~{Dunkley}, G.~{Hinshaw}, N.~{Jarosik}, E.~{Komatsu}, L.~{Page}, H.~V.
  {Peiris}, L.~{Verde}, M.~{Halpern}, R.~S. {Hill}, A.~{Kogut}, M.~{Limon},
  S.~S. {Meyer}, N.~{Odegard}, G.~S. {Tucker}, J.~L. {Weiland}, E.~{Wollack},
  and E.~L. {Wright}.
\newblock {Three-Year Wilkinson Microwave Anisotropy Probe (WMAP) Observations:
  Implications for Cosmology}.
\newblock {\em \apjs}, 170:377--408, June 2007.

\bibitem{stark07}
D.~{Stark}, R.~{Ellis}, and J.~{Richard}.
\newblock {The Case for an Abundant Population of Feeble Lyman-alpha Emitting
  Galaxies at z > 8}.
\newblock In {\em American Astronomical Society Meeting Abstracts}, volume~38
  of {\em Bulletin of the American Astronomical Society}, pages 143.02--+,
  December 2007.

\bibitem{steidel01}
C.~C. {Steidel}, M.~{Pettini}, and K.~L. {Adelberger}.
\newblock {Lyman-Continuum Emission from Galaxies at Z \~{}= 3.4}.
\newblock {\em \apj}, 546:665--671, January 2001.

\bibitem{sugiyama95}
N.~{Sugiyama}.
\newblock {Cosmic Background Anisotropies in Cold Dark Matter Cosmology}.
\newblock {\em \apjs}, 100:281--+, October 1995.

\bibitem{sun09}
X.~H. {Sun} and W.~{Reich}.
\newblock {Simulated square kilometre array maps from Galactic 3D-emission
  models}.
\newblock {\em \aap}, 507:1087--1105, November 2009.

\bibitem{sun08}
X.~H. {Sun}, W.~{Reich}, A.~{Waelkens}, and T.~A. {En{\ss}lin}.
\newblock {Radio observational constraints on Galactic 3D-emission models}.
\newblock {\em \aap}, 477:573--592, January 2008.

\bibitem{sunyaev80}
R.~A. {Sunyaev} and I.~B. {Zeldovich}.
\newblock {The velocity of clusters of galaxies relative to the microwave
  background - The possibility of its measurement}.
\newblock {\em \mnras}, 190:413--420, February 1980.

\bibitem{sunyaev81}
R.~A. {Sunyaev} and I.~B. {Zeldovich}.
\newblock {Intergalactic gas in clusters of galaxies, the microwave background,
  and cosmology}.
\newblock {\em Astrophysics and Space Physics Reviews}, 1:1--60, 1981.

\bibitem{sunyaev72}
R.~A. {Sunyaev} and Y.~B. {Zeldovich}.
\newblock {The Observations of Relic Radiation as a Test of the Nature of X-Ray
  Radiation from the Clusters of Galaxies}.
\newblock {\em Comments on Astrophysics and Space Physics}, 4:173--+, November
  1972.

\bibitem{susa06}
H.~{Susa}.
\newblock {Smoothed Particle Hydrodynamics Coupled with Radiation Transfer}.
\newblock {\em \pasj}, 58:445--460, April 2006.

\bibitem{tashiro08}
H.~{Tashiro}, N.~{Aghanim}, M.~{Langer}, M.~{Douspis}, and S.~{Zaroubi}.
\newblock {The cross-correlation of the CMB polarization and the 21-cm line
  fluctuations from cosmic reionization}.
\newblock {\em \mnras}, 389:469--477, September 2008.

\bibitem{tashiro10}
H.~{Tashiro}, N.~{Aghanim}, M.~{Langer}, M.~{Douspis}, S.~{Zaroubi}, and
  V.~{Jelic}.
\newblock {Detectability of the 21-cm CMB cross-correlation from the epoch of
  reionization}.
\newblock {\em \mnras}, 402:2617--2625, March 2010.

\bibitem{tashiro11}
H.~{Tashiro}, N.~{Aghanim}, M.~{Langer}, M.~{Douspis}, S.~{Zaroubi}, and
  V.~{Jeli{\'c}}.
\newblock {Second order cross-correlation between kinetic Sunyaev-Zel'dovich
  effect and 21-cm fluctuations from the epoch of reionization}.
\newblock {\em \mnras}, pages 638--+, May 2011.

\bibitem{taylor99}
G.~B. {Taylor}, C.~L. {Carilli}, and R.~A. {Perley}, editors.
\newblock {\em {Synthesis Imaging in Radio Astronomy II}}, volume 180 of {\em
  Astronomical Society of the Pacific Conference Series}, 1999.

\bibitem{theuns98}
T.~{Theuns}, A.~{Leonard}, G.~{Efstathiou}, F.~R. {Pearce}, and P.~A. {Thomas}.
\newblock {P\^{}3M-SPH simulations of the Lyalpha forest}.
\newblock {\em \mnras}, 301:478--502, December 1998.

\bibitem{theuns02}
T.~{Theuns}, J.~{Schaye}, S.~{Zaroubi}, {T.-S.} {Kim}, P.~{Tzanavaris}, and
  B.~{Carswell}.
\newblock {Constraints on Reionization from the Thermal History of the
  Intergalactic Medium}.
\newblock {\em \apjl}, 567:L103--L106, March 2002.

\bibitem{theuns02b}
T.~{Theuns}, S.~{Zaroubi}, T.-S. {Kim}, P.~{Tzanavaris}, and R.~F. {Carswell}.
\newblock {Temperature fluctuations in the intergalactic medium}.
\newblock {\em \mnras}, 332:367--382, May 2002.

\bibitem{thomas08}
R.~M. {Thomas} and S.~{Zaroubi}.
\newblock {Time-evolution of ionization and heating around first stars and
  miniqsos}.
\newblock {\em \mnras}, 384:1080--1096, March 2008.

\bibitem{thomas11}
R.~M. {Thomas} and S.~{Zaroubi}.
\newblock {On the spin-temperature evolution during the epoch of reionization}.
\newblock {\em \mnras}, 410:1377--1390, January 2011.

\bibitem{thomas09}
R.~M. {Thomas}, S.~{Zaroubi}, B.~{Ciardi}, A.~H. {Pawlik}, P.~{Labropoulos},
  V.~{Jeli{\'c}}, G.~{Bernardi}, M.~A. {Brentjens}, A.~G. {de Bruyn}, G.~J.~A.
  {Harker}, L.~V.~E. {Koopmans}, G.~{Mellema}, V.~N. {Pandey}, J.~{Schaye}, and
  S.~{Yatawatta}.
\newblock {Fast large-scale reionization simulations}.
\newblock {\em \mnras}, 393:32--48, February 2009.

\bibitem{thompson01}
A.~R. {Thompson}, J.~M. {Moran}, and G.~W. {Swenson}, Jr.
\newblock {\em {Interferometry and Synthesis in Radio Astronomy, 2nd Edition}}.
\newblock 2001.

\bibitem{valdes10}
M.~{Vald{\'e}s}, C.~{Evoli}, and A.~{Ferrara}.
\newblock {Particle energy cascade in the intergalactic medium}.
\newblock {\em \mnras}, 404:1569--1582, May 2010.

\bibitem{hulst45}
H.C. {van de Hulst}.
\newblock {}.
\newblock {\em Nederlands Tijdschrift voor Natuuurkunde}, 11:210--221, October
  1945.

\bibitem{waelkens09}
A.~{Waelkens}, T.~{Jaffe}, M.~{Reinecke}, F.~S. {Kitaura}, and T.~A.
  {En{\ss}lin}.
\newblock {Simulating polarized Galactic synchrotron emission at all
  frequencies. The Hammurabi code}.
\newblock {\em \aap}, 495:697--706, February 2009.

\bibitem{whalen06}
D.~{Whalen} and M.~L. {Norman}.
\newblock {A Multistep Algorithm for the Radiation Hydrodynamical Transport of
  Cosmological Ionization Fronts and Ionized Flows}.
\newblock {\em \apjs}, 162:281--303, February 2006.

\bibitem{wild52}
J.~P. {Wild}.
\newblock {The Radio-Frequency Line Spectrum of Atomic Hydrogen and its
  Applications in Astronomy.}
\newblock {\em \apj}, 115:206--+, March 1952.

\bibitem{wilman08}
R.~J. {Wilman}, L.~{Miller}, M.~J. {Jarvis}, T.~{Mauch}, F.~{Levrier}, F.~B.
  {Abdalla}, S.~{Rawlings}, H.-R. {Kl{\"o}ckner}, D.~{Obreschkow},
  D.~{Olteanu}, and S.~{Young}.
\newblock {A semi-empirical simulation of the extragalactic radio continuum sky
  for next generation radio telescopes}.
\newblock {\em \mnras}, 388:1335--1348, August 2008.

\bibitem{wood00}
K.~{Wood} and A.~{Loeb}.
\newblock {Escape of Ionizing Radiation from High-Redshift Galaxies}.
\newblock {\em \apj}, 545:86--99, December 2000.

\bibitem{wouthuysen52}
S.~A. {Wouthuysen}.
\newblock {On the excitation mechanism of the 21-cm (radio-frequency)
  interstellar hydrogen emission line.}
\newblock {\em \aj}, 57:31--32, 1952.

\bibitem{wyithe08}
J.~S.~B. {Wyithe}, J.~S. {Bolton}, and M.~G. {Haehnelt}.
\newblock {Reionization bias in high-redshift quasar near-zones}.
\newblock {\em \mnras}, 383:691--704, January 2008.

\bibitem{wyithe04a}
J.~S.~B. {Wyithe} and A.~{Loeb}.
\newblock {A characteristic size of \~{}10Mpc for the ionized bubbles at the
  end of cosmic reionization}.
\newblock {\em \nat}, 432:194--196, November 2004.

\bibitem{wyithe04c}
J.~S.~B. {Wyithe} and A.~{Loeb}.
\newblock {A large neutral fraction of cosmic hydrogen a billion years after
  the Big Bang}.
\newblock {\em \nat}, 427:815--817, February 2004.

\bibitem{xu09}
Y.~{Xu}, X.~{Chen}, Z.~{Fan}, H.~{Trac}, and R.~{Cen}.
\newblock {The 21 cm Forest as a Probe of the Reionization and The Temperature
  of the Intergalactic Medium}.
\newblock {\em \apj}, 704:1396--1404, October 2009.

\bibitem{yatawatta09}
S.~{Yatawatta}, S.~{Zaroubi}, G.~{de Bruyn}, L.~{Koopmans}, and J.~{Noordam}.
\newblock {Radio Interferometric Calibration Using The SAGE Algorithm}.
\newblock In {\em Digital Signal Processing Workshop and 5th IEEE Signal
  Processing Education Workshop, 2009. DSP/SPE 2009. IEEE 13th}, January 2009.

\bibitem{yoshida03}
N.~{Yoshida}, T.~{Abel}, L.~{Hernquist}, and N.~{Sugiyama}.
\newblock {Simulations of Early Structure Formation: Primordial Gas Clouds}.
\newblock {\em \apj}, 592:645--663, August 2003.

\bibitem{zahn07}
O.~{Zahn}, A.~{Lidz}, M.~{McQuinn}, S.~{Dutta}, L.~{Hernquist},
  M.~{Zaldarriaga}, and S.~R. {Furlanetto}.
\newblock {Simulations and Analytic Calculations of Bubble Growth during
  Hydrogen Reionization}.
\newblock {\em \apj}, 654:12--26, January 2007.

\bibitem{zahn11}
O.~{Zahn}, A.~{Mesinger}, M.~{McQuinn}, H.~{Trac}, R.~{Cen}, and L.~E.
  {Hernquist}.
\newblock {Comparison of reionization models: radiative transfer simulations
  and approximate, seminumeric models}.
\newblock {\em \mnras}, 414:727--738, June 2011.

\bibitem{zaldarriaga97}
M.~{Zaldarriaga}.
\newblock {Polarization of the microwave background in reionized models}.
\newblock {\em \prd}, 55:1822--1829, February 1997.

\bibitem{zaldarriaga02}
M.~{Zaldarriaga}.
\newblock {Searching for Fluctuations in the Intergalactic Medium Temperature
  Using the Ly{$\alpha$} Forest}.
\newblock {\em \apj}, 564:153--161, January 2002.

\bibitem{zaldarriaga04}
M.~{Zaldarriaga}, S.~R. {Furlanetto}, and L.~{Hernquist}.
\newblock {21 Centimeter Fluctuations from Cosmic Gas at High Redshifts}.
\newblock {\em \apj}, 608:622--635, June 2004.

\bibitem{zaldarriaga97b}
M.~{Zaldarriaga} and U.~{Seljak}.
\newblock {All-sky analysis of polarization in the microwave background}.
\newblock {\em \prd}, 55:1830--1840, February 1997.

\bibitem{zaroubi10}
S.~{Zaroubi}.
\newblock {Probing the Epoch of Reionization with Low Frequency Arrays}.
\newblock In {S.A. Torchinsky, A. van Ardenne, T. van den Brink-Havinga, A. van
  Es, A.J. Faulkner}, editor, {\em Widefield Science and Technology for the
  SKA}, page~75, 2010.

\bibitem{zaroubi05}
S.~{Zaroubi} and J.~{Silk}.
\newblock {LOFAR as a probe of the sources of cosmological reionization}.
\newblock {\em \mnras}, 360:L64--L67, June 2005.

\bibitem{zaroubi07}
S.~{Zaroubi}, R.~M. {Thomas}, N.~{Sugiyama}, and J.~{Silk}.
\newblock {Heating of the intergalactic medium by primordial miniquasars}.
\newblock {\em \mnras}, 375:1269--1279, March 2007.

\bibitem{zhang06}
L.~{Zhang}, X.~{Chen}, Y.-A. {Lei}, and Z.-G. {Si}.
\newblock {Impacts of dark matter particle annihilation on recombination and
  the anisotropies of the cosmic microwave background}.
\newblock {\em \prd}, 74(10):103519--+, November 2006.

\bibitem{zhang95}
Y.~{Zhang}, P.~{Anninos}, and M.~L. {Norman}.
\newblock {A Multispecies Model for Hydrogen and Helium Absorbers in
  Lyman-Alpha Forest Clouds}.
\newblock {\em \apjl}, 453:L57+, November 1995.

\bibitem{zhang97}
Y.~{Zhang}, P.~{Anninos}, M.~L. {Norman}, and A.~{Meiksin}.
\newblock {Spectral Analysis of the Ly alpha Forest in a Cold Dark Matter
  Cosmology}.
\newblock {\em \apj}, 485:496--+, August 1997.

\bibitem{zygelman05}
B.~{Zygelman}.
\newblock {Hyperfine Level-changing Collisions of Hydrogen Atoms and Tomography
  of the Dark Age Universe}.
\newblock {\em \apj}, 622:1356--1362, April 2005.

\end{thebibliography}
\end{document}